\documentclass{elsart}

\usepackage{amsmath,amssymb,epsfig,bm}

\numberwithin{equation}{subsection}

\journal{Physics Reports}

\begin{document}



\newcommand{\eV}{\text{eV}}
\newcommand{\GeV}{\text{GeV}}
\newcommand{\arcmin}{\text{arcmin}}
\newcommand{\Mpc}{\text{Mpc}}
\newcommand{\Hunit}{$\text{km\,s}^{-1}\,\text{Mpc}^{-1}$}
\newcommand{\muK}{$\mu\text{K}$}


\newcommand{\D}{\text{D}}
\newcommand{\ud}{\text{d}}
\newcommand{\curl}{\text{curl}}


\newcommand{\alt}{\lesssim}
\newcommand{\agt}{\gtrsim}


\newcommand{\cla}{\mathcal{A}}
\newcommand{\clb}{\mathcal{B}}
\newcommand{\clc}{\mathcal{C}}
\newcommand{\cld}{\mathcal{D}}
\newcommand{\cle}{\mathcal{E}}
\newcommand{\clf}{\mathcal{F}}
\newcommand{\clg}{\mathcal{G}}
\newcommand{\clh}{\mathcal{H}}
\newcommand{\cli}{\mathcal{I}}
\newcommand{\clj}{\mathcal{J}}
\newcommand{\clk}{\mathcal{K}}
\newcommand{\cll}{\mathcal{L}}
\newcommand{\clm}{\mathcal{M}}
\newcommand{\cln}{\mathcal{N}}
\newcommand{\clo}{\mathcal{O}}
\newcommand{\clp}{\mathcal{P}}
\newcommand{\clq}{\mathcal{Q}}
\newcommand{\clr}{\mathcal{R}}
\newcommand{\cls}{\mathcal{S}}
\newcommand{\clt}{\mathcal{T}}
\newcommand{\clu}{\mathcal{U}}
\newcommand{\clv}{\mathcal{V}}
\newcommand{\clw}{\mathcal{W}}
\newcommand{\clx}{\mathcal{X}}
\newcommand{\cly}{\mathcal{Y}}
\newcommand{\clz}{\mathcal{Z}}


\newcommand{\ve}{\mathbf{e}}
\newcommand{\vehat}{\hat{\mathbf{e}}}
\newcommand{\vn}{\mathbf{n}}
\newcommand{\vnhat}{\hat{\mathbf{n}}}
\newcommand{\vv}{\mathbf{v}}
\newcommand{\vx}{\mathbf{x}}
\newcommand{\vp}{\mathbf{p}}
\newcommand{\vk}{\mathbf{k}}


\newcommand{\Omtot}{\Omega_{\mathrm{tot}}}
\newcommand{\Omb}{\Omega_{\mathrm{b}}}
\newcommand{\Omc}{\Omega_{\mathrm{c}}}
\newcommand{\Omm}{\Omega_{\mathrm{m}}}
\newcommand{\omb}{\omega_{\mathrm{b}}}
\newcommand{\omc}{\omega_{\mathrm{c}}}
\newcommand{\omm}{\omega_{\mathrm{m}}}
\newcommand{\omnu}{\omega_{\nu}}
\newcommand{\Omnu}{\Omega_{\nu}}
\newcommand{\Oml}{\Omega_\Lambda}
\newcommand{\OmK}{\Omega_K}




\newcommand{\Al}{{A_l}}
\newcommand{\TT}{\text{TT}}


\newcommand{\Bc}{\vec{\mathcal B}_k}
\newcommand{\Ec}{\vec{\mathcal E}_k}


\newcommand{\sigt}{\sigma_{\mbox{\scriptsize T}}}


\newcommand{\annphys}{\rm Ann.~Phys.~}
\newcommand{\araa}{\rm Ann.~Rev.~Astron.~\&~Astrophys.~}
\newcommand{\aap}{\rm Astron.~\&~Astrophys.~}
\newcommand{\aaps}{\rm Astron.~\&~Astrophys.~Suppl.~}
\newcommand{\apj}{\rm Astrophys.~J.~}
\newcommand{\apjl}{\rm Astrophys.~J.~Lett.~}
\newcommand{\apjs}{\rm Astrophys.~J.~Supp.~}
\newcommand{\apss}{\rm Astrophys.~Space~Sci.~}
\newcommand{\an}{Astron. Nach.}
\newcommand{\cqg}{\rm Class.~Quant.~Grav.~}
\newcommand{\cmp}{\rm Commun.~Math.~Phys.~}
\newcommand{\EPJC}{\rm Eur.~Phys.~J.~C~}
\newcommand{\grg}{\rm Gen.~Rel.~Grav.~}
\newcommand{\ijmpb}{\rm Int.~J.~Mod.~Phys.~B~}
\newcommand{\ijmpd}{\rm Int.~J.~Mod.~Phys.~D~}
\newcommand{\jcap}{\rm JCAP~}
\newcommand{\jhep}{\rm JHEP~}
\newcommand{\jetpl}{\rm J.~Exp.~Theor.~Phys.~Lett.~}
\newcommand{\jfm}{\rm J.~Fluid~Mech.~}
\newcommand{\jmp}{\rm J.~Math.~Phys.~}
\newcommand{\jpcs}{\rm J.~Phys.~Conf.~Ser.~}
\newcommand{\jpg}{\rm J.~Phys.~G~}
\newcommand{\mnras}{\rm Mon.~Not.~R.~Astron.~Soc.~}
\newcommand{\nat}{\rm Nature~}
\newcommand{\npb}{\rm Nucl.~Phys.~B~}
\newcommand{\plb}{\rm Phys.~Lett.~B~}
\newcommand{\prd}{\rm Phys.~Rev.~D~}
\newcommand{\pre}{\rm Phys.~Rev.~E~}
\newcommand{\prl}{\rm Phys.~Rev.~Lett.~}
\newcommand{\physrep}{\rm Phys.~Rep.~}
\newcommand{\progthp}{\rm Prog.~Theor.~Phys.~}
\newcommand{\rmp}{\rm Rev.~Mod.~Phys.~}

\let\Oldsection\section
\renewcommand{\section}[1]{\Oldsection{\bf #1}}

\newcommand{\be} {\begin{equation}}
\newcommand{\ee} {\end{equation}}
\newcommand{\bea} {\begin{eqnarray}}
\newcommand{\eea} {\end{eqnarray}}


\begin{frontmatter}


\title{Primordial magnetogenesis}


\author{Alejandra Kandus}
\address{LATO-DCET, Universidade Estadual de Santa Cruz, Rodovia Ilh\'eus-Itabuna km 16 s/n, Salobrinho CEP-05508-900,
Ilh\'eus - BA, Brazil}

\author{Kerstin E. Kunze}
\address{Departmento de F\'isica Fundamental and  IUFFyM, Universidad de Salamanca, Plaza de la Merced s/n, E-37008 Salamanca, Spain}

\author{Christos G. Tsagas}
\address{Section of Astrophysics, Astronomy and Mechanics, Department of Physics, Aristotle University of Thessaloniki, Thessaloniki 54124, Greece.}\vspace{1cm}

\tableofcontents

\newpage

\begin{abstract}
Magnetic fields appear everywhere in the universe. From stars and  galaxies, all the way to galaxy clusters and remote protogalactic clouds, magnetic fields of considerable strength and size have been repeatedly observed. Despite their widespread presence, however, the origin of cosmic magnetic fields is still a mystery. The galactic dynamo is believed capable of amplifying weak magnetic seeds to strengths like those measured in ours and other galaxies. But the question is where do these seed fields come from? Are they a product of late, post-recombination, physics or are they truly cosmological in origin? The idea of primordial magnetism is attractive because it makes the large-scale magnetic fields, especially those found in early protogalactic systems, easier to explain. As a result, a host of different scenarios have appeared in the literature. Nevertheless, early magnetogenesis is not problem free, with a number of issues remaining open and matter of debate. We review the question of the origin of primordial magnetic fields and consider the limits set on their strength by the current observational data. The various mechanisms of pre-recombination magnetogenesis are presented and their advantages and shortcomings are debated. We consider both classical and quantum scenarios, that operate within as well as outside the standard model, and also discuss how future observations could be used to decide whether the large-scale magnetic fields we see in the universe today are truly primordial or not.
\end{abstract}


\end{frontmatter}

\newpage


\section{Introduction}\label{sI}
Observations have well established the widespread presence of magnetic fields in the universe~\cite{1988A&A...190...41K,%
1989Natur.341..720K,1994RPPh...57..325K,2002ChJAA...2..293H,%
2002ARA&A..40..319C,2004NewAR..48..763V,2005A&A...444..739B}. In fact, as the technology and the detection methods improve, it seems that magnetic fields are everywhere. The Milky Way, for example, possesses a coherent $B$-field of $\mu$G strength over the plain of its disc. These fields are a very important component of the interstellar medium, since they govern the gas-cloud dynamics, determine the energy of cosmic rays and affect star formation. Similar magnetic fields have also been detected in other spiral and barred galaxies. Cosmic magnetism is not confided to galaxies only however. Observations have repeatedly verified $B$-fields of $\mu$G-order strength in galaxy clusters and also in high redshift protogalactic structures. Recently, in particular, Kronberg~et~al and Bernet~et~al reported organized, strong $B$-fields in galaxies with redshifts close to 1.3~\cite{2008ApJ...676...70K,%
2008Natur.454..302B}. Also, Wolfe~et~al have detected a coherent magnetic field of approximately 100~$\mu$G in a galaxy at $z\simeq0.7$~\cite{2008Natur.455..638W}. All these seem to suggest that magnetic fields similar to that of the Milky Way are common in remote, high-redshift galaxies. This could imply that the time needed by the galactic dynamo to build up a coherent $B$-field is considerably less than what is usually anticipated. On the other hand, the widespread presence of magnetic fields at high redshifts may simply mean that they are cosmological (pre-recombination) in origin. Although it is still too early to reach a conclusion, the idea of primordial magnetism gains ground, as more fields of micro-Gauss strength are detected in remote protogalaxies. Further support may also come from very recent reports indicating the presence of coherent magnetic fields in the low density intergalactic space, where typical dynamo mechanisms cannot operate, with strengths close to $10^{-15}$~G~\cite{2010MNRAS.406L..70T,%
2010ApJ...722L..39A,2010Sci...328...73N,2010arXiv1009.1048T,%
2010arXiv1012.5313E}. The measurements of~\cite{2010ApJ...722L..39A}, in particular, are based on halos detected around Active Galactic Nuclei (AGN) observed by the Fermi Gamma-Ray Space Telescope. Complementary studies seem to limit the strength of these $B$-fields between $\sim10^{-17}$ and $\sim10^{-14}$ Gauss~\cite{2010arXiv1012.5313E}. Analogous lower limits were also reported by~\cite{2010MNRAS.406L..70T,%
2010arXiv1009.1048T} and~\cite{2010Sci...328...73N}, after measuring radiation in the GeV band ($\gamma$-rays) produced by the interaction of TeV photons from distant blazars with those of the Cosmic Microwave Background (CMB). If supported by future surveys, these measurements will render considerable credence to the idea of primordial magnetism. It is possible, however, that the matter will not settle unless magnetic imprints are found in the CMB spectrum~\cite{2006CQGra..23R...1G}.

Among the attractive aspects of cosmological magnetic fields, is that they can in principle explain all the large-scale fields seen in the universe today~\cite{1964MNRAS.128..345P,%
1997ApJ...483..648H,1998IJMPD...7..331E,2001PhR...348..163G,%
2002RvMP...74..775W,2004IJMPD..13..391G}. Nevertheless, early magnetogenesis is not problem free. The galactic dynamo needs an initial magnetic field in order to operate. Such seed fields must satisfy two basic requirements related to their coherence scale and strength~\cite{1978mfge.book.....M,1979cmft.book.....P,%
1983flma....3.....Z,1992ApJ...396..606K,1993IAUS..157..487K,%
1997A&A...322...98H,2005PhR...417....1B}. The former should not drop below 10~Kpc, otherwise it will destabilise the dynamo. The latter typically varies between $10^{-12}$ and $10^{-22}$~Gauss. It is conceivable, however, that in open, or dark-energy dominated, Friedmann-Robertson-Walker (FRW) cosmologies the minimum required magnetic strength could be pushed down to $\sim10^{-30}$~G~\cite{1999PhRvD..60b1301D}. Producing magnetic seeds that comply with the above mentioned specifications, however, has so far proved a rather difficult theoretical exercise. There are problems with both the scale and the strength of the initial field. Roughly speaking, magnetic seeds generated between inflation and recombination have too small coherence lengths. The reason is causality, which confines the scale of the $B$-field within the size of the horizon at the time of magnetogenesis. This is typically well below the dynamo requirements. If we generate the seed at the electroweak phase transition, for example, the size of the horizon is close to that of the Solar System. Assuming that some degree of turbulence existed in the pre-recombination plasma, one can increase the coherence scale of the initial field by appealing to a mechanism that in hydrodynamics is known as `inverse cascade'. In magnetohydrodynamics (MHD), the process results from the conservation of magnetic helicity and effectively transfers magnetic energy from small to successively larger scales. The drawback is that inverse cascade seems to require rather large amounts of magnetic helicity in order to operate efficiently~\cite{1996PhRvD..54.1291B,1997PhRvD..56.6146C,%
1999PhRvD..59f3008S,2001PhRvE..64e6405C}. Inflation can solve the scale problem, since it naturally creates superhorizon-sized correlations. There, however, we have a serious strength issue. Magnetic fields that  were generated during  a period of typical de Sitter-type inflation are thought to be too weak to seed the galactic dynamo~\cite{1988PhRvD..37.2743T}. The solution to the strength problem is usually sought outside the realm of classical electromagnetic theory, or of conventional FRW cosmology. There is a plethora of articles that do exactly that, although mechanisms operating within standard electromagnetism and the Friedmann models have also been reported in the literature.

The aim of this review is to present the various mechanisms of early magnetogenesis, outline their basic features and discuss their advantages and weaknesses. The next section starts with a brief overview of the observation techniques that have established the ubiquitous presence of large-scale magnetic fields in the universe. We then provide the limits on cosmological $B$-fields, imposed by primordial nucleosynthesis and the isotropy of the CMB. Section three sets the mathematical framework for the study of large-scale magnetic fields in relativistic cosmological models. There, for completeness, we also outline the typical magnetic effects on structure formation and how the latter could have backreacted on the $B$-field itself. In section four we discuss cosmic magnetogenesis within the realm of standard electrodynamics and within the limits of conventional Friedmannian cosmology. After a brief review of the FRW dynamics, we explain why it is theoretically difficult to generate and sustain astrophysically relevant $B$-fields in these models. More specifically, why inflationary magnetic fields in FRW cosmologies are generally expected to have residual strengths less than $10^{-50}$~G (far below the current galactic dynamo requirements) at the epoch of galaxy formation. At the same time, it is also pointed out that the standard picture can change when certain general relativistic aspects of the magnetic evolution are accounted for. More specifically, it is shown that curvature effects can in principle slow down the standard `adiabatic' magnetic decay and thus lead to $B$-fields with residual strengths much stronger than previously anticipated.

In section five we describe the generation of magnetic fields by nonlinear and out-of-equilibrium processes, which are believed to have taken place in the early universe. We begin by analyzing several mechanisms of magnetogenesis that could have operated during the reheating epoch of the universe, namely parametric resonance, generation of stochastic electric currents and the breaking of the conformal invariance of the electromagnetic field by cosmological perturbations. Then, we address the generation of magnetic fields during cosmological phase transitions. It is believed that at least two of such phase transitions have occurred in the early universe: the EW (Electroweak) and the QCD (Quantum Chromodynamical). In general, the problem with the (post-inflationary) early-universe magnetogenesis is that the generated $B$-fields have high intensity but very short coherent scale (in contrast to what happens during inflation), which amounts to performing certain line averages to obtain the desired large-scale intensities. This procedure generally results in weak magnetic fields. To a certain extent, the uncertainty in the obtained residual magnetic values reflects our limited knowledge of the dissipative processes operating at those times. Thus, better understanding of the reheating physics is required, if we are to make more precise predictions. Phase transitions, on the other hand, are better understood. Note, however, that despite the fact that the EW phase transition in the standard model is a second order process, extensions to other particle physics models treat it as first order. The QCD phase transition, on the other hand, was recently established to be a smooth crossover~\cite{2006Natur.443..675A}. To the best of our knowledge, however, no work on primordial magnetogenesis in this scenario has been reported in the literature.

Section six provides an overview of magnetic generation mechanisms operating outside the standard model. In all scenarios the magnetic fields are created in the very early universe, during inflation. Then, subhorizon quantum fluctuations in the electromagnetic field become classical superhorizon perturbations, manifesting themselves as current supported magnetic fields during the subsequent epochs of standard cosmology. In order to overcome the problem of not creating strong enough magnetic seeds, which is known to plague standard electrodynamics, different theories are explored. There are two basic classes of models, depending on whether electrodynamics is linear or nonlinear. In the first case, magnetic fields of astrophysically relevant strengths are usually achieved after breaking the conformal invariance of electromagnetism. This can be achieved by coupling the electromagnetic with a scalar field (as it naturally happens with the dilaton in string cosmology), by introducing dynamical extra dimensions, through quantum corrections leading to coupling with the curvature tensor, by inducing symmetry breaking, or by means of trace anomaly. When dealing with nonlinear electrodynamics, on the other hand, the conformal invariance of Maxwell's equations is naturally broken in four dimensions. Recall that the concept of nonlinear electrodynamics was first introduced by Born, in his search for a classical, singularity-free theory of the electron. Another example is provided by the description of virtual electron pair creation, which induces a self-coupling of the electromagnetic field. Linear and nonlinear models of electrodynamics are discussed in detail and the parameter space, for which strong enough magnetic seeds are generated, is determined.

Finally, in section seven, we briefly summarise the current state of research on primordial magnetogenesis and take look at the future expectations.

\newpage

\section{Magnetic fields in the universe}\label{sMFU}
Magnetic fields have long established their ubiquitous presence in the universe. They are a major component of the interstellar medium, contributing to the total pressure, affecting the gas dynamics, the distribution of cosmic rays and star formation. It also seems very likely that large-scale magnetic fields have played a fundamental role during the formation of galaxy clusters. Despite our increasing knowledge, however, many key questions related to the origin and the role of these fields remain as yet unanswered.

\subsection{Large-scale magnetic fields in the
universe}\label{ssLsMFU}
Most galaxies, including the Milky Way, carry coherent large-scale magnetic fields of $\mu$G-order strength. Analogous fields have also been detected in galaxy clusters and in young, high-redsift protogalactic structures. In short, the deeper we look for magnetic fields in the universe, the more widespread we find them to be.

\subsubsection{Detection and measuring methods}\label{sssDMM}
The key to magnetic detection is polarized emission at the optical, the infrared, the submillimeter and the radio wavelengths. Optical polarization is due to extinction along the line of sight, caused by elongated dust grains aligned by the interstellar magnetic field. The net result is that the electromagnetic signal has a polarisation direction parallel to the intervening $B$-field. This physical mechanism is sometimes referred to as the Davis-Greenstein effect~\cite{1951ApJ...114..206D}. Although optical polarization is of limited value, it has unveiled the magnetic structure in the spiral arms of the Milky Way and in other nearby galaxies~\cite{1970MmRAS..74..139M,1987MNRAS.224..299S,%
1991MNRAS.249P..16S,1998A&A...335..123F}.

Most of our knowledge about galactic and intergalactic magnetic fields comes from radio-wave signals. The intensity of synchrotron emission is a measure of the strength of the total magnetic field component in the sky plane. Note that polarized emission is due to ordered $B$-fields and unpolarized comes from turbulent ones. The Zeeman splitting of radio spectral lines is the best method to directly measure the field strength in gas clouds of our galaxy~\cite{1999ApJ...520..706C}, OH masers in starburst galaxies~\cite{2008ApJ...680..981R} and in dense HI clouds in distant galaxies on the line of sight towards bright quasars \cite{2008Natur.455..638W}. The drawback is that the Zeeman effect is very weak and can mainly be used for detecting intersellar magnetic fields. This is due to the small line shift, which given by
\begin{equation}
{\Delta\nu\over\nu}= 1.4g\,{B\over\nu}\,,  \label{Zeeman}
\end{equation}
and is extremely difficult to observe at large distances. Note that in the above the $B$-field is measured in $\mu$G and the line frequency in Hz. Also, the parameter $g$ represents the Land\'e factor that relates the angular momentum of an atom to its magnetic moment.

When polarized electromagnetic radiation crosses a magnetized plasma its orientation is changed by Faraday rotation. The latter is caused by the left and right circular polarisation states traveling with different phase velocities. For linearly polarised radiation, the rotation measure (RM) associated with a source at redshift $z_{s}$ is (cf. , e.g., \cite{2005LNP...664....9K})
\begin{equation}
RM\left(z_s\right)\simeq 8\times10^{5} \int_0^{z_s}\frac{n_eB_{\parallel}(z)}{(1+z)^2}\,dL(z)
\hspace{10mm} ({\rm rad}/{\rm m^2})\,,  \label{rm-1}
\end{equation}
where $n_{e}$ is the electron density of the intervening plasma (in cm$^{-3}$), $B_{\parallel}$ is the magnetic intensity along the line of sight (in $\mu$G) and $dL$ is the distance traveled by the radio signal. The latter is given by
\begin{equation}
dL\left(z\right)= 10^{-6}\frac{1+z}{H_0\sqrt{1+\Omega_0z}}\, {\rm dz} \hspace{10mm} {\rm Mpc}\,,  \label{rm-2}
\end{equation}%
with $H_{0}$ and $\Omega_0$ representing the present values of the Hubble constant of the density parameter respectively.\footnote{Conventionally, positive RM values indicate magnetic fields directed towards the observer and negative ones correspond to those pointing away.} As the rotation angle is sensitive to the sign of the field direction, only ordered $B$-fields can give rise to Faraday rotation. Multi-wavelength observations determine the strength and the direction of the line-of-sight magnetic component. Then, the total intensity and the polarization vectors yields the three-dimensional picture of the field and allow us to distinguish between its regular, anisotropic and random components.

Some novel detection methods try to exploit the effects that an intervening magnetic field can have upon the highly energetic photons emitted by distant active sources~\cite{1994ApJ...423L...5A,1995Natur.374..430P,%
2002ApJ...580.1013D,2009PhRvD..80l3012N}. Using such techniques, together with data from state-of-the-art instruments (like the  Fermi Gamma-Ray Space Telescope for example), three independent groups have recently reported the detection of intergalactic magnetic fields with strengths close to $10^{-15}$~G (see \S~\ref{sssGEMFs} next).

\subsubsection{Galactic and extragalactic magnetic
fields}\label{sssGEMFs}
The strength of the total magnetic field in galaxies can be determined from the intensity of the total synchrotron emission, assuming equipartition between the magnetic energy density and that of the cosmic rays\footnote{Determining the magnetic strength from the synchrotron intensity requires information about the number density of the cosmic-ray electrons. The latter can be obtained via X-ray emission, by inverse-Compton scattering, or through $\gamma$-ray bremsstrahlung. When such data is unavailable, an assumption must be made about the relation between cosmic-ray electrons and magnetic fields. This is usually the aforementioned principle of energy equipartition~\cite{2005AN....326..414B,%
2008AIPC.1085...83B}}. This seems to hold on large scales (both in space and time), though deviations occur locally. Typical equipartition strengths in spiral galaxies are around $10\mu$G. Radio-faint galaxies, like the M31 and M33, have weaker total fields (with $B\sim5\mu$G), while gas-rich galaxies with high star-formation rates, such as the M51, M83 and NGC6946, have magnetic strengths of approximately $15\mu$G. The strongest fields, with values between 50 $\mu$G and 100 $\mu$G, are found in starburst and merging galaxies, like the M82 and NGC4038/39 respectively~\cite{2004A&A...417..541C}.

Spiral galaxies observed in total radio emission appear very similar to those seen in the far-infrared. The equipartition magnetic strength in the arms can be up to $30\mu$G and shows a low degree of polarization. The latter indicates that the fields are randomly oriented there. On the other hand, synchrotron radio-emission from the inter-arm regions has a higher degree of polarization. This is due to stronger (10$\mu$G -- 15$\mu$G) and more regular $B$-fields, oriented parallel to the adjacent optical arm. The ordered fields form spiral patterns in almost every galaxy, even in ringed and flocculent galaxies. Therefore, the magnetic lines do not generally follow the gas flow (which is typically almost circular) and dynamo action is needed to explain the observed radial magnetic component. In galaxies with massive bars, however, the field lines appear to follow the gas flow. As the gas rotates faster than the bar pattern of the galaxy, a shock occurs in the cold gas. At the same time, the warm gas is only slightly compressed. Given that the observed magnetic compression in the spiral arms and the bars is also small, it seems that the ordered field is coupled to the warm diffuse gas and is strong enough to affect its flow~\cite{2005A&A...444..739B}.

Spiral dynamo modes can be identified from the pattern of polarization angles and Faraday rotation measures from multi-wavelength radio observations of galaxy disks~\cite{1992A&AS...94..587E}, or from RM data of polarized background sources~\cite{2008A&A...480...45S}. The disks of some spiral galaxies show large-scale RM patterns, but many galaxy disks posses no clear patterns of Faraday rotation. Faraday rotation in the direction of QSOs helps to determine the field pattern along the line of sight of an intervening galaxy~\cite{2008A&A...480...45S,%
1992ApJ...387..528K}. Recently, high resolution spectra have unambiguously associated quasars with strong MgII absorption lines to large Faraday rotation measures. As MgII absorption occurs in the haloes of normal galaxies lying along the line of sight to the quasars, this implies that organized strong $B$-fields are also present in high-redshift galaxies~\cite{2008ApJ...676...70K,%
2008Natur.454..302B,2008Natur.455..638W}.

Magnetic fields have also been detected within clusters of galaxies, where X-ray observations have revealed the presence of hot gas~\cite{2002ARA&A..40..319C}. There are several indications that favour the existence of cluster magnetic fields. In particular, galaxy clusters are known to have radio halos that trace the spatial distribution of the intra-cluster gas found in the X-ray observations. The radio signals are due to synchrotron emission from relativistic electrons spiralling along the field lines. In addition, there have been reports of Faraday rotation measurements of linearly polarized emissions crossing the intracluster medium. The first detection of a cluster magnetic field was made in the Coma cluster~\cite{1990ApJ...355...29K}. The Very Large Array (VLA) was used to compare Faraday rotation measures of radio sources within and directly behind the Coma cluster with radiation not crossing the cluster. Since then, there have been more analogous detections. It turns out that the observed cluster-field strengths vary slightly with the type of  cluster. In particular, the magnetic field strength depends on whether we are dealing with cooling flow or non-cooling flow clusters. Faraday observations indicate turbulent field strengths of $\mu$G-order in non-cooling flow clusters, such as the Coma. For cooling flow clusters, like the Hydra for example, the $B$-fields are of the order of a few 10~$\mu$G~\cite{2003A&A...412..373V}. In fact, the cool core region of the Hydra A cluster is associated with a magnetic field of 7$\mu$G with correlation length of 3~Kpc. Non-cooling flow clusters like the Coma, on the other hand, have weaker fields of the order of 3$\mu$G but with larger correlation lengths (between 10~Kpc and 30~Kpc)~\cite{2005mpge.conf..231E}. In general, the magnetic structure is not homogeneous but is rather patchy on small scales (5~Kpc -- 20~Kpc), indicating the presence of tangled magnetic fields~\cite{2003A&A...412..373V}. An alternative way of determining the strength of cluster $B$-fields is to compare the radio synchrotron emission with inverse Compton X-ray emission~\cite{2002ARA&A..40..319C}. The former comes from spiraling electrons along the cluster magnetic field. The latter is mainly due to CMB photons being upwardly scattered by the relativistic electrons of the intracluster gas.

In view of the accumulating observational evidence for magnetic presence on all scales up to that of a galaxy cluster, the idea of a  truly cosmological origin for cosmic magnetism gains ground. The potential detection of such primordial $B$-fields in the intergalactic medium may also change our understanding of the way structure formation has progressed. Note that an intergalactic magnetic field ordered on very large scales would pick out a preferred direction, which should then manifest itself in Faraday rotation measurements from distant radio sources. This puts an upper limit on any cosmological intergalactic magnetic field of $B_{\rm IGM}\lesssim10^{-11}$~\cite{1975Natur.254...23V}. Assuming that such a field has a characteristic scale, galaxy rotation measures suggest a size of 1~Mpc and an upper limit of the order of 1~nG~\cite{2005LNP...664....9K}. Indications of intergalactic magnetic fields have come from observations of radio-galaxy groupings near the Coma cluster, suggesting the presence of $B$-fields with strengths between 0.2~$\mu$G and 0.4~$\mu$G and a coherence scale close to $4$~Mpc~\cite{2007ApJ...659..267K}. There is also evidence for an intergalactic magnetic field around 0.3~$\mu$G on scales of the order of 500~Kpc, from excess rotation measures towards the Hercules and the Perseus-Pisces superclusters~\cite{2006ApJ...637...19X,2009arXiv0912.2918B}. In addition, intergalactic $B$-fields close to 30~nG and spanning scales of approximately 1$h^{-1}$~Mpc were recently suggested after cross-correlating the galaxy density field, obtained from the 6th Data Release of the Sloan Digital Sky Survey, with a large sample of Faraday rotation measures supplied by the NRAO-VLA Sky Survey~\cite{2009arXiv0906.1631L}.

Additional reports of intergalactic magnetic fields have appeared within the last year, using techniques that exploit the magnetic effects on the highly energetic photons emitted by distant sources (e.g.~see~\cite{1994ApJ...423L...5A,1995Natur.374..430P,%
2002ApJ...580.1013D,2009PhRvD..80l3012N}). More specifically, TeV-energy photons from a distant AGNs interact with the low frequency photons of the extragalactic background and lead to electron-positron pair creation. These produce (GeV-level) $\gamma$-rays through the inverse Compton scattering of the CMB photons. Observation wise, the key point is that a magnetic presence, even a very weak one, can affect the profile of the the resulting $\gamma$-ray spectra. For instance, the $B$-field can cause the formation of an extended halo around the $\gamma$-ray images of distant AGNs. Such halos were first reported by Ando and Kusenko, using combined data from the Atmospheric Cherenkov Telescopes and Fermi Gamma-Ray Space Telescope~\cite{2010ApJ...722L..39A}. Subsequent, complementary analysis indicated the presence of an intergalactic magnetic field with strength between $10^{-17}$ and $10^{-14}$ Gauss~\cite{2010arXiv1012.5313E}. In addition to halo formation, the $B$-field can also reduce the observed flux of the secondary GeV-photons by deflecting them into larger solid angles. Using observations of the Fermi/Large Area Telescope and assuming that the original TeV-photons were strongly beamed, a lower limit of $\sim10^{-15}$~G was imposed on the intergalactic magnetic field~\cite{2010MNRAS.406L..70T,2010arXiv1009.1048T}. A similar lower limit of $\sim10^{-16}$~G was also obtained assuming that the blazar source radiated isotropically~\cite{2010Sci...328...73N}.

\subsection{Limits on primordial magnetic fields}\label{ssLPMFs}
Any primordial magnetic field must comply with a number of astrophysical constraints, the most significant of which come from Big Bang Nucleosynthesis (BBN) and the isotropy of the CMB. The latter probes $B$-fields with coherence scales larger than the particle horizon during nucleosynthesis, while the BBN limits apply in principle to all scales.

\subsubsection{Nucleosynthesis limits}\label{sssNL}
The main effects of a magnetic presence on the output of primordial nucleosynthesis are related with: (a) the proton-to-neutron conversion ratio; (b) the expansion and cooling of the universe; and (c) the electron thermodynamics. Here we will only provide a very brief summary of these effects. For a detailed review, the reader is referred to~\cite{2001PhR...348..163G}.

(a) In the early universe, the weak interaction is responsible for maintaining chemical equilibrium between protons and neutrons. The main effect of a strong magnetic presence at the time of nucleosynthesis is to enhance the conversion rate of neutrons into protons. As a result, the neutron-to-proton ratio would freeze-out at a lower temperature. This in turn would lead to a less efficient production of $^{4}$He and of heavier elements~\cite{1969PhRv..180.1289M,1969Natur.222..649O}. In fact, the magnetic effect would be catastrophic if $B\gg m_{p}^{2}/e\sim 10^{17}$~G at the time of nucleosynthesis.

(b) The temperature at which the proton-to-neutron ratio freezes-out is determined by the balance between the timescale of the weak interaction and the expansion rate of the universe~\cite{1969Natur.223..938G}. Equilibrium is attained when $\Gamma_{n\rightarrow p}\sim H$, where $\Gamma_{n\rightarrow p}$ is the cross-section of the interaction and $H$ is the Hubble constant at the time. The latter is proportional to the total energy density of the universe, where the $B$-field contributes as well. Thus, a strong magnetic presence will increase the value of the Hubble parameter. This would cause an earlier freeze-out of the proton-to-neutron ratio and result into larger residual amounts of $^{4}$He~\cite{1994PhRvD..49.5006C,1996PhLB..379...73G}.

(c) The magnetic presence will also change the phase-space volume of electrons and positrons, since their momentum component normal to the $B$-field will become discrete (Landau levels). Therefore, the energy density, the number density and the pressure of the electron gas augment, relative to their magnetic-free values~\cite{1969Natur.222..649O}. The rise happens at the expense of the background photons, which transfer energy to the lowest Landau level. This delays the electron-positron annihilation, which in turn increases the photon-to-baryon ratio and finally leads to lower $^{3}$He and D abundances~\cite{1996PhRvD..54.7207K}.

All of the above need to be accounted for when calculating the BBN limits on primordial magnetic fields. This is done by means of numerical methods, which seem to conclude that the main magnetic effect on the light-element abundances comes from the field's contribution to the expansion rate of the universe (i.e.~case (b)). The overall constraint on the magnetic strength is $B\lesssim10^{11}$~G at the time of nucleosynthesis, which (roughly) translates to $B\lesssim7\times10^{-7}$~G at the time of galaxy formation~\cite{2001PhR...348..163G}.

\subsubsection{Cosmic microwave limits}\label{sssCML}
Observations of the  CMB temperature anisotropies and polarization provide valuable tools to constrain  cosmological models. As such, they also play an important role in the diagnostic of early universe magnetic fields.

In comparison to the data of the angular power spectra of polarization, $C_{\ell}^{EE}$,  and temperature-polarization, $C_{\ell}^{TE}$, the temperature angular power spectrum, $C_{\ell}^{TT}$, is known at higher precision. For example, the 7-year WMAP ({\it Wilkinsion Microwave Anisotropy Probe}) power spectrum is limited only by cosmic variance up to $\ell\approx548$~\cite{2010arXiv1001.4635L}. Moreover, on smaller-scales, observations from CBI ({\it Cosmic Background Imager})~\cite{2009arXiv0901.4540S} and VSA ({\it Very Small Array})~\cite{2005MNRAS.363.1125R}, ACBAR ({\it Arcminute Cosmology Bolometer Array Receiver})~\cite{2009ApJ...694.1200R} and the forthcoming SPT ({\it South Pole Telescope})~\cite{2004SPIE.5498...11R} missions will determine the $C_{\ell}^{TT}$ to even higher accuracies. The PLANCK satellite is expected to extend the region limited only by cosmic variance to $\ell\approx1500$.

At the moment, the high isotropy of the Cosmic Microwave Background appears to exclude homogeneous cosmological magnetic fields much stronger than $\sim10^{-9}$~G~\cite{1997PhRvL..78.3610B}. A similar limit is found for stochastic magnetic fields as well~\cite{2009PhRvD..79f3007G,2009PhRvD..79l1302G,%
2010arXiv1005.0148P,2010arXiv1006.4242S}. It has  been shown that the temperature angular power  spectrum, $C_{\ell}^{TT}$, from magnetically-induced vector and scalar perturbations increases slightly across all angles. The extra pressure that the $B$-field adds into the system can change the position and the magnitude of the acoustic peaks, thus producing a potentially observable effect~\cite{1996PhLB..388..253A,2010arXiv1006.2985C}. The presence of small-scale magnetic fields appears to leave undamped features on small angular scales and may also lead to distinctive polarisation structures~\cite{1998PhRvL..81.3575S,2003MNRAS.344L..31S}. In addition, large-scale primordial fields could be related to the low-quadrupole moment problem~\cite{2006PhRvL..97m1302C}. Nevertheless, the magnetic signal remains subdominant to that from standard scalar perturbations until around $\ell\approx2000$, depending on the field strength and spectral index~\cite{2006ApJ...646..719Y,2008PhRvD..77f3003G,%
2008PhRvD..78b3510F,2010PhRvD..81d3517S,2010arXiv1007.3163K}. Magnetic fields also source tensor modes, which however are of relatively low-amplitude. The signal is similar to that of inflationary gravitational wave, but probably weaker in strength~\cite{2004PhRvD..70d3011L}. Therefore, the  direct magnetic impact onto the CMB $\left\langle TT\right\rangle$ correlation does not generally provide an ideal probe of primordial magnetism. However, the CBI mission observed a weak increase of power on small scales, as compared to the concordance model~\cite{2005MNRAS.363.1125R}. Provided this is real and not a statistical or systematic artifact, it could be partly explained by the presence of a cosmological $B$-field~\cite{2006ApJ...646..719Y,2005ApJ...625L...1Y}. Nucleosynthesis bounds, however, imply that a primordial field is unlikely to account for all the increase.

Besides contributing to the CMB temperature fluctuations, a primordial magnetic field also produces $E$-mode polarisation that can significantly change the angular power spectrum of the standard $\Lambda$CDM model~\cite{2008PhRvD..77f3003G}. However, the polarisation limits are not as strong as those coming from the temperature anisotropy. Due to the presence of both vector and tensor perturbations, the magnetic field also leads to $B$-mode polarisation. Moreover, $B$-modes are also induced in the scalar sector by Faraday rotation if a magnetic field is present at decoupling~\cite{1996ApJ...469....1K,1997PhRvD..55.1841H,%
2004ApJ...616....1C,2004PhRvD..70f3003S,2005PhRvD..71d3006K,%
2008PhRvD..78b3010G}. Taking into account that in the standard picture $B$-modes are produced only by lensed $E$-modes and by inflationary gravitational waves, in principle, the observation of a distinct $B$-mode power spectrum would be the clearest indication of a primordial magnetic field. However, the CMB polarisation maps are poorly known, compared to the temperature ones. While we currently possess a power spectrum $C^{TE}_{\ell}$, this is by no means cosmic-variance limited on any scale. The observations of the $B$-modes yield bounds consistent with zero~\cite{2005ApJ...624...10L,2004Sci...306..836R}. These are on relatively small scales, directly observing the region at which magnetic effects may come to dominate. Nevertheless, we are far from the required accuracy, particularly for the $B$-modes. Given the limitations of the power spectra, the non-Gaussianity of the temperature map is a reasonable place to look for further constraints on primordial magnetic fields
\cite{2005PhRvD..72f3002B,2009PhRvL.103h1303S,2009JCAP...06..021C,%
2010PhRvD..82l3006T,2010PhRvD..82j3505S,2010arXiv1010.4543K,%
2010arXiv1012.2892B}. Although up to now the observations are entirely consistent with Gaussian initial conditions, there are non-Gaussian features in the WMAP maps~\cite{2005MNRAS.357..994L}. Also, the number of non-Gaussian features could well increase with the next generation of CMB experiments.

\subsubsection{Limits from gravitational waves}\label{sssLGW}
A strong limit on stochastic magnetic fields produced before nucleosynthesis has been derived in~\cite{2002PhRvD..65b3517C}. The anisotropic stress of the magnetic field acts as a source term in the evolution equation of gravity waves. This causes the conversion of magnetic field energy into gravity waves above a certain critical value of the magnetic field strength. In particular, the field strength smoothed over a scale $\lambda$ of magnetic fields generated during inflation must be smaller than $B_{\lambda}\sim10^{-20}$~G for spectral indices $n_{\rm B}>-2$, where $n_{\rm B}=-3$ corresponds to a scale invariant magnetic field energy spectrum. If the magnetic field is produced by a causal mechanism, for example during the electroweak phase transition, $n_{\rm B}>2$, its strength has to be below $10^{-27}$ G in order not to loose all its energy density to gravitational waves.

The magnetic-strength limits asserted in~\cite{2002PhRvD..65b3517C} are the strongest reported in the literature, far more restrictive than those coming from nucleosynthesis or the CMB. However, analogous studies of magnetically produced gravity waves have reached different conclusions. It has been claimed, in particular, that the limits on cosmological magnetic fields set by the latest LIGO~S5 data lie close to those obtained by BBN and the CMB~\cite{2010PhRvD..81b3002W}.

Finally, we should also note the possibility of constraining primordial $B$-fields using the ionisation history of the post-recombination universe and, in particular, the observed re-ionisation depth. Thus, based on the 5-year WMAP data, upper limits of nGauss order have been reported in the literature~\cite{2008PhRvD..78h3005S}.

\newpage

\section{Magnetic fields in cosmology}\label{sMFC}
\subsection{Relativistic magnetised cosmologies}\label{ssRMC}
Although the study of large-scale magnetic fields goes a long way back into the past, the first systematic attempts to incorporated magnetism in cosmology appeared in the late 60s and the early 70s~\cite{1968ApJ...153..661J,1969ApJ...155..379J,%
1970ApJ...160..147H,1972CMaPh..27...37C}. Next, we will provide the basic background for the relativistic study of cosmological $B$-fields. For the details and a recent review the reader is referred to~\cite{1997CQGra..14.2539T,1998CQGra..15.3523T,%
2000PhRvD..61h3519T,2000CQGra..17.2215T,2005CQGra..22..393T,%
2007PhR...449..131B}.

\subsubsection{The gravitational field}\label{sssGF}
In the geometrical framework of general relativity, gravity is a manifestation of the non-Euclidean geometry of the spacetime. The gravitational field is therefore described by the Riemann curvature tensor ($R_{abcd}$), which satisfies the Ricci identities
\begin{equation}
2\nabla_{[a}\nabla_{b]}v_c= R_{abcd}v^d\,,  \label{Ricci}
\end{equation}
applied here to an arbitrary vector $v_a$ (with $\nabla_a$ representing the familiar covariant derivative operator). The Riemann tensor also assumes the invariant decomposition
\begin{equation}
R_{abcd}= {1\over2}\,\left(g_{ac}R_{bd}+g_{bd}R_{ac}-g_{bc}R_{ad} +g_{ad}R_{bc}\right)-{1\over6}\,R\left(g_{ac}g_{bd} -g_{ad}g_{bc}\right)+ C_{abcd}\,,  \label{Riemann}
\end{equation}
and obeys the symmetries $R_{abcd}=R_{cdab}$ and $R_{abcd}=R_{[ab][cd]}$. Note that $g_{ab}$ is the spacetime metric, $R_{ab}=R^c{}_{acb}$ is the Ricci tensor, $R=R^a{}_a$ is the associated Ricci scalar and $C_{abcd}$ is the Weyl (or conformal curvature) tensor.\footnote{Unless stated otherwise, we consider a general 4-dimensional (pseudo) Riemannian spacetime with a Lorentzian metric of signature ($-,+,+,+$). Also, throughout this review, Latin indices take values between 0 and 3, while their Greek counterparts vary from 1 to 3.} The Ricci component of the Riemann tensor determines the local gravitational field through the Einstein field equations
\begin{equation}
R_{ab}- {1\over2}\,Rg_{ab}=\kappa T_{ab}-\Lambda g_{ab}\,,  \label{EFE}
\end{equation}
where $T_{ab}$ is the energy-momentum tensor of all the matter fields involved, $\kappa=8\pi G$ and $\Lambda$ the cosmological constant.\footnote{We use the Heaviside-Lorentz units for the electromagnetic field in this section. Natural units, with $c=1=k_B=\hbar$ and energy as the fundamental dimension, are used throughout this review.} The Weyl tensor, on the other hand, has to do with the long-range component of the gravitational field (i.e.~tidal forces and gravity waves), shares the same symmetries with $R_{abcd}$ and it is also trace-free (i.e.~$C^c{}_{acb}=0$).

\subsubsection{Kinematics}\label{sssKs}
We introduce a family of observers with worldlines tangent to the timelike 4-velocity field $u_a$ (i.e.~$u_au^a=-1$). These are the fundamental observers that define the direction of time. Then, the tensor $h_{ab}=g_{ab}+u_au_b$ projects orthogonal to $u_a$ and into the observers' instantaneous 3-dimensional rest-space.\footnote{By construction, $h_{ab}$ is a symmetric spacelike tensor, with $h_a{}^a=3$ and $h_{ab}h^b{}_c=h_{ac}$. The projector coincides with the metric of the observers' 3-dimensional space in non-rotating spacetimes.} Together, $u_a$ and $h_{ab}$ introduce an 1+3 `threading' of the spacetime into time and space, which decomposes physical quantities, operators and equations into their irreducible timelike and spacelike parts (see~\cite{1999toc..conf....1E,%
2008PhR...465...61T} for further details).

For example, splitting the covariant derivative of the observers' 4-velocity, leads to the irreducible kinematic variables of the motion. In particular, we arrive at
\begin{equation}
\nabla_bu_a= {1\over3}\,\Theta h_{ab}+ \sigma_{ab}+ \omega_{ab}- A_au_b\,,  \label{Nbua}
\end{equation}
where $\Theta=\nabla^au_a=\D^au_a$ is the volume scalar, $\sigma_{ab}=\D_{\langle b}u_{a\rangle}$ is the shear tensor, $\omega_{ab}=\D_{[b}u_{a]}$ is the vorticity tensors and $A_a=\dot{u}_a=u^b\nabla_bu_a$ is the 4-acceleration vector.\footnote{Overdots indicate (proper) time derivatives along the $u_a$-field, while the gradient ${\rm D}_a=h_a{}^b\nabla_b$ defines the 3-dimensional covariant derivative operator. Round brackets denote symmetrisation, square antisymmetrisation and angled ones indicate the symmetric and traceless part of projected tensors and vectors. For example, $\sigma_{ab}=\D_{\langle b}u_{a\rangle}= {\rm D}_{(b}u_{a)}-({\rm D}^cu_c/3)h_{ab}$ and $\dot{E}_{\langle a\rangle}=h_a{}^b\dot{E}_b$ -- see Eq.~(\ref{emTab}).} The volume scalar describes changes in the average separation between neighbouring observes. When $\Theta$ is positive this separation increases, implying that the associated fluid element expands. In the opposite case we have contraction. The volume scalar also defines a representative length scale ($a$) by means of $\dot{a}/a=\Theta/3$. In cosmological studies, $a$ is commonly referred to as the `scale factor'. We use the shear to monitor changes in the shape of the moving fluid under constant volume, while the vorticity traces its rotational behaviour. Note that we can replace the vorticity tensor with the vorticity vector $\omega_a=\varepsilon_{abc}\omega^{bc}/2$, where $\varepsilon_{abc}$ represents the 3-D Levi-Civita tensor. Finally, the 4-acceleration reflects the presence of non-gravitational forces and vanishes when the observers worldlines are timelike geodesics.

The time evolution of the volume scalar, the vorticity vector and the shear tensor is determined by a set of three propagation equations, supplemented by an equal number of constraints. Both sets are onbtained after applying the Ricci identities (see (\ref{Ricci}) in \S~\ref{sssGF}) to the fundamental 4-velocity field~\cite{1999toc..conf....1E,2008PhR...465...61T}.

\subsubsection{Matter fields}\label{sssMFs}
Analogous decompositions apply to the rest of the kinematical and dynamical variables. Thus, relative to the $u_a$-frame, the energy-momentum tensor of a general (imperfect) fluid splits as
\begin{equation}
T^{(m)}_{ab}= \rho u_au_b+ ph_{ab}+ 2q_{(a}u_{b)}+ \pi_{ab}\,.  \label{ifTab}
\end{equation}
Here, $\rho=T_{ab}u^au^b$ represents the energy density, $p=T_{ab}h^{ab}/3$ the isotropic pressure, $q_a=-h_a{}^bT_{bc}u^c$ the total energy flux and $\pi_{ab}=h_{\langle a}{}^ch_{b\rangle}{}^dT_{cd}$ the anisotropic pressure of the matter, as measured by the fundamental observers~\cite{1999toc..conf....1E,2008PhR...465...61T}. When dealing with a perfect fluid, both $q_a$ and $\pi_{ab}$ vanish and the above reduces to
\begin{equation}
T^{(m)}_{ab}= \rho u_au_b+ ph_{ab}\,.  \label{pfTab}
\end{equation}
The remaining degrees of freedom are determined by the equation of state, which for a barotropic medium takes the simple $p=p(\rho)$ form.

\subsubsection{Electromagnetic fields}\label{sssEMFs}
Magnetic and electromagnetic fields introduce new features to any cosmological model through their energy density and pressure contributions and due to their generically anisotropic nature. The Maxwell field is invariantly described by the antisymmetric Faraday tensor. Relative to the fundamental observers introduced in \S~\ref{sssKs}, the latter decomposes as
\begin{equation}
F_{ab}= 2u_{[a}E_{b]}+ \varepsilon_{abc}B^c\,,  \label{Fab}
\end{equation}
where $E_a=F_{ab}u^b$ and $B_a=\varepsilon_{abc}F^{bc}/2$ are respectively the electric and magnetic components.

The inherit anisotropy of the electromagnetic field is reflected in the form of its energy-momentum tensor. The latter has the invariant form $T_{ab}^{(em)}=-F_{ac}F^c{}_b+(F_{cd}F^{cd}/4)g_{ab}$, which relative to the $u_a$-frame recasts to
\begin{equation}
T_{ab}^{(em)}= {1\over2}\left(B^2+E^2\right)u_au_b+ {1\over6}\left(B^2+E^2\right)h_{ab}+ 2\mathcal{P}_{(a}u_{b)}+ \Pi_{ab}\,.  \label{emTab}
\end{equation}
with $E^2=E_aE^a$, $B^2=B_aB^a$, $\mathcal{P}_a= \varepsilon_{abc}E^bB^c$ and
\begin{equation}
\Pi_{ab}= -E_{\langle a}E_{b\rangle}- B_{\langle a}B_{b\rangle}= {1\over3}\,\left(E^2+B^2\right)h_{ab}- E_aE_b- B_aB_b\,,  \label{Pi}
\end{equation}
Comparing the above to (\ref{ifTab}) in \S~\ref{sssMFs}, we conclude that the Maxwell field corresponds to an imperfect fluid with energy density $(E^2+B^2)/2$, isotropic pressure $(E^2+B^2)/6$, energy flux given by the Poynting vector $\mathcal{P}_a$ and anisotropic stresses represented by the symmetric and trace-free $\Pi_{ab}$-tensor. This fluid-like description of the Maxwell field has been proved particularly helpful in many applications~\cite{1997CQGra..14.2539T,1998CQGra..15.3523T,%
2000PhRvD..61h3519T,2000CQGra..17.2215T,2005CQGra..22..393T,%
2007PhR...449..131B}.

\subsubsection{Conservation laws}\label{sssCLs}
In the case of charged matter, the total energy-momentum tensor is $T_{ab}=T_{ab}^{(m)}+T_{ab}^{(em)}$, with the individual components given by (\ref{ifTab}) and (\ref{emTab}) respectively. The total stress-tensor satisfies the familiar conservation law $\nabla^bT_{ab}=0$, while the electromagnetic stress-energy tensor obeys the additional constraint $\nabla^bT_{ab}^{(em)}=-F_{ab}J^b$, with the right-hand side representing the Lorentz 4-force. Combining the two, we obtain the conservation laws of the total energy and momentum. These are given by the continuity equation
\begin{equation}
\dot{\rho}= -\Theta(\rho+p)- \D^aq_a- 2A^aq_a- \sigma^{ab}\pi_{ab}+ E_a\clj_a\,,  \label{tec}
\end{equation}
and by the Navier-Stokes equation
\begin{eqnarray}
(\rho+p)A_a&=& -\D_ap- \dot{q}_{\langle a\rangle}- {4\over3}\,\Theta q_a- (\sigma_{ab}+\omega_{ab})q^b- \D^b\pi_{ab}- \pi_{ab}A^b \nonumber\\ &&+\mu E_a- \varepsilon_{abc}B^b\clj^c\,,  \label{tmc}
\end{eqnarray}
respectively~\cite{2007PhR...449..131B}. Note that $\mu=-J_au^a$ is the electric charge density and $\mathcal{J}_a=h_a{}^bJ_b$ is the associated 3-current, so that $J_a=\mu u_a+\clj_a$.

An additional conservation law is that of the 4-current, which satisfies the invariant constraint $\nabla^aJ_a=0$. The latter translates into the conservation law for the charge density, given by~\cite{2007PhR...449..131B}
\begin{equation}
\dot{\mu}= -\Theta\mu-\D^a\clj_a- A^a\clj_a\,.  \label{ccl}
\end{equation}

\subsection{Evolution of the electromagnetic field}\label{ssEEMF}
The vector nature of the electromagnetic components and the geometrical approach to gravity that general relativity introduces, mean that the Maxwell field is the only known energy source that couples directly to the spacetime curvature through the Ricci identities as well as the Einstein Field Equations. Both sets are therefore necessary for the full relativistic treatment of electromagnetic fields.

\subsubsection{Maxwell's equations}\label{sssMEs}
We monitor the evolution of the electromagnetic field using Maxwell's formulae. In their invariant form these read
\begin{equation}
\nabla_{[c}F_{ab]}= 0 \hspace{15mm} {\rm and} \hspace{15mm} \nabla^bF_{ab}= J_a\,, \label{Max}
\end{equation}
with the first manifesting the existence of a 4-potential. Relative to the $u_a$-frame, the above set splits into two pairs of propagation and constraint equations. The former consists of
\begin{equation}
\dot{E}_{\langle a\rangle}= -{2\over3}\,\Theta E_a+ \left(\sigma_{ab}+\omega_{ab}\right)E^b+ \varepsilon_{abc}A^bB^c+ \curl{B}_a- \clj_a\,,  \label{M1}
\end{equation}
and
\begin{equation}
\dot{B}_{\langle a\rangle}= -{2\over3}\,\Theta B_a+ \left(\sigma_{ab}+\omega_{ab}\right)B^b- \varepsilon_{abc}A^bE^c- \curl{E}_a\,,  \label{M2}
\end{equation}
which may be seen as the 1+3 covariant analogues of the Ampere and the Faraday laws respectively. The constrains, on the other hand, read
\begin{equation}
\D^aE_a= \mu- 2\omega^aB_a \hspace{15mm} {\rm and} \hspace{15mm}
\D^aB_a= 2\omega^aE_a\,.  \label{M34}
\end{equation}
These provide the 1+3 forms of Coulomb's and Gauss' laws respectively. Note that Eqs.~(\ref{M1})-(\ref{M34}) contain relative motion effects, in addition to the standard `curl' and 'divergence' terms of their more traditional versions. This is an essentially built-in property of the 1+3 formalism, which should be always kept in mind when applying expressions like the above.

\subsubsection{The electromagnetic wave equations}
Maxwell's equations also provide the wave equation of the electromagnetic tensor. This can be obtained by applying the Ricci identities to the Faraday tensor and takes the invariant form (e.g.~see~\cite{2005CQGra..22..393T,1989ApJ...341..786N})
\begin{equation}
\nabla^2F_{ab}= -2R_{abcd}F^{cd}- 2R_{[a}{}^cF_{b]c}- 2\nabla_{[a}J_{b]}\,,  \label{N2Fab}
\end{equation}
where $\nabla^2=\nabla^a\nabla_a$ is the d'Alembertian and $R_{abcd}$, $R_{ab}$ are the Riemann and Ricci tensors respectively (see \S~\ref{sssGF}). The above results from the vector nature of the electromagnetic components and from the geometrical interpretation of gravity that general relativity advocates. The two guarantee that the Maxwell field is the only known source of energy that couples directly to gravity through both the Einstein equations and the Ricci identities.

Expression (\ref{N2Fab}), which clearly shows how spacetime curvature drives electromagnetic disturbances, can also provide the wave-equations of the individual components of the Maxwell field. Alternatively, one may obtain these relations using the set (\ref{M1})-(\ref{M34}) together with the Ricci identities. Assuming matter in the form of a single perfect fluid, we obtain~\cite{2005CQGra..22..393T}
\begin{eqnarray}
\ddot{E}_{\langle a\rangle}- \D^2E_a&=& \left(\sigma_{ab}-\varepsilon_{abc}\omega^c-{5\over3}\,\Theta h_{ab}\right)\dot{E}^b+ {1\over3}\,\kappa\rho(1+3w)E_a \nonumber\\ &&+{1\over3}\left(\sigma_{ab}+\varepsilon_{abc}\omega^c- {4\over3}\,\Theta h_{ab}\right)\Theta E_a- \sigma_{\langle a}{}^c\sigma_{b\rangle c}E^b+ \varepsilon_{abc}E^b\sigma^{cd}\omega_d \nonumber\\ &&+{4\over3}\left(\sigma^2-{2\over3}\,\omega^2\right)E_a+ {1\over3}\,\omega_{\langle a}\omega_{b\rangle}E^b+ A^bA_bE_a- {5\over2}\,\varepsilon_{abc}A^b\curl{E}^c+ \D_{\langle a}E_{b\rangle}A^b \nonumber\\ &&+{2\over3}\,\varepsilon_{abc}B^b\D^c\Theta+ \varepsilon_{abc}B_d\D^b\sigma^{cd}+ \D_{\langle a}\omega_{b\rangle}B^b+ {3\over2}\,\varepsilon_{abc}B^b\curl{\omega}^c+ 2\D_{\langle a}B_{b\rangle}\omega^b \nonumber\\ &&-2\varepsilon_{abc}\sigma^b{}_d\D^{\langle c}B^{d\rangle}+ \varepsilon_{abc}\dot{A}^bB^c+ {7\over3}\,A^b\omega_bB_a+ {4\over3}\,B^b\omega_bA_a- 3A^bB_b\omega_a \nonumber\\ &&+3\varepsilon_{abc}A^b\sigma^{cd}B_d- \mathcal{R}_{ab}E^b- E_{ab}E^b+ H_{ab}B^b \nonumber\\ &&+{1\over3}\,\mu A_a- \D_a\mu- \dot{\mathcal{J}}_a- \Theta\mathcal{J}_a\,,  \label{ddotEa}
\end{eqnarray}
for the electric field, and
\begin{eqnarray}
\ddot{B}_{\langle a\rangle}- \D^2B_a&=& \left(\sigma_{ab}-\varepsilon_{abc}\omega^c-{5\over3}\,\Theta h_{ab}\right)\dot{B}^b+ {1\over3}\,\kappa\rho(1+3w)B_a \nonumber\\ &&+{1\over3}\left(\sigma_{ab}+\varepsilon_{abc}\omega^c- {4\over3}\,\Theta h_{ab}\right)\Theta B_a- \sigma_{\langle a}{}^c\sigma_{b\rangle c}B^b+ \varepsilon_{abc}B^b\sigma^{cd}\omega_d \nonumber\\ &&+{4\over3}\left(\sigma^2-{2\over3}\,\omega^2\right)B_a+ {1\over3}\,\omega_{\langle a}\omega_{b\rangle}B^b+ A^bA_bB_a- {5\over2}\,\varepsilon_{abc}A^b\curl{B}^c+ \D_{\langle a}B_{b\rangle}A^b \nonumber\\ &&-{2\over3}\,\varepsilon_{abc}E^b\D^c\Theta- \varepsilon_{abc}E_d\D^b\sigma^{cd}- \D_{\langle a}\omega_{b\rangle}E^b- {3\over2}\,\varepsilon_{abc}E^b\curl{\omega}^c- 2\D_{\langle a}E_{b\rangle}\omega^b \nonumber\\ &&+2\varepsilon_{abc}\sigma^b{}_d\D^{\langle c}E^{d\rangle}- \varepsilon_{abc}\dot{A}^bE^c- {7\over3}\,A^b\omega_bE_a- {4\over3}\,E^b\omega_bA_a+ 3A^bE_b\omega_a \nonumber\\ &&-3\varepsilon_{abc}A^b\sigma^{cd}E_d- \mathcal{R}_{ab}B^b- E_{ab}B^b- H_{ab}E^b \nonumber\\ &&-{2\over3}\,\mu\omega_a+ \curl{\mathcal{J}}_a+ 2\varepsilon_{abc}A^b\mathcal{J}^c\,, \label{ddotBa}
\end{eqnarray}
for its magnetic counterpart. Here, $\mathcal{R}_{ab}$ is the Ricci tensor of the observer's 3-D rest-space, while $E_{ab}$ and $H_{ab}$ are the electric and the magnetic parts of the Weyl tensor respectively (see~\cite{2008PhR...465...61T} for details). Note the curvature terms in the right-hand side of the above, which show explicitly how the different parts of the spacetime geometry can affect the propagation of electromagnetic signals.

\subsection{Cosmological magnetohydrodynamics}\label{ssCMHD}
The evolution and the implications of cosmological electromagnetic fields depend on a number of parameters. A decisive factor is the electrical properties of the medium that fills the universe, which in turn are determined by the specific form of Ohm's law.

\subsubsection{Ohm's law}\label{sssOL}
The literature contains various expressions of Ohm's law, which in principle is the propagation equation of the electric 3-current (e.g~see~\cite{1973ppp..book.....K,2004ApJ...605..340M,%
2008MNRAS.385..883K}). For a single fluid at the limit of resistive MHD, Ohm's law takes the simple form~\cite{1971ApJ...164..589G,%
1975clel.book.....J}
\begin{equation}
\clj_a= \varsigma E_a\,,  \label{Ohm}
\end{equation}
with $\varsigma$ representing the electric conductivity of the matter. In highly conducting environments, $\varsigma\rightarrow\infty$ and the electric field vanishes. This is the familiar ideal-MHD approximation, where the electric currents keep the magnetic field frozen-in with the charged fluid. At the opposite end, namely when the conductivity is very low, $\varsigma\rightarrow0$. Then, the 3-currents vanish despite the presence of nonzero electric fields.

\subsubsection{The ideal MHD limit}\label{sssIMHDL}
At the ideal MHD limit Maxwell's formulae reduce to one propagation equation and three constraints. The former comes from (\ref{M2}) and is the familiar magnetic induction equation
\begin{equation}
\dot{B}_{\langle a\rangle}= -{2\over3}\,\Theta B_a+ \left(\sigma_{ab}+\omega_{ab}\right)B^b\,.  \label{MHDM2}
\end{equation}
The constraints, on the other hand, are obtained from Eqs.~(\ref{M1}) and (\ref{M34}). In particular, eliminating the electric field form these relations, we arrive at
\begin{equation}
\clj_a= \varepsilon_{abc}A^bB^c+ \curl{B}_a\,, \hspace{15mm} \mu= 2\omega^aB_a \hspace{10mm} \mathrm{and} \hspace{10mm} \D^aB_a= 0\,,  \label{MHDM134}
\end{equation}
respectively. Following (\ref{MHDM134}b), the inner product $\omega^aB_a$ corresponds to an effective charge density, triggered by the relative motion of the $B$-field.

In the absence of the electric field, the electromagnetic energy-momentum tensor simplifies as well. To be precise, expression (\ref{emTab}) reduces to
\begin{equation}
T_{ab}^{(B)}= {1\over2}\,B^2u_au_b+ {1\over6}\,B^2h_{ab}+ \Pi_{ab}\,,  \label{MHDTab}
\end{equation}
with $\Pi_{ab}=-B_{\langle a}B_{b\rangle}$. This means that the magnetic field can be seen as an imperfect fluid with energy density given by $B^2/2$, isotropic pressure equal to $B^2/6$, zero energy flux and anisotropic stresses given by
\begin{equation}
\Pi_{ab}= {1\over3}\,B^2h_{ab}- B_aB_b\,.  \label{PIab}
\end{equation}

At the MHD limit, the matter energy and that of the residual magnetic field are conserved separately, with the induction equation (see formula (\ref{MHDM2}) above) providing the conservation law of the magnetic energy. At the same time, the momentum conservation law reads
\begin{eqnarray}
\left(\rho+p+{2\over3}\,B^2\right)A_a&=& -\D_ap- \dot{q}_{\langle a\rangle}- {4\over3}\,\Theta q_a- (\sigma_{ab}+\omega_{ab})q^b- \D^b\pi_{ab}- \pi_{ab}A^b \nonumber\\ &&- \varepsilon_{abc}B^b\curl{B}^c- \Pi_{ab}A^b\,.  \label{MHDmc1}
\end{eqnarray}

When dealing with a perfect fluid with zero pressure, we may set $p=0=q_a=\pi_{ab}$. Then, starting from (\ref{MHDmc1}), one can show that $A_aB^a=0$ and recast the latter into the form
\begin{equation}
\left(\rho+B^2\right)A_a= -\varepsilon_{abc}B^b\curl{B}^c= -{1\over2}\,{\rm D}_aB^2+ B^b{\rm D}_bB_a\,.  \label{MHDmc2}
\end{equation}
where in the right-hand side we have two expressions for the Lorentz force. Note that the first term in the last equality is due to the field's pressure, while the second carries the effects of the magnetic tension. The former reflects the tendency of the field lines to push each other apart and the latter their elasticity and tendency to remain `straight'. As we will explain below, the majority of studies analysing the magnetic effects on structure formation do not account for the tension contribution to the Lorentz force.

\subsection{Magnetism and structure formation}\label{ssMSF}
Despite the widespread presence of large-scale magnetic fields in the universe, their role and implications during structure formation are still not well understood. Here, we will briefly summarise the way $B$-fields could have altered the linear and the mildly non-linear stages of galaxy formation and also how the latter might have backreacted on the magnetic evolution.

\subsubsection{The linear regime}\label{sssLR}
Scenarios of magnetised structure formation typically work within the ideal-MHD approximation, looking at the effects of the magnetic Lorentz force on density inhomogeneities. The bulk of the available inhomogeneous treatments are Newtonian, with the relativistic approaches being a relatively recent addition to the literature. Although the role of the magnetism as a source of density and vorticity perturbations was established early on~\cite{1978ApJ...224..337W,1996ApJ...468...28K,1997A&A...326...13B}, the complicated action of the $B$-field did not allow for analytic solutions (with the exception of~\cite{1971SvA....14..963R} for the case of dust). Full solutions were provided by means of covariant techniques, which considerably simplified the mathematics~\cite{1997CQGra..14.2539T,1998CQGra..15.3523T,%
2000PhRvD..61h3519T,2000CQGra..17.2215T,2005CQGra..22..393T,%
2007PhR...449..131B}.

Magnetic fields generate and affect all three types of density inhomogeneities, namely scalar, vector and (trace-free) tensor inhomogeneities. The former are those commonly referred to as density perturbations and represent overdensities or underdensities in the matter distribution. Vector inhomogeneities describe rotational (i.e.~vortex-like) density perturbations. Finally, tensor-type inhomogeneities correspond to shape distortions. Following~\cite{1997CQGra..14.2539T,1998CQGra..15.3523T,%
2000PhRvD..61h3519T,2000CQGra..17.2215T,2005CQGra..22..393T,%
2007PhR...449..131B}, the scalar
\begin{equation}
\Delta= {a^2\over\rho}\,{\rm D}^2\rho\,,  \label{divs}
\end{equation}
describes linear perturbations in the density ($\rho$) of the matter and corresponds to the more familiar density contrast $\delta\rho/\rho$. Note that positive values for $\Delta$ indicate overdensitites and negative ones underdensities. In a perturbed, weakly magnetised and spatially flat Friedmann-Robertson-Walker (FRW) universe, the above defined scalar evolves according to\footnote{Basic aspects of the FRW dynamics and evolution are discussed in \S~\ref{sssFRWDs} and \S~\ref{sssSfEFRWMs}.}
\begin{equation}
\dot{\Delta}= 3wH\Delta- (1+w)\mathcal{Z}+ {3\over2}\,c_{\rm a}^2(1+w)H\mathcal{B}\,,  \label{dotDelta}
\end{equation}
where $\mathcal{Z}=a^2{\rm D}^2\Theta$ and $\mathcal{B}= (a^2/B^2){\rm D}^2 B^2$. The first of the last two variables describes linear inhomogeneities in the smooth Hubble expansion and the second represents perturbations in the magnetic energy density. Then, to linear order,
\begin{equation}
\dot{\mathcal{Z}}= -2H\mathcal{Z}- {1\over2}\,\rho\Delta+ {1\over4}\,c_{\rm a}^2(1+w)\rho\mathcal{B}- {c_s^2\over1+w}\,{\rm D}^2\Delta- {1\over2}\,c_{\rm a}^2{\rm D}^2\mathcal{B}  \label{dotcZ}
\end{equation}
and
\begin{equation}
\dot{\mathcal{B}}= {4\over3(1+w)}\,\dot{\Delta}\,,  \label{dotcB}
\end{equation}
respectively. In the above $w=p/\rho$ is the (constant) barotropic index of the matter, $H=\dot{a}/a$ is the background Hubble parameter, $c_s^2=\dot{p}/\dot{\rho}$ is the square of the adiabatic sound speed and $c_{\rm a}^2=B^2/\rho(1+w)$ is that of the Alfv\'en speed. We have also assumed that $B^2\ll\rho$, given the relative weakness of the magnetic field.

Expression (\ref{dotDelta}) verifies that $B$-fields are generic sources of linear density perturbations. Indeed, even when $\Delta$ and $\mathcal{Z}$ are zero initially, $\dot{\Delta}$ will generally take nonzero values due to the magnetic presence. Also, Eq.~(\ref{dotcB}) shows that linear perturbations in the magnetic energy density evolve in tune with those in the density of the matter (i.e.~$\mathcal{B}\propto\Delta$). This means that a $B$-field residing in an overdense region of an Einstein-de Sitter universe will grow by approximately two to three orders of magnitude (see solution (\ref{dDelta}) below). Note that the aforementioned growth occurs during the linear regime of structure formation and is independent of the (nonlinear) increase in the field's strength due to the adiabatic compression of a protogalactic cloud (see \S~\ref{sssAN-lR} for more details). Finally, we should emphasise that only the pressure part of the Lorentz force contributes to Eqs.~(\ref{dotDelta}) and (\ref{dotcZ}). To account for the tension effects at the linear level, one needs to allow for FRW backgrounds with non-Euclidean spatial geometry.\footnote{The Lorentz force splits according to $\varepsilon_{abc}B^b{\rm curl}B^c=(1/2){\rm D}_aB^2-B^b{\rm D}_bB_a$, with the former term carrying the effects of the magnetic pressure and the latter those of the field's tension (see Eq.~(\ref{MHDmc2}) in \S~\ref{sssIMHDL}). Given that ${\rm D}^aB_a=0$ at the ideal-MHD limit, only the pressure part affects linear density perturbations, unless FRW backgrounds with nonzero spatial curvature are involved.}

The system (\ref{dotDelta})-(\ref{dotcB}) has analytical solutions in the radiation and the dust eras. Before equipartition, when $w=1/3=c_s^2$, $H=1/2t$, $\rho=3/4t^2$ and $c_{\rm a}^2=3B^2/4\rho=\,$constant, large-scale magnetised density perturbations obey the power-law solution. In particular, keeping only the dominant growing and decaying modes one arrives at~\cite{2000PhRvD..61h3519T,2007PhR...449..131B}
\begin{equation}
\Delta= \mathcal{C}_1t^{-1/2+10c_{\rm a}^2/9}+ \mathcal{C}_2t^{1-4c_{\rm a}^2/9}\,.  \label{lsrDelta}
\end{equation}
In the absence of the $B$-field, we recover the standard growing and decaying modes of $\Delta\propto t$ and $\Delta\propto t^{-1/2}$ respectively. So, the magnetic presence has reduced the growth rate of the density contrast by $4c_{\rm a}^2/9$. Also, since  $\mathcal{B}\propto\Delta$ -- see Eq.~(\ref{dotcB}), the above describes the linear evolution of the magnetic energy-density perturbations as well.

Well inside the horizon we can no longer ignore the role of the pressure gradients. There, the $k$-mode oscillates like a magneto-sonic wave with
\begin{equation}
\Delta_{(k)}\propto \sin\left[c_s\left(1+{2\over3}\,c_{\rm a}^2\right) \left({\lambda_H\over\lambda_k}\right)_0\sqrt{t\over t_0}\, \right]\,,  \label{ssrDelta}
\end{equation}
where $\lambda_k=a/k$ is the perturbed scale and $\lambda_H=1/H$ the Hubble horizon~\cite{2000PhRvD..61h3519T,2007PhR...449..131B}. Here, the magnetic pressure increases the effective sound speed and therefore the oscillation frequency. The former makes the Jeans length larger than in non-magnetised models. The latter brings the peaks of short-wavelength oscillations in the radiation density closer, leaving a potentially observable signature in the CMB spectrum~\cite{1996PhLB..388..253A}.

When dust dominates, $w=0=c_s^2$, $H=2/3t$, $\rho=4/3t^2$ and $c_{\rm a}^2=B^2/\rho\propto t^{-2/3}$. Then, on superhorizon scales, the main growing and decaying modes of the density contrast are~\cite{1997CQGra..14.2539T,2007PhR...449..131B}
\begin{equation}
\Delta= \mathcal{C}_1t^{\alpha_1}+ \mathcal{C}_2t^{\alpha_2}\,,
\label{dDelta}
\end{equation}
with $\alpha_{1,2}= -[1\pm5\sqrt{1-(32/75) (c_{\rm a}\,\lambda_H/\lambda_k)_0^2}]/6$. In the absence of the $B$-field we recover again the standard solution with $\alpha_1=2/3$ and $\alpha_2=-1$. Thus, as with the radiation era before, the magnetic presence slows down the growth rate of density perturbations. In addition, the field's pressure leads to a magnetically induced Jeans length, below which density perturbations cannot grow. As a fraction of the Hubble radius, this purely magnetic Jeans scale is
\begin{equation}
\lambda_J\sim c_{\rm a}\lambda_H\,.  \label{mJeans}
\end{equation}
Setting $B\sim10^{-9}$~G, which is the maximum homogeneous field allowed by the CMB~\cite{1997PhRvL..78.3610B}, we find that $\lambda_J\sim10$~Kpc. Alternative, magnetic fields close to $10^{-7}$~G, like those found in galaxies and galaxy clusters, give $\lambda_J\sim1$~Mpc. The latter lies intriguingly close to the size of a cluster of galaxies.

Overall, the magnetic effect on density perturbations is rather negative. Although $B$-fields generate this type of distortions, they do not help them to grow. Instead, the magnetic presence either suppresses the growth rate of density perturbations, or increases the effective Jeans length and therefore the domain where these inhomogeneities cannot grow. The negative role of the $B$-field, which was also observed in the Newtonian treatment of~\cite{1971SvA....14..963R}, reflects the fact that only the pressure part of the Lorentz force has been incorporated into the equations. When the tension component (i.e.~the elasticity of the field lines) is also accounted for, the overall magnetic effect can change and in some cases it could even reverse~\cite{2000CQGra..17.2215T}.

Magnetic fields also induce and affect rotational, vortex-like, density inhomogeneities. To linear order, these are described by the vector $\mathcal{W}_a=-(a^2/2\rho)\varepsilon_{abc}{\rm D}^b{\rm D}^c\rho$. Then, on an spatially flat FRW background,
\begin{equation}
\ddot{\mathcal{W}}_a= -4H\dot{\mathcal{W}}_a- {1\over2}\,\rho\mathcal{W}_a+ {1\over3}\,c_{\rm a}^2{\rm D}^2\mathcal{W}_a\,,  \label{ddotW}
\end{equation}
after matter-radiation equality~\cite{2000PhRvD..61h3519T,2007PhR...449..131B}. Defining $\lambda_{\rm a}=c_{\rm a}\lambda_H$ as the `Alfv\'en horizon', we may write the associated solution in the form
\begin{equation}
\mathcal{W}_{(k)}= \mathcal{C}_1t^{\alpha_1}+ \mathcal{C}_2t^{\alpha_2}\,,  \label{dW}
\end{equation}
where $\alpha_{1,2}= -[5\pm\sqrt{1-(48/9) (\lambda_{\rm a}/\lambda_k)^2_0}]/6$. On scales well outside the Alfv\'en horizon, namely for $\lambda_{\rm a}\ll\lambda_k$, the perturbed mode decays as $\mathcal{W}\propto t^{-2/3}$. This rate is considerably slower than $\mathcal{W}\propto t^{-1}$, the decay rate associated with magnetic-free dust cosmologies. In other words, the $B$-field has reduced the standard depletion rate of the vortex mode. An analogous effect is also observed on $\omega_a$, namely on the vorticity proper~\cite{2000PhRvD..61h3519T,2007PhR...449..131B}. Hence, magnetised cosmologies rotate faster than their magnetic-free counterparts. In contrast to density perturbations, the field seems to favour the presence of vorticity. This qualitative difference should probably be attributed to the fact that the tension part of the Lorentz force also contributes to Eq.~(\ref{ddotW}).

In addition to scalar and vector perturbations, magnetic fields also generate and affect tensor-type inhomogeneities that describe shape-distortions in the density distribution~\cite{2000PhRvD..61h3519T}. An initially spherically symmetric inhomogeneity, for example, will change shape due to the magnetically induced anisotropy. All these are the effects of the Lorentz force. Even when the latter is removed from the system, however, the $B$-field remains active. Due to its energy density and anisotropic nature, for example, magnetism affects both the local and the long-range gravitational field. The anisotropic magnetic pressure, in particular, leads to shear distortions and subsequently to gravitational-wave production. Overall, magnetic fields are a very versatile source. They are also rather unique in nature, since $B$-fields are the only known vector source of energy. An additional unique magnetic feature, which remains relatively unexplored, is its tension. When we add to all these the widespread presence of magnetic fields, it makes sense to say that no realistic structure formation scenario should a priori exclude them.

\subsubsection{Aspects of the nonlinear regime}\label{sssAN-lR}
The evolution of large-scale magnetic fields during the nonlinear stages of structure formation has been addressed primarily by means of numerical methods. The reason is the high complexity of the nonlinear MHD equations, which makes analytical studies effectively impossible, unless certain simplifying assumptions are imposed.

The simplest approximation is to assume spherically symmetric compression. Realistic collapse, however, is not isotropic. In fact, when a magnetic field is present, its generically anisotropic nature makes the need to go beyond spherical symmetry greater. Anisotropic contraction can be analytically studied within the Zeldovich approximation~\cite{1970Afz.....6..319Z,1989RvMP...61..185S}. The latter is based on a simple ansatz, which extrapolates to the nonlinear regime a well known linear result. The assumption is that the irrotational and acceleration-free linear motion of the dust component, also holds during the early nonlinear stages of galaxy formation. This approximation allows for the analytical treatment of the nonlinear equations, the solution of which describe anisotropic (one dimensional) collapse and lead to the formation of the well-known Zeldovich `pancakes'.

Suppose that a magnetic field is frozen into a highly conductive protogalactic cloud that is falling into the (Newtonian) potential wells formed by the Cold Dark Matter (CMB) sector.\footnote{The Newtonian theory is a very good approximation, since the scales of interest are well below the curvature radius of the universe and we are dealing with non-relativistic matter.} Relative to the physical coordinate system $\{r^{\alpha}\}$, the motion of the fluid velocity is $u_{\alpha}=3Hr_{\alpha}+v_{\alpha}$, where $H=\dot{a}/a$ is the Hubble parameter of the unperturbed FRW background and $v_{\alpha}$ is the peculiar velocity of the fluid (with $\alpha=1,2,3$). Then, the magnetic induction equation reads~\cite{2003MNRAS.338..785B}
\begin{equation}
\dot{B}_{\alpha}= -2HB_{\alpha}- {2\over3}\,\vartheta B_{\alpha}+ \sigma_{\alpha\beta}B^{\beta}\,,  \label{ZdotB}
\end{equation}
where overdots now indicate convective derivatives (i.e.~$\dot{}=\partial_t+u^{\beta}\partial_{\beta}$). Also, $\vartheta=\partial^{\alpha}v_{\alpha}$ and $\sigma_{\alpha\beta}=\partial_{\langle\beta}v_{\alpha\rangle}$ are the peculiar volume scalar and the peculiar shear tensor respectively.\footnote{When dealing with purely baryonic collapse, the Zeldovich ansatz only holds during the early stages of the nonlinear regime, when the effects of the fluid pressure are negligible. Assuming that the contraction is driven by non-baryonic CDM, means that we can (in principle) extend the domain of the Zeldovich approximation beyond the above mentioned mildly nonlinear stage.} The former takes negative values (i.e.~$\vartheta<0$), since we are dealing with a protogalactic cloud that has started to `turn around' and collapse. Note that the first term in the right-hand side of (\ref{ZdotB}) represents the background expansion, the second is due to the peculiar contraction and the last reflects the anisotropy of the collapse. Introducing the rescaled magnetic field $\mathcal{B}_{\alpha}=a^2B_{\alpha}$, the above expression recasts into
\begin{equation}
\mathcal{B}^{\prime}_{\alpha}= -{2\over3}\,\tilde{\vartheta}\mathcal{B}_{\alpha}+ \tilde{\sigma}_{\alpha\beta}\mathcal{B}^{\beta}\,,  \label{ZrB}
\end{equation}
with primes indicating differentiation with respect to the scale factor. Also $\vartheta=aH\tilde{\vartheta}$ and $\sigma_{\alpha\beta}=aH\tilde{\sigma}_{\alpha\beta}$, where $\tilde{\vartheta}=\partial^{\alpha}\tilde{v}_{\alpha}$ and $\tilde{\sigma}_{\alpha\beta}= \partial_{\langle\beta}\tilde{v}_{\alpha\rangle}$ (with $\tilde{v}_{\alpha}=ax^{\prime}_{\alpha}$ and $v_{\alpha}=aH\tilde{v}_{\alpha}$). Relative to the shear eigenframe, $\tilde{\sigma}_{\alpha\beta}=(\tilde{\sigma}_{11},\, \tilde{\sigma}_{22},\,\tilde{\sigma}_{33})$ and expression (\ref{ZrB}) splits into
\begin{equation}
\mathcal{B}^{\prime}_1= -{2\over3}\,\tilde{\vartheta}\mathcal{B}_1+ \tilde{\sigma}_{11}\mathcal{B}_1\,, \hspace{25mm} \mathcal{B}^{\prime}_2= -{2\over3}\,\tilde{\vartheta}\mathcal{B}_2+ \tilde{\sigma}_{22}\mathcal{B}_2  \label{ZrB12}
\end{equation}
and
\begin{equation}
\mathcal{B}^{\prime}_3= -{2\over3}\,\tilde{\vartheta}\mathcal{B}_3+ \tilde{\sigma}_{33}\mathcal{B}_3\,.  \label{ZrB3}
\end{equation}
This system describes the second-order evolution of a magnetic field, which is frozen in with the highly conductive matter of a collapsing protogalaxy, within the limits of the Zeldovich approximation.

In order to solve the set of Eqs. (\ref{ZrB12}), (\ref{ZrB3}), we recall that in the absence of rotation and acceleration, the peculiar volume scalar is given by
\begin{equation}
\tilde{\vartheta}= {\lambda_1\over1+a\lambda_1}+ {\lambda_2\over1+a\lambda_2}+ {\lambda_3\over1+a\lambda_3}\,,  \label{vartheta}
\end{equation}
Similarly, the shear eigenvalues are
\begin{equation}
\tilde{\sigma}_{11}= {\lambda_1\over1+a\lambda_1}- {1\over3}\,\vartheta\,, \hspace{25mm} \tilde{\sigma}_{22}= {\lambda_2\over1+a\lambda_2}- {1\over3}\,\vartheta  \label{tsigma12}
\end{equation}
and
\begin{equation}
\tilde{\sigma}_{33}= {\lambda_3\over1+a\lambda_3}- {1\over3}\,\vartheta\,.  \label{tsigma3}
\end{equation}
where $\lambda_1$, $\lambda_2$ and $\lambda_3$ are the eigenvalues of the initial tidal field and determine the nature of the collapse~\cite{1996dmu..conf..601M,1996ASPC...94...31B}. In particular, one-dimensional collapse along, say, the third eigen-direction is characterised by $\lambda_1=0=\lambda_2$ and by $\lambda_3<0$. In that case, the pancake singularity is reached as $a\rightarrow-1/\lambda_3$. Spherically symmetric collapse, on the other hand, has $\lambda_1=\lambda_2=\lambda_3=\lambda<0$. Then, we have a point-like singularity when $a\rightarrow-1/\lambda$.

Substituting, the above expressions into the right-hand side of Eqs.~(\ref{ZrB12}) and (\ref{ZrB3}), we obtain the solutions
\begin{equation}
B_1= B^0_1\left[{(1+a_0\lambda_2)(1+a_0\lambda_3)\over (1+a\lambda_2)(1+a\lambda_3)}\right]\left({a_0\over a}\right)^2\,, \label{ZB1}
\end{equation}
\begin{equation}
B_2= B^0_2\left[{(1+a_0\lambda_1)(1+a_0\lambda_3)\over (1+a\lambda_1)(1+a\lambda_3)}\right]\left({a_0\over a}\right)^2  \label{ZB2}
\end{equation}
and
\begin{equation}
B_3= B^0_3\left[{(1+a_0\lambda_1)(1+a_0\lambda_2)\over (1+a\lambda_1)(1+a\lambda_2)}\right]\left({a_0\over a}\right)^2\,,  \label{ZB3}
\end{equation}
with the zero suffix corresponding to a given time during the protogalactic collapse. Note that the ratios in parentheses reflect the magnetic dilution due to the background expansion, while the terms in brackets monitor the increase in the field's strength caused by the collapse of the protogalactic cloud. According to the above solution, when dealing with pancake collapse along the third eigen-direction, the $B_3$-component decays as $a^{-2}$, while the other two increase arbitrarily. Alternatively, during a spherically symmetric contraction the $B$-field evolves as
\begin{equation}
B= B_0\left({1+a_0\lambda\over1+a\lambda}\right)^2\left({a_0\over a}\right)^2\,.  \label{sphB}
\end{equation}
Here, all the magnetic components diverge as we approach the point singularity (i.e.~for $a\rightarrow-1/\lambda$). Comparing the two results, we deduce that the anisotropic (pancake) collapse leads to a stronger increase as long as $\lambda_3<\lambda$. The latter is always satisfied, provided that the initial conditions are the same for both types of collapse, given that $\lambda_3=\tilde{\vartheta} /(1-a_0\tilde{\vartheta})$ and $\lambda_3=\tilde{\vartheta} /(3-a_0\tilde{\vartheta})$ -- see expression (\ref{vartheta}) above.

The above given qualitative analysis indicates that a magnetic field trapped in an anisotropically contracting protogalactic cloud will increase beyond the limits of the idealized spherically symmetric scenario. Note that this type of amplification mechanism appears to be the only alternative left if the galactic dynamo (see \S~\ref{sssGDP} below) fails to operate. Quantitatively, the achieved final strength depends on when exactly the backreaction of the $B$-field becomes strong enough to halt the collapse~\cite{1983flma....3.....Z}. Thus, the longer the anisotropic collapse lasts, the stronger the residual $B$-field. The analytical study of~\cite{2003MNRAS.338..785B}, in particular, showed that (realistically speaking) the anisotropy could add one or two orders of magnitude to the magnetic strengths achieved through conventional isotropic compression. These results are in very good agreement with numerical studies simulating shear and tidal effects on the magnetic evolution in galaxies and galaxy clusters~\cite{1999ApJ...518..594R,1999A&A...348..351D,%
2002A&A...387..383D}.

\newpage

\section{Magnetogenesis in conventional FRW models}\label{sMCFRWMs}
In order to operate successfully, the galactic dynamo needs magnetic seeds that satisfy two specific requirements. The first refers to the (comoving) coherence length of the initial $B$-field and the second is related to its strength. The scale must not drop below $\sim10$~Kpc. The strength typically varies between $\sim10^{-12}$~G and $\sim10^{-22}$~G, depending on the efficiency of the dynamo amplification. At first, these requirements may seem relatively straightforward to fulfill. Nevertheless, within classical electromagnetism and conventional FRW cosmology, magnetic seeds with the aforementioned desired properties are very difficult to produce.

\subsection{The Friedmann models}\label{ssFMs}
Current observations, primarily the isotropy of the CMB, strongly support a universe that is homogeneous and isotropic on cosmological scales. In other words our universe seems to be described by the simplest cosmological solution of the Einstein field equations, the FRW models. Before proceeding to discuss the magnetic evolution on FRW backgrounds, it helps to summarise some basic features of these models.

\subsubsection{The FRW Dynamics}\label{sssFRWDs}
The high symmetry of the Friedmann models means that all kinematical and dynamical variables are functions of time only, while every quantity that represents anisotropy or inhomogeneity vanishes identically. Thus, in covariant terms, an FRW model has $\Theta=3H(t)\neq0$, $\sigma_{ab}=0=\omega_a=A_a$, $E_{ab}=0=H_{ab}$, where $H=\dot{a}/a$ is the familiar Hubble parameter. The isotropy of the Friedmann models also constrains their matter content, which can only have the perfect-fluid form (with $\rho=\rho(t)$ and $p=p(t)$). In addition,
due to the spatial homogeneity, all orthogonally projected gradients (e.g.~${\rm D}_a\rho$, ${\rm D}_ap$, etc -- see \S~\ref{sMFC}) are by definition zero. This means that the only nontrivial equations left, are the FRW version of Raychaudhuri's formula, the equation of continuity and the Friedmann equation. These are given by~\cite{2008PhR...465...61T}
\begin{equation}
\dot{H}= -H^2- {1\over6}\,\kappa(\rho+3p)+ {1\over3}\,\Lambda\,,
\hspace{20mm} \dot{\rho}= -3H(\rho+p)  \label{FLRWeqs1}
\end{equation}
and
\begin{equation}
H^2= {1\over3}\,\kappa\rho- {K\over a^2}+ {1\over3}\,\Lambda\,,
\label{FLRWeqs2}
\end{equation}
respectively. Note that $K=0,\pm1$ is the 3-curvature index. The latter is  associated to the Ricci scalar ($\clr$) of the spatial sections by means of the relation $\clr=6K/a^2$~\cite{2008PhR...465...61T}.

In FRW spacetimes with non-Euclidean spatial geometry, the scale factor also defines the curvature scale ($\lambda_K=a$) of the model. This marks the threshold at which the effects of spatial curvature start becoming important (e.g.~see~\cite{1995PhRvD..52.3338L}). Lengths smaller than the curvature scale are termed subcurvature, while those exceeding
$\lambda_K$ are referred to as supercurvature. The former are essentially immune to the effects of spatial geometry, with the latter dominating on supercurvature lengths. The relation between the curvature scale and the Hubble radius is determined by Eq.~(\ref{FLRWeqs2}). In the absence of a cosmological constant, the latter reads
\begin{equation}
\left({\lambda_K\over\lambda_H}\right)^2=
-{K\over1-\Omega_{\rho}}\,,  \label{K/H}
\end{equation}
where $\lambda_H=H^{-1}$ and $\Omega_{\rho}=\kappa\rho/3H^2$ are the Hubble radius and the density parameter respectively. Hence, hyperbolic 3-geometry (i.e.~$K=-1$) ensures that $\lambda_K>\lambda_H$ always, with $\lambda_K\rightarrow\infty$ as $\Omega_{\rho}\rightarrow1$ and $\lambda_K\rightarrow\lambda_H$ for $\Omega_{\rho}\rightarrow0$. In practice, this means that supercurvature scales in spatially open FRW cosmologies are always outside the Hubble radius.\footnote{In Friedmann universes with hyperbolic spatial sections, the particle horizon can exceed the Hubble scale (e.g.~see~\cite{1972gcpa.book.....W}). This means that wavelengths larger than the curvature radius of an open FRW cosmology can be causally connected, despite the fact that they always lie outside the Hubble length. Here, to keep things simple, we will treat the Hubble scale as our causal horizon as well.} This is not the case in closed models, where $\lambda_K>\lambda_H$ when $\Omega_{\rho}<2$ and $\lambda_K\leq\lambda_H$ if $\Omega_{\rho}\geq2$. Finally, we note that, since the curvature scale simply redshifts with the expansion, the importance of spatial geometry within a comoving region does not change with time.

\subsubsection{Scale-factor evolution in FRW
models}\label{sssSfEFRWMs}
In order to close the system (\ref{FLRWeqs1}), one needs to introduce an equation of state for the matter. Here, we will only consider barotropic perfect fluids, mainly in the form of non-relativistic `dust' or isotropic radiation (with $p=0$ and $p=\rho/3$ respectively). When $w=p/\rho$ is the (constant) barotropic index of the medium, the continuity equation (see (\ref{FLRWeqs1}b)) gives $\rho\propto a^{-3(1+w)}$. Then, setting $K=0=\Lambda$ and normalising so that $a(t=0)=0$, we obtain
\begin{equation}
a= a_0\left({t\over t_0}\right)^{2/3(1+w)}\,,  \label{sfFLRWa}
\end{equation}
when $w\neq-1$. For non-relativistic matter with $w=0$ (e.g.~baryonic dust or non-baryonic cold dark matter), we have the Einstein-de Sitter universe with $a\propto t^{2/3}$. Alternatively, $a\propto t^{1/2}$ in the case of relativistic species (e.g.~isotropic radiation) and $a\propto t^{1/3}$ for a stiff medium with $w=1$. When $w=-1/3$, which corresponds to matter with zero gravitational mass, the above leads to `coasting' expansion with $a\propto t$. Solution (\ref{sfFLRWa}) does not apply to the $w=-1$ case. There, both $\rho$ and $H$ are constant to ensure de Sitter-type inflation with $a\propto{\rm e}^{H_0(t-t_0)}$.

When the FRW spacetime has non-Euclidean spatial geometry it helps to use conformal rather than proper time. Then, for $K=+1$, $\Lambda=0$ and $w\neq-1/3$ Eqs.~(\ref{FLRWeqs1}), (\ref{FLRWeqs2}) combine to~\cite{2008PhR...465...61T}
\begin{equation}
a= a_0\left\{{\sin[(1+3w)\eta/2]\over
\sin[(1+3w)\eta_0/2]}\right\}^{2/(1+3w)}\,,  \label{scFLRWa}
\end{equation}
where $\eta$ is the conformal time (with $\dot{\eta}=1/a$). Also, $(1+3w)\eta/2\in(0,\pi)$ and we have normalised so that
$a(\eta\rightarrow0)\rightarrow0$. For non-relativistic matter $w=0$ and the above solution reduces to $a\propto\sin^2(\eta/2)$~\cite{1983itc..book.....N}, while we obtain $a\propto\sin\eta$ if radiation dominates. When $w=-1/3$ one can no longer use solution (\ref{scFLRWa}). Instead, Eq.~(\ref{FLRWeqs1}a) leads immediately to the familiar coasting-expansion phase with $a\propto t$. Finally, in the $w=-1$ case, expression (\ref{FLRWeqs1}b) ensures that $\rho=\rho_0=$~constant and (\ref{FLRWeqs2}) leads to $a(1+\sqrt{3/\rho_0}H)\propto {\rm e}^{\sqrt{(\rho_0/3)}\,t}$.

Applied to FRW cosmologies with hyperbolic spatial geometry, zero cosmological constant and $w\neq-1/3$, the same analysis leads to
\begin{equation}
a= a_0\left\{{\sinh[(1+3w)\eta/2]\over
\sinh[(1+3w)\eta_0/2]}\right\}^{2/(1+3w)}\,,  \label{soFLRWa}
\end{equation}
where now $(1+3w)\eta/2>0$~\cite{2008PhR...465...61T}. Assuming pressure-free dust and normalising as before, we find $a\propto\sinh^2(\eta/2)$~\cite{1983itc..book.....N}. On the other hand, solution (\ref{soFLRWa}) implies $a\propto\sinh\eta$ for a open FRW universe dominated by relativistic species. We finally note that, similarly to the $K=+1$ case, the system (\ref{FLRWeqs1}), (\ref{FLRWeqs2}) ensures that $a\propto t$ when $w=-1/3$ and $a(1+\sqrt{3/\rho_0}H)\propto {\rm e}^{\sqrt{(\rho_0/3)}\,t}$ for $w=-1$. A special spacetime with open spatial geometry is the vacuum Milne universe, where the absence of matter guarantees that $a=t$.

\subsection{Late vs early-time magnetogenesis}\label{ssL-EtM}
The various mechanisms of magnetogenesis have been traditionally classified into those operating at late times, that is after recombination, and the ones that advocate an early (pre-recombination) origin for the $B$-field. In either case, the aim of the proposed scenarios is to produce the initial magnetic fields that will successfully seed the galactic dynamo.

\subsubsection{The galactic dynamo paradigm}\label{sssGDP}
The belief that some kind of nonlinear dynamo action is responsible for amplifying and sustaining the galactic magnetic fields has long roots in the astrophysical community~\cite{1978mfge.book.....M,%
1979cmft.book.....P,1983flma....3.....Z}. Dynamos provide the means of converting kinetic energy into magnetic energy and the reader is referred to~\cite{2005PhR...417....1B} for a recent extended review. Nevertheless, one can get a quick insight of how the mechanism in principle works by looking at the magnetic induction equation. In the Newtonian limit and assuming resistive MHD, the latter reads
\begin{equation}
\dot{B}_{\alpha}= -{2\over3}\,\vartheta B_{\alpha}+ \left(\sigma_{\alpha\beta}+\omega_{\alpha\beta}\right)B^{\beta}- \varsigma^{-1}\mathrm{curl}\mathcal{J}_{\alpha}\,,  \label{Nmie1}
\end{equation}
where overdots indicate convective derivatives (see also \S~\ref{sssAN-lR}) and $\alpha$, $\beta=$1,2,3. Contracting the above along the magnetic field vector and recalling that $\mathrm{curl}B_{\alpha}= \mathcal{J}_{\alpha}$, leads to
\begin{equation}
\left(B^2\right)^{\cdot}= -{4\over3}\,\vartheta B^2+ 2\sigma_{\alpha\beta}B^{\alpha}B^{\beta}+ 2\varsigma^{-1}\partial^{\alpha}F_{\alpha}- 2\varsigma^{-1}\mathcal{J}^2\,,  \label{Nmie2}
\end{equation}
with $F_{\alpha}= \varepsilon_{\alpha\beta\mu}B^{\beta}\mathcal{J}^{\mu}$ representing the magnetic Lorentz force. The latter contributes to the kinetic energy of the fluid via the Navier-Stokes equation (see expression (\ref{tmc}) in \S~\ref{sssCLs}). Following (\ref{Nmie2}), the action of the Lorentz force can in principle enhance the magnetic energy at the expense of the fluid's kinetic energy. The amplification can happen provided that the dissipative effects, carried by the last term on the right-hand side of Eq.~(\ref{Nmie2}), are subdominant. The first term, on the other hand gives the magnetic increase caused by the adiabatic galactic collapse (typically $\vartheta\lesssim0$ in gravitationally bounded systems), while the second conveys the effects of the shearing stresses (see \S~\ref{sssAN-lR}).

Dynamos are typically powered by the differential rotation of the galaxy. The latter combines with the small-scale turbulent motion of the ionised gas causing the exponential increase of the large-scale mean $B$-field in the plane of the galactic disc. The growth continues until it reaches saturation, which typically occurs when $B\sim10^{-6}$~G and the backreaction of the magnetic stresses suppresses any further increase. The amplification factor, however is quite sensitive to the specific parameters of the dynamo model. This sensitivity leads to serious uncertainties regarding the total ammount of magnetic amplification and has been the subject of ongoing discussions.

The pattern and the orientation of the galactic magnetic field, especially of those seen in spiral galaxies, seem to support the dynamo idea. On the other hand, the detection of strong magnetic fields in high-redshift protogalactic structures has raised a number of question regarding the role of dynamos~\cite{2008ApJ...676...70K,2008Natur.454..302B}. In any case, the galactic dynamo needs the presence of an initial magnetic seed in order to operate. These seeds must satisfy certain requirements regarding their coherence length and strength. The minimum required scale for the magnetic seed is comparable to the size of the smallest turbulent eddy. This lies close to 100~pc at the time the host galaxy is formed, which translates to a comoving length of approximately 10~Kpc before the collapse of the protogalactic cloud. The strength of the seed-field, on the other hand, varies, depending on the efficiency of the dynamo amplification and on the cosmological model it operates  in~\cite{1992ApJ...396..606K,%
1993IAUS..157..487K,1997A&A...322...98H,2005PhR...417....1B}. Typically, the required values range between $10^{-12}$~G and $10^{-22}$~G. It is conceivable, however, that the lower limit could be brought down to $10^{-30}$~G in spatially open or dark-energy dominated FRW universes~\cite{1999PhRvD..60b1301D}. Note that, in the absence of the dynamo, protogalactic collapse (spherically symmetric or anisotropic -- see \S~\ref{sssAN-lR}) seems the only alternative means of magnetic amplification. Then, $B$-seeds as strong as $10^{-9}$~G may be needed in order to meet the observations. So, provided that galactic dynamos work, the question is where do the initial magnetic seeds come from?

\subsubsection{Late-time magnetogenesis}\label{sssL-tM}
Post recombination mechanisms of magnetic generation appeal to astrophysical processes and battery-type effects. It has been proposed, in particular, that the Biermann-battery mechanism can produce seed $B$-fields, which the dynamo could subsequently amplify on galactic scales and to the observed strengths. The Biermann effect~\cite{1950ZNatA...5...65B}, which was originally discussed in the stellar context, exploits differences between the electron and the ion acceleration that are triggered by pressure gradients. These will first give rise to electric currents and subsequently lead to magnetic fields by induction.

The literature contains several alternative scenarios using battery-type mechanisms to generate magnetic seed-fields in the post recombination era. Supernovae explosions of the first stars, for example, could eject into the interstellar medium $B$-fields that could seed the galactic dynamo~\cite{1998MNRAS.301..547M,%
2005ApJ...633..941H}. Active galaxies and Active Galactic Nuclei (AGN) can also channel away jets of magnetised plasma~\cite{1990ApJ...364..451D,2000IAUS..195..255C}. Thermal-battery processes operating in (re)ionisation fronts can also lead to magnetic seeds that can sustain the dynamo~\cite{1994MNRAS.271L..15S,2000ApJ...539..505G}. Analogous results could be achieved through turbulent motions or shocks developed in collapsing protogalactic clouds~\cite{1997ApJ...480..481K,2000ApJ...540..755D}.

Nevertheless, while Biermann battery effects can produce the seed fields that the dynamo will subsequently amplify to the observed strengths, the whole process operates on galactic scales. For this reason, it is less straightforward to invoke the Biermann mechanism when trying to explain the magnetic fields found in galaxy clusters. To a certain extent, this also weakens the overall position of the Biermann battery as a likely candidate for generating the galactic magnetic fields. Indeed, the possibility that the galactic and the cluster $B$-fields have a different origin seems rather unlikely, in view of their similarities.

\subsubsection{Early-time magnetogenesis}\label{sssE-tM}
The idea that cosmic magnetism might have pre-recombination origin is attractive because it makes the widespread presence of magnetic fields in the universe easier to explain. Especially the origin of the fields observed in high-redshift proto-galactic condensations. However, generating cosmological $B$-fields that will also successfully seed the galactic dynamo is not a problem-free exercise.

In the early 1970s, Harrison proposed that battery-type effects, operating during the radiation era, could generate $B$-fields with strengths capable of sustaining the galactic dynamo~\cite{1970MNRAS.147..279H}. The mechanism is based on conventional physics and does not need any new postulates. The disadvantage in Harrison's idea is that it requires significant amounts of primordial vorticity. The latter is essentially absent from the standard cosmological model. Note that the possibility of simultaneously generating both vorticity and magnetic fields in the late radiation era and around recombination (when the tight-coupling between photons and baryons is relaxed) was recently investigated in~\cite{2005PhRvD..71d3502M,2007PhRvD..75j3501K}.

An alternative approach is to generate the magnetic seeds during phase transitions early in the radiation era (see \S~\ref{sM-GSM} below). There are problems, however, primarily related to the coherence length of the initial $B$-field. The difficulties arise because the size of the post-inflationary magnetic seeds, namely those created between inflation and (roughly) recombination is typically too small and will destabilise the dynamo. The reason is causality, which confines the scale of the field within that of the horizon at the time of magnetogenesis. For example, $B$-fields produced during the electroweak phase transition have coherence lengths of the order of the astronomical unit. The size of the magnetic field can increase if the host plasma has some degree of MHD turbulence. In such environments ``cascade'' processes are known to occur, whereby certain ideally conserved quantities flow from larger towards smaller scales (direct cascade) or the other way around (inverse cascade). In three
dimensional MHD turbulence, the total (kinetic plus magnetic) energy cascades toward smaller scales, where it is dissipated by viscosity and resistivity. However, the other important ideal invariant, the {\em magnetic helicity}, inverse-cascades towards larger scales. The magnetic helicity is defined by the integral~\cite{1975JFM....68..769F}
\begin{equation}
\mathcal{H}_M ={\frac{1}{V}}\int A^{a}B_{a}\mathrm{d}V\,,  \label{helicity}
\end{equation}
where $A_{a}$ is the electromagnetic vector potential (recall that $B_{a}=\mathrm{curl}A_{a}$) and is equivalent to the Chern-Simon number of particle physics~\cite{1997PhRvD..56.6146C}. Besides being an ideal invariant, the magnetic helicity is also asymptotically conserved within the resistive MHD approximation. Physically, $\mathcal{H}_M$ describes the topology of the field lines, that is their degree of withering and twisting~\cite{1984JFM....147..133B}. As mentioned before, magnetic helicity inverse-cascades and evolves from smaller to larger scales~\cite{1975JFM....68..769F}, while its conservation has profound effects in the operation of MHD dynamos~\cite{2005LNP...664..219B}. The aforementioned inverse-cascade effect makes primordial helicity very important, because it allows the magnetic energy to shift from smaller to larger scales, as the system tries to minimize its energy while conserving magnetic helicity. For example, Pouquet et al carried out a study in which nonhelical kinetic energy and maximally helical magnetic energy were injected into the plasma at a constant rate~\cite{1976JFM....77..321P}. The outcome was a well defined wave of magnetic energy and helicity, propagating from smaller to larger scales. Similar results were also obtained in the case of steady turbulence~\cite{1996PhRvD..54.1291B,1997PhRvD..56.6146C,%
1999PhRvD..59f3008S,2001PhRvE..64e6405C,1981PhRvL..47.1060M,%
1999PhPl....6...89B,2001ApJ...550..824B} and for freely decaying MHD turbulence~\cite{2007PhRvL..98y1302C}.

Although helical magnetic fields can enhance their original length,  inflation seems to be the only effective solution to the scale problem faced by fields generated during the early radiation era. The reason is that inflation can naturally generate correlations on superhorizopn lengths. There are still problems, however, this time with the magnetic strength. In particular, $B$-fields that has survived a period of standard de Sitter inflation are typically to weak to sustain the galactic dynamo.

\subsection{Typical inflation produced magnetic 
fields}\label{ssTIPMFs}
Inflation is known to produce long wavelength effects from microphysical processes that operate well inside the Hubble radius. For this reason, inflation has long been seen as the best candidate for producing large-scale, cosmological magnetic fields. Here, we will look at scenarios operating within standard electromagnetic theory and conventional FRW models. Alternative approaches are given in \S~\ref{sM-GBSM}.

\subsubsection{Quantum-mechanically produced magnetic seeds}
The inflationary paradigm provides the dynamical means of producing long-wavelength electromagnetic fluctuations, by stretching subhorizon-sized quantum mechanical fluctuations to superhorizon scales. Roughly speaking, quantum fluctuations in the Maxwell field are excited inside the horizon and cross the Hubble horizon approximately
\begin{equation}
N= N(\lambda)= 45+ \ln(\lambda)+{2\over3}\,\ln\left({M\over10^{14}}\right)+ {1\over3}\,\left({T_{RH}\over10^{10}}\right)\,,  \label{N}
\end{equation}
e-folds before the end of the de Sitter phase~\cite{1990eaun.book.....K}. In the above $\lambda$ is the comoving scale of the mode (measured in Mpc and normalised to coincide with the mode's current physical length), $M$ is the scale of inflation and $T_{RH}$ is the reheat temperature (both measured in GeV). Assuming that $\rho$ is the energy density of the electromagnetic mode, then
\begin{equation}
{\mathrm{d}\rho\over\mathrm{d}k}\sim H^3\,,  \label{krhoEM}
\end{equation}
at the first horizon crossing. Once outside the Hubble radius, the aforementioned quantum-mechanically excited modes are expected to freeze-out as classical electromagnetic waves. The latter, which initially appear like static electric and magnetic fields, can subsequently lead to current-supported magnetic fields. This happens after the modes have re-entered the horizon in the radiation era, or later during the dust-dominated epoch. Note that, after the second horizon crossing, the currents of the highly conductive plasma will also eliminate the electric component of the Maxwell field, leaving the universe permeated by a large-scale $B$-field of primordial origin.

The fast expansion of the de Sitter phase means that, by the end of inflation, the initial electromagnetic quantum fluctuations have achieved correlation lengths much larger than the current size of the observable universe. Thus, inflation produced $B$-fields have no scale problem whatsoever. Nevertheless, magnetic seeds that have survived a period of de Sitter expansion are generally too weak to sustain the dynamo. In particular, the typical strength of the residual $B$-field (in today's values) is less than $10^{-50}$~G~\cite{1988PhRvD..37.2743T}. To understand why and how this happens, we first need to consider the linear magnetic evolution on FRW backgrounds.

\subsubsection{The adiabatic magnetic decay}\label{ssAMD}
The evolution of large-scale electromagnetic fields on FRW backgrounds depends on the electric properties of the medium that fills the universe. Here, we will consider the two limiting cases of poorly and highly conductive matter. For any intermediate case, one needs a model for the electrical conductivity of the cosmic medium.

In poorly conductive environments, $\varsigma\rightarrow0$ and the electric currents vanish despite the presence of nonzero electric fields (see Ohm's law (\ref{Ohm}) in \S~\ref{ssCMHD}). Then, the wave equation (\ref{ddotBa}) linearises to
\begin{equation}
\ddot{B}_a- \D^2B_a= -5H\dot{B}_a+ {1\over3}\,\kappa\rho(1+3w)B_a- 4H^2B_a- \clr_{ab}B^b\,,  \label{lddBa}
\end{equation}
where $H=\dot{a}/a$ is the Hubble parameter of the unperturbed model. To simplify the above we introduce the rescaled magnetic field $\clb_a=a^2B_a$ and employ conformal, rather than proper, time~\cite{1988PhRvD..37.2743T}. Then, on introducing the harmonic splitting $\mathcal{B}_a= \sum_n\mathcal{B}_{(n)}\mathcal{Q}_a^{(n)}$ -- with ${\rm D}_a\mathcal{B}_{(n)}=0$, expression (\ref{lddBa}) takes the compact form\footnote{We use pure-vector harmonics that satisfy the constrains $\dot{\mathcal{Q}}_a^{(n)}=0= {\rm D}^a\mathcal{Q}_a^{(n)}$ and the associated Laplace-Beltrami equation, namely ${\rm D}^2\mathcal{Q}_a^{(n)}= -(n/a)^2\mathcal{Q}_a^{(n)}$. Following~\cite{1963AdPhy..12..185L}, the (comoving) eigenvalues depend on the background spatial curvature according to $n^2=\nu^2+2K$, where $\nu$ represents the associated wavenumber. Also, $n$ has a continuous spectrum, with $n^2\geq0$, when $K=0,-1$ and a discrete one, with $n^2\geq2$, if $K=+1$.}
\begin{equation}
\mathcal{B}_{(n)}^{\prime\prime}+ n^2\mathcal{B}_{(n)}= -2K\mathcal{B}_{(n)}\,,  \label{cB''}
\end{equation}
with the primes denoting conformal-time derivatives and $K=0,\pm1$~\cite{2005CQGra..22..393T}. The above describes the linear evolution of the rescaled magnetic field on a Friedmannian background with any type of spatial curvature. Note the magneto-curvature term on the right-hand side of (\ref{cB''}), which results form the direct coupling between the $B$-field and the geometry of the 3-space. The interaction is monitored by the Ricci identities and reflects the fact that we are dealing with an energy source of vector nature within a geometrical theory of gravity. We will discuss the implications of this interaction, which is largely bypassed in the literature, for the evolution of large-scale magnetic fields in \S~\ref{ssMACFRWM}.

When the FRW host has Euclidean spatial hypersurfaces, the 3-curvature index is zero (i.e.~$K=0$) and expression (\ref{cB''}) assumes the Minkowski-like form
\begin{equation}
\mathcal{B}_{(n)}^{\prime\prime}+ n^2\mathcal{B}_{(n)}=0\,.  \label{EcB''}
\end{equation}
The latter accepts the oscillatory solution
\begin{equation}
\mathcal{B}_{(n)}= \mathcal{C}_1\sin(n\eta)+ \mathcal{C}_2\sin(n\eta)\,,  \label{EcB}
\end{equation}
with the integration constants depending on the initial conditions. Then, recalling that $\mathcal{B}_{(n)}=a^2B_{(n)}$, the above given  solution translates into
\begin{eqnarray}
B_{(n)}&=& \left({a_0\over a}\right)^2 \left\{B_0^{(n)}\sin(n\eta_0) +{1\over n}\left[B_0^{\prime\,(n)}+2\left({a^{\prime}\over a}\right)_0B_0^{(n)}\right]\cos(n\eta_0)\right\}\sin(n\eta) \nonumber\\ &&+\left({a_0\over a}\right)^2 \left\{B_0^{(n)}\cos(n\eta_0) -{1\over n}\left[B_0^{\prime\,(n)} +2\left({a^{\prime}\over a}\right)_0B_0^{(n)}\right] \sin(n\eta_0)\right\}\cos(n\eta)\,.  \label{adB}
\end{eqnarray}
This guarantees an adiabatic ($B_{(n)}\propto a^{-2}$) depletion for the magnetic field, irrespective of the equation of state of the matter, as long as the background spacetime is a spatially flat FRW model and the electrical conductivity remains very poor.

The adiabatic magnetic decay is also guaranteed in highly conductive environments, namely at the ideal-MHD limit. There, $\varsigma\rightarrow\infty$ and, according to Ohm's law -- see Eq.~(\ref{Ohm}) in \S~\ref{ssCMHD}, the electric field vanishes in the frame of the fluid. As a result, when linearised around a FRW background, Faraday's law (see Eq.~(\ref{M2}) in \S~\ref{ssRMC}) reduces to
\begin{equation}
\dot{B}_a= -2HB_a\,.  \label{MHDadB}
\end{equation}
The latter guarantees that $B_a\propto a^{-2}$ on all scales, regardless of the equation of state of the matter and of the background 3-curvature.

The universe is believed to have been a very good electrical conductor throughout its classical Big-Bang evolution, at least on subhorizon scales. During inflation, on the other hand, the conductivity is expected to be very low. However, as the universe leaves the inflationary phase and starts reheating, its conductivity grows. So, by the time we have entered the radiation era, the currents have eliminated the electric fields and frozen their magnetic counterparts in with the matter.\footnote{Causality ensures that the post-inflationary currents are confined within the horizon. This means that outside the Hubble radius the low conductivity assumption still holds.} These arguments essentially guarantee that the set (\ref{EcB''}) and (\ref{MHDadB}) monitors the evolution of cosmological magnetic fields throughout the universe's lifetime. This in turn has led to the widespread belief that the adiabatic decay-rate of cosmological magnetic fields is ensured at all times, unless classical electromagnetism is modified or the FRW symmetries are abandoned. As we will see in \S~\ref{ssMACFRWM}, however, this is not necessarily the case.

\subsubsection{The residual magnetic field}\label{sssRMF}
The immediate implication of (\ref{adB}) is that magnetic fields that survived a period of typical de Sitter-type inflation have been drastically diluted by the accelerated expansion of the universe. Together with (\ref{MHDadB}), this means that $B$-fields of primordial origin are too weak to be of astrophysical relevance today. To demonstrate the dramatic magnetic depletion during the de Sitter phase we follow~\cite{1988PhRvD..37.2743T}. As a first step, recall that the relative energy density stored in the $n$-th magnetic mode at the (first) horizon crossing is $(\rho_B/\rho)_{HC}\simeq(H/M_{Pl})^2$. Here, $\rho_B=B^2_{(n)}$, $\rho$ is the energy density of the background universe and $M_{Pl}$ is the Planck mass. During inflation the total energy density is dominated by that of the vacuum (i.e.~$\rho\simeq M^4$, with $M$ representing the energy scale of the adopted inflationary scenario). Consequently, the relative strength of the $n$-th magnetic mode at horizon crossing is $(\rho_B/\rho)_{HC}\simeq(M/M_{Pl})^4$. Also, throughout inflation the universe is believed to be a very poor electrical conductor. This means that any magnetic field that may be present at the time decays adiabatically (see solution (\ref{adB}) in \S~\ref{ssAMD}). As a result, $B^2_{(n)}=(B^2_{(n)})_{HC}{\rm e}^{-4N}$ by the end of inflation, with $N=\ln(a_{IN}/a_{HC})$ representing the number of e-folds between horizon crossing and the end of the de Sitter era. This number depends on the scale of the mode in question and, in typical inflationary scenarios, is given by Eq.~(\ref{N}). Using the latter and recalling that $(\rho_B/\rho)_{RH}=(\rho_B/\rho)_{IN}(T_{RH}/M)^{4/3}$ is the relative change of the magnetic energy density between the end of inflation proper and that of reheating, we find that~\cite{1988PhRvD..37.2743T}
\begin{equation}
\left({\rho_B\over\rho_{\gamma}}\right)_{RH}\simeq 10^{-104} \lambda^{-4}\,,  \label{adr1}
\end{equation}
at the onset of the radiation era. Note that $\rho_{\gamma}\simeq\rho_{RH}\simeq T_{RH}^4$ represents the energy density of the relativistic species and $\lambda$ is the current (comoving) scale of the $B$-field. Also, the above ratio is independent of the energy scale of the adopted inflationary scenario and of the associated reheat temperature. Moreover, given that $\rho_B$, $\rho_{\gamma}\propto a^{-4}$ throughout the subsequent evolution of the universe, the same ratio remains unchanged until the time of galaxy formation.

Once the scale of the magnetic mode is given, we can use (\ref{adr1}) to evaluate the residual strength of any primordial $B$-field that underwent an era of (typical) de Sitter inflation. For example, in order to operate successfully, the galactic dynamo requires magnetic seeds with a minimum coherence scale of approximately 10~Kpc. Substituting this scale into Eq.~(\ref{adr1}) and recalling that $\rho_{\gamma}\simeq10^{-51}$~GeV today, we find that the corresponding magnetic field has strength of $\sim10^{-53}$~G~\cite{1988PhRvD..37.2743T}. This value is well below the galactic dynamo requirements, which leads to the conclusion that magnetic fields that have survived a period of standard, de Sitter-type inflationary expansion are (for all practical purposes) astrophysically irrelevant.

\subsection{Magnetic amplification in conventional FRW
models}\label{ssMACFRWM}
The ``negative'' results of the previous section have been been widely attributed to the conformal invariance of Maxwell's equations and to the conformal flatness of the Friedmannian spacetimes. The two have been thought to guarantee an adiabatic decay-rate for all large-scale magnetic fields at all times. This, in turn, has led to the widespread perception that inflation produced $B$-fields are astrophysically unimportant, unless standard electromagnetism is modified or the FRW symmetries are abandoned.

\subsubsection{Superadiabatic amplification}\label{sssS-AA}
Strictly speaking, the adiabatic magnetic depletion seen in solution (\ref{adB}) of \S~\ref{ssTIPMFs} has only been proved on Friedmannian backgrounds with Euclidean spatial sections. Although it is true that all three FRW universes are conformally flat, they are not the same. There are differences in their 3-curvature, which mean that only the spatially flat model is globally conformal to  Minkowski space. For the rest, the conformal mappings are local~\cite{1990grit.book.....S,2000CQGra..17..201K}. Another way of putting it is that, when dealing with spatially curved Friedmann universes, the conformal factor is no longer the cosmological scale factor but has an additional spatial dependence. All these imply that the wave equation of the rescaled magnetic field ($\mathcal{B}_a=a^2B_a$) takes the simple Minkowski-like form (\ref{EcB''}), which guarantees an adiabatic decay for the actual $B$-field, only on FRW backgrounds with zero 3-curvature. In any other case, there is an additional curvature-related term
(see expressions (\ref{lddBa}) and (\ref{cB''}) in \S~\ref{ssTIPMFs}), reflecting the non-Euclidean spatial geometry of the host spacetime. As a result, when linearised around an FRW background with nonzero spatial curvature, the magnetic wave equation reads~\cite{2005CQGra..22..393T}
\begin{equation}
\mathcal{B}_{(n)}^{\prime\prime}+ \left(n^2\pm2\right)\mathcal{B}_{(n)}= 0\,,  \label{pm1cB''}
\end{equation}
with the plus and the minus signs indicating the spatially closed and the spatially open model respectively. Recall that in the former case the eigenvalue is discrete (with $n^2\geq2$), while in the latter it is continuous (with $n^2\geq0$). In either case, the curvature-related effects fade away as we move down to successively smaller scales (i.e.~for $n^2\gg2$).

According to (\ref{pm1cB''}), on FRW backgrounds with spherical spatial hypersurfaces, the $B$-field still decays adiabatically. The curvature term only modifies the frequency of the magnetic oscillation in accord with the solution~\cite{2005PhRvD..71l3506T}
\begin{equation}
B_{(n)}= \left[\mathcal{C}_1\sin\left(\sqrt{n+2}\,\eta\right)+ \mathcal{C}_2\cos\left(\sqrt{n+2}\,\eta\right)\right]\left({a_0\over a}\right)^2\,.  \label{+1B}
\end{equation}
Overall, the adiabatic decay-rate of the $B$-field remains. Also, as expected, the smaller the scale the less important the role of the background 3-geometry.

The standard picture, and the adiabatic-decay law, change when the background FRW model has open spatial sections. In particular, the hyperbolic geometry of the 3-D hypersurfaces alters the nature of the magnetic wave equation on large enough scales (i.e.~when $0<n^2<2$). These wavelengths include what we may regard as the largest subcurvature modes (i.e.~those with $1\leq n^2<2$) and the supercurvature lengths (having $0<n^2<1$). Recall that eigenvalues with $n^2=1$ correspond to the curvature scale with physical wavelength $\lambda=\lambda_K=a$ (see \S~\ref{sssFRWDs}).

Following~\cite{2005PhRvD..71l3506T,2008PhRvD..77j7302B,%
2011arXiv1101.2390B}, we introduce the scale-parameter $k^2=2-n^2$, with $0<k^2<2$. Then, $k^2=1$ indicates the curvature scale, the range $0<k^2<1$ corresponds to the largest subcurvature modes and their supercurvature counterparts are contained within the $1<k^2<2$ interval. In the new notation and with $K=-1$, Eq.~(\ref{pm1cB''}) reads
\begin{equation}
\mathcal{B}_{(n)}^{\prime\prime}- k^2\mathcal{B}_{(n)}= 0\,,  \label{-1cB''}
\end{equation}
while its solution leads to
\begin{equation}
\mathcal{B}_{(k)}= \mathcal{C}_1\sinh(|k|\eta) +\mathcal{C}_2\cosh(|k|\eta)\,.  \label{-1cB1}
\end{equation}
Written with respect to the actual magnetic field, the above takes the form
\begin{eqnarray}
B_{(k)}&=& {1\over2}\,\left\{B_0^{(k)} +{1\over|k|}\left[B_0^{\prime\,(k)} +2\left({a^{\prime}\over a}\right)_0B_0^{(k)}\right]\right\}\left({a\over a_0}\right)^{-2} {\rm e}^{|k|(\eta-\eta_0)} \nonumber\\ &&+ {1\over2}\,\left\{B_0^{(k)} -{1\over|k|}\left[B_0^{\prime\,(k)} +2\left({a^{\prime}\over a}\right)_0B_0^{(k)}\right]\right\} \left({a\over a_0}\right)^{-2} {\rm e}^{-|k|(\eta-\eta_0)}\,.  \label{-1B1}
\end{eqnarray}
Magnetic fields that obey the above evolution laws can experience superadiabatic amplification without modifying conventional electromagnetism and despite the conformal flatness of the FRW host.\footnote{Superadiabatic amplification does not imply amplification per se, but decay at a slower than the adiabatic pace. The concept was originally introduced in gravitational-wave studies~\cite{1974JETP...40..409G,1979JETPL..30..682S}.} For instance, throughout the radiation and the dust eras (as well as during reheating), the scale factor of a FRW universe with hyperbolic spatial geometry evolves as $a\propto\sinh(\eta)$ and $a\propto\sinh^2(\eta/2)$ respectively (see solution (\ref{soFLRWa}) in \S~\ref{sssSfEFRWMs}). Focusing on the curvature scale, for simplicity, we may set $|k|=1$ in (\ref{-1cB1}). It is then clear that, on that scale, the magnetic mode never decays faster than $B_{(1)}\propto a^{-1}$~\cite{2008PhRvD..77j7302B}. In other words, large-scale $B$-fields are superadiabatically amplified throughout the post-inflationary evolution of an open Friedmann universe.

Although in the above examples we only considered the cases of radiation and dust, the amplification effect is essentially independent of the type of matter that fills the universe. In particular, $B$-fields in open FRW models containing a barotropic medium with $p/\rho\neq-1/3$ are superadiabatically amplified on large scales~\cite{2011arXiv1101.2390B}.\footnote{Mathematically, the easiest way of demonstrating the amplification effect is by adopting the Milne universe as our background spacetime. The latter corresponds to an empty spacetime with hyperbolic spatial geometry (see \S~\ref{sssSfEFRWMs}) and can be used to describe a low density open universe. There, the scale factor evolves as $a\propto{\rm e}^{\eta}$, which substituted into solution (\ref{-1B1}) leads to~\cite{2008PhRvD..77j7302B}
\begin{equation}
B_{(k)}= \mathcal{C}_5\left({a_0\over a}\right)^{|k|-2}+ \mathcal{C}_6\left({a_0\over a}\right)^{-|k|-2}\,.  \label{MilneB}
\end{equation}
Consequently, all magnetic modes spanning scales with $0<k^2<2$ are superadiabatically amplified. Close to the curvature scale, that is for $k^2\rightarrow1$, the dominant magnetic mode is $B_{(1)}\propto a^{-1}$; a rate considerably slower than the adiabatic $a^{-2}$-law. The latter is only restored at the $k^2=0$ limit, namely on small enough scales where the curvature effects are no longer important. Stronger amplification is achieved on supercurvature lengths, with $B_{(k)}\propto a^{\sqrt{2}-2}$ at the homogeneous limit (i.e.~as $k^2\rightarrow2$).} This means that the mechanism also operates during reheating (when $p=0$) and also throughout a phase of slow-roll inflation, namely in spatially open FRW models with a false-vacuum equation of state (i.e.~when $p=-\rho$). In the latter case, the background scale factor evolves as~\cite{2008PhR...465...61T}
\begin{equation}
a= a_0\left({1-{\rm e}^{2\eta_0}\over1-{\rm e}^{2\eta}}\right) {\rm e}^{\eta-\eta_0}\,,  \label{-1infFRW}
\end{equation}
where $\eta,\eta_0<0$. Substituting the above into Eq.~(\ref{-1B1}), we find that near the curvature scale (i.e.~for $|k|\rightarrow1$) the magnetic evolution is given by
\begin{equation}
B_{(1)}= \mathcal{C}_3\left(1-{\rm e}^{2\eta}\right)\left({a\over a_0}\right)+ \mathcal{C}_4{\rm e}^{-\eta}\left({a_0\over a}\right)^2\,,  \label{-1B2}
\end{equation}
with $\mathcal{C}_3$, $\mathcal{C}_4$ depending on the initial conditions~\cite{2005PhRvD..71l3506T,2008PhRvD..77j7302B}. This result also implies a superadiabatic type of amplification for the $B$-field, since the dominant magnetic mode never decays faster than $B_{(1)}\propto a^{-1}$. The adiabatic decay rate is only recovered at the end of inflation, as $\eta\rightarrow0^-$. It should be noted that the magnetogeometrical interaction triggering the above described effect is possible because,  when applied to spatially curved FRW models, inflation does not lead to a globally flat de Sitter space. Although the inflationary expansion dramatically increases the curvature radius of the universe, it does not change its spatial geometry. Unless the universe was perfectly flat from the beginning, there is always a scale where the 3-curvature effects are important. It is near and beyond these lengths that primordial $B$-fields are superadiabatically amplified.

The magnitude of the residual magnetic field is calculated in a way analogous to that given in \S~\ref{sssRMF}. Now, however, there is an additional parameter due to the non-Euclidean geometry of the FRW background.  In particular, near the curvature scale -- where $B\propto a^{-1}$, we find that
\begin{equation}
r_{RH}= \left({\rho_B\over\rho_{\gamma}}\right)_{RH}\simeq 10^{-54} \left({M\over10^{14}}\right)^4 \left({T_{RH}\over10^{10}}\right)^{-2}\lambda^{-2}\,,  \label{SAr}
\end{equation}
at the beginning of the radiation era~\cite{2011arXiv1101.2390B}. Note that, in deriving the above, we also used the auxiliary relation $(\rho_B/\rho)_{RH}=(\rho_B/\rho)_{IN}(M/T_{RH})^{4/3}$. The latter provides the relative change in the energy density of the (superadiabatically amplified) magnetic field between the end of inflation and that of reheating. Comparing (\ref{SAr}) to expression (\ref{adr1}), one can see that the (superadiabatic) magnetic amplification is already substantial by the end of reheating, despite its dependence on the energy scale of the inflationary model and of the corresponding reheat temperature. Moreover, large-scale $B$-fields are superadiabatically amplified during the subsequent evolution of the universe. This means that on scales close to the curvature radius of our background model, the ratio $r=\rho_B/\rho_{\gamma}$ is no longer constant but increases as $r\propto a^2\propto T^{-2}$. Consequently, recalling that $\lambda_K=\lambda_H/\sqrt{1-\Omega}$ is the curvature scale of a spatially open FRW cosmology, we obtain
\begin{equation}
B_0\sim 10^{-13}\left({M\over10^{14}}\right)^2 \left(1-\Omega_0\right)^{1/2} \hspace{3mm} {\rm G}\,.  \label{saB1}
\end{equation}
for the present strength of the residual $B$-field~\cite{2011arXiv1101.2390B}. Therefore, the higher the energy scale of inflation the stronger the superadiabatic amplification. On the other hand, the closer the density parameter to unity, the weaker the final field.

Currently, the WMAP reports indicate that $1-\Omega_0\lesssim10^{-2}$~\cite{2007ApJS..170..377S,%
2009ApJS..180..330K,2010arXiv1001.4538K}. On these grounds, and provided that the universe is spatially open, expression (\ref{saB1}) gives
\begin{equation}
B_0\sim 10^{-14} \hspace{2mm} {\rm G}\,,  \label{saB2}
\end{equation}
when $M\sim10^{14}$~GeV and $1-\Omega_0\sim10^{-2}$~\cite{2011arXiv1101.2390B}. The last parameter choice implies a curvature radius of the order of $10^{4}$~Mpc at present. These lengths are far larger than $10$~Kpc, which is the minimum magnetic size required for the dynamo to work. Nevertheless, once the galaxy formation starts, the field lines should break up and reconnect on scales similar to that of the collapsing protogalaxy. According to (\ref{saB1}), the above quoted magnetic strength will increase if the energy scale of inflation is greater than $10^{14}$~GeV. On the other hand, the magnitude of the residual $B$-field will drop if the current curvature scale of the universe is much larger than the Hubble horizon (i.e.~for $1-\Omega_0\ll1$). Nevertheless, the $\Omega$-dependence in Eq.~(\ref{saB1}) is relatively weak, which means that $B$-fields capable of seeding the galactic dynamo (i.e.~with $B_0\gtrsim10^{-22}$~G) are possible even when $1-\Omega_0\sim10^{-18}$ (or lower -- when the scale of inflation is higher than $10^{14}$~GeV). Overall, FRW universes with hyperbolic spatial geometry seem capable of sustaining astrophysically important magnetic fields under a fairly broad range of initial conditions.\footnote{If the results of~\cite{1999PhRvD..60b1301D} are taken at face value, the $1-\Omega_0$ difference can drop down to $\sim10^{-34}$ (or even lower) and still produce magnetic fields able to sustain the galactic dynamo (i.e.~with $B_0\gtrsim10^{-30}$~G). We also note that to these magnitudes one should add the magnetic amplification during the linear and the nonlinear regime of structure formation -- see \S~\ref{sssLR} and \S~\ref{sssAN-lR} respectively.}

Large-scale, primordial magnetic fields with residual magnitudes like those quoted above are far stronger than any other of their conventionally produced counterparts. Such strengths are usually achieved outside classical electromagnetism or beyond the standard model (see \S~\ref{sM-GSM} and \S~\ref{sM-GBSM} below).\footnote{Primordial magnetic fields can be superadiabatically, or even resonantly, amplified through their interaction with cosmological gravitational waves (see~\cite{2010PhRvD..81d3501T} and references therein). The former type of amplification is typically associated with highly conductive environments, but requires rather large amounts of shear anisotropy to operate efficiently. Resonant amplification, on the other hand, occurs in media of poor electrical conductivity and can lead to substantially strong $B$-fields with relatively minimal shear anisotropy. Both mechanisms are essentially nonlinear in nature and their detailed discussion lies beyond the limits of this review.} Moreover, although magnetic seeds with strengths $\lesssim10^{-10}$~G cannot affect primordial nucleosynthesis or the CMB spectrum, their strength lies within the galactic dynamo requirements (see \S~\ref{sssGDP}) and are therefore of astrophysical interest. Finally, we should also note very recent reports indicating the presence of coherent magnetic fields in empty intergalactic space with strengths intriguingly close to those quoted here~\cite{2010ApJ...722L..39A,2010Sci...328...73N,%
2010arXiv1009.1048T,2010arXiv1012.5313E}.

\newpage

\section{Magnetogenesis in the standard model}\label{sM-GSM}
In this section we review mechanisms of primordial magnetic field generation
in the framework of the standard model of particle physics and of non-linear
and out of equilibrium processes that may have happened in the very early
universe. In the first subsection we address magnetogenesis during
reheating and in the second subsection we review magnetogenesis due to
the EW (electroweak) and QCD (quantum chromodynamical) phase transitions.
Although the EW phase transition is likely to have taken place during
reheating, due to its importance and to the fact that it is framed in
particle physics field as well as the QCD transition, we treat it together
with the latter, in a specific subsection.

\subsection{Magnetogenesis during reheating}\label{ssMDR}
Accepting the inflationary paradigm, the reheating phase of the Universe was
one of the richest epochs in its evolution.\footnote{%
The term ``reheating'' was coined in the first inflationary models, in
which the Universe was hot before the onset of inflation, and was ``reheated'' again, after super-cooling during inflation.} It is usually
treated as the intermediate phase between the exponential expansion
and the radiation dominated expansion, during which almost all the matter
that constitutes the Universe was produced. This period can be divided
roughly into two or three stages: \textsl{preheating}, \textsl{heating} and
\textsl{thermalisation}. Of these the most interesting ones are the first and
the third. During the first stage, the dominant effect is parametric
particle creation. The importance of this process for reheating
was first realized in 1990 by Traschen and Brandenberger \cite{1990PhRvD..42.2491T}
and also by Dolgov and Kirilova \cite{1990SJNP...51..172D} and later developed in Refs.
\cite{1997PRD..56.3258K,1995PhRvD..51.5438S,%
1994PRL..73.12K,1996PRD..54.7570B}. The thermalization process is a
difficult and complex one. The interested reader is referred to specific
works, such as, e.g.,~\cite{2006RvMP...78..537B}, with references therein.

In few words, the process of reheating the Universe is due to the
profuse creation of particles, caused by inflaton oscillations around the
minimum of the effective potential. Those particles self-interact and
ultimately reach a state of thermal equilibrium, when all (or almost all)
the inflaton energy has been transformed into thermal energy of the created
elementary particles, with temperature $T_{RH}$, the so-called
\textsl{reheating temperature}. Being a strong out of equilibrium process
(and also turbulent, according to theoretical and numerical studies \cite{1996PhRvL..77..219K,2008CoPhC.178..929F,2001PhRvL..87a1601F,%
2001..PhRvD63.103503F,2002PhRvD..65f3522G,2010arXiv1002.3039J}),
the reheating period is therefore a suitable scenario for primordial
magnetogenesis.

It is important to observe that, irrespective of the mechanism that may have
generated the magnetic field during inflation, the
quantity $r=\rho _{B}/\rho _{\gamma }$ can be diluted during reheating because, during that
phase, the radiation density increases by a factor of at least $e^{4N}$%
, with $N$ being the number of e-foldings. So, unless the gauge field is also
amplified by the same amount, $r$ is likely to decrease during reheating.

\subsubsection{Parametric resonance}

Although the conformal invariance of the gauge fields is the main drawback for their
amplification by the expansion of the universe, it also opens up the possibility
of amplification by parametric resonance, if the conditions in the early universe are favourable. In this sense, the preheating
stage of reheating offers a suitable scenario of magnetic amplification, through its parametric resonance with a scalar field
\cite{2001PhLB..502..216F}. From the Lagrangian density of scalar
electrodynamics conformally coupled to gravity,%
\begin{equation}
\mathcal{L}=-\frac{1}{16\pi }F_{ab}F^{ab}-\left( \partial
_{a}+ieA_{a}\right) \Phi ^{\dagger }\left( \partial ^{a}-ieA^{a}\right) \Phi
-V\left( \Phi ^{\dagger }\Phi \right)\,,  \label{FiGru-1}
\end{equation}%
where $A_{a}$ is the gauge potential, $F_{ab}=\partial _{a}A_{b}-\partial
_{b}A_{a}$ and $\Phi $ a complex (i.e., charged) scalar field, we obtain the
evolution equation for the gauge field. Working in the Coulomb gauge ($%
\nabla \cdot \mathbf{A}=0$) and writing the homogenous part of the scalar
field as $\overline{\Phi }=\exp \left[ i\Theta \left( t\right) \right] \rho
\left( t\right) /\sqrt{2}$, the evolution equation of the fluctuations in
the spatial component of the gauge potential, $\delta \mathbf{A}$, reads%
\begin{equation}
\delta \mathbf{\ddot{A}+}H\delta \mathbf{\dot{A}}-\frac{\nabla ^{2}}{a^{2}}%
\delta \mathbf{A}+4\pi e^{2}\rho ^{2}\delta \mathbf{A}=4\pi \delta \mathbf{j}\,,
\label{FiGru-2}
\end{equation}%
where $\delta \mathbf{j}$ is a source term that can be non null when
statistical correlations of the electric currents are considered \cite%
{1998PhRvD..57.7139C,2000PhRvD..62.103512G}. In Fourier space and in terms od conformal
time, $\eta$ with $d\eta =dt/a\left( t\right) $, the homogeneous part of eq. (\ref%
{FiGru-2}) is%
\begin{equation}
\delta \mathbf{A}_{k}^{\prime \prime }+\omega _{k}^{2}\left( \eta \right)
\delta \mathbf{A}_{k}^{\prime \prime }=0\,,  \label{FiGru-3}
\end{equation}%
with $\omega _{k}^{2}\left( \eta \right) =k^{2}+4\pi e^{2}a^{2}\left( \eta
\right) \rho ^{2}\left( \eta \right) $, and where the primes denote
derivatives with respect to $\eta $. Expression (\ref{FiGru-3}) describes a
harmonic oscillator with time dependent frequency, which during the
oscillation of the complex scalar field can be rewritten as a Mathieu-like
equation. The solutions of that equation exhibit exponential increase,
i.e., they are proportional to $e^{\mu _{k}\eta }$ ($\mu _{k}$ is the Floquet
exponent) for some frequency intervals known as resonant bands. The
features of the parametric resonance depend strongly on the time evolution
of the homogeneous scalar field, which in turn depends on the form of $%
V\left( \Phi ^{\dagger }\Phi\right) $. For potentials of the form $%
V=\lambda _{n}\left( \Phi ^{\dagger }\Phi\right) ^{2n}$, Finelli and
Gruppuso found that, for a quadratic potential, parametric resonance is
efficient when $4\pi e^{2}\rho ^{2}\gg \lambda _{1}$ and it is stochastic and
broad, with the largest $\mu _{k}$ occurring for small $k$ \cite%
{1997PRD..56.3258K}. For a quartic potential, eq. (\ref{FiGru-3}) becomes a
Lame equation \cite{1997PhRvD..56.617G}. The resonance features and $\mu
_{k}$ depend on $e^{2}/\lambda _{2}$ \cite{1997PhRvD..56.617G}. In
particular modes with $k^{2}\ll \lambda _{2}\tilde{\rho}_{0}^{2}$ are
resonant for $1/2\pi <e^{2}/\lambda _{2}<3/2\pi $ ($\tilde{\rho}_{0}$ being
the initial value of $\tilde{\rho}$). For a symmetry-breaking
potential, i.e., $V=m^{2}\Phi ^{\dagger }\Phi +\lambda \left( \Phi ^{\dagger
}\Phi \right) ^{2}$ and in the case of $m^{2}<0$, the gauge field acquires an
effective mass propotional to $\sqrt{m^{2}/\lambda }$, a fact that can
completely inhibit the resonance in an expanding universe. For a general
value of $m^{2}$, the gauge coupling affects the resonance structure of the
scalar field and it is not possible to determine the resonant bands for the
imaginary part of the scalar field, as it would be the case if the charged
scalar fields were not coupled to the gauge field.

Another possible coupling is the one described by the Lagrangian density%
\begin{equation}
\mathcal{L}=-\frac{1}{16\pi }F_{ab}F^{ab}- \partial_a\phi\partial^a\phi- V\left(\phi\right)- \frac{g}{4}\phi F_{ab}\tilde{F}^{ab}\,,  \label{FiGru-4}
\end{equation}%
where $\phi $ may represent an axion or a general pseudo-Goldstone boson. If
the scalar field performs coherent oscillations, the evolution equation of
the transverse circularly polarized photons, $\bar{A}_{\pm k}$, is again
given by a Mathieu-like equation%
\begin{equation}
\delta \bar{A}_{\pm k}^{\prime \prime }+\left( k^{2}\pm 4\pi g\phi ^{\prime
}k\right) \delta \bar{A}_{\pm k}=0\,.  \label{FiGru-5}
\end{equation}
In this case the resonance occurs when $k\sim 4\pi g\phi \omega $, with $\omega $ being the oscillation frequency of $\phi $ (which is very small). For $V=\lambda\phi^{4}/4$ and $4\pi gf=1$ (with $f$ being the Peccei-Quinn symmetry scale), $\bar{A}_{\pm k}$ grows linearly for $k/\omega\ll4\pi gf$~\cite{2001PhLB..502..216F}.

All the previous description did not take into account dissipation due to
the presence of other charged fields, i.e., plasma effects. When they are
taken into account (basically in the form of electric conductivity),
their effect on the magnetic field depends on the wavelengths considered.
For wavelengths larger than the plasma collision length, the equations
acquire a damping term proportional to the conductivity, $4\pi a\varsigma
\delta A_{k}^{\prime }$, with $a=a(\eta)$ being the scale factor of the universe and $\varsigma$ the conductivity. If $%
a\varsigma $ is constant (as is the case considered in the literature, where $%
\varsigma \propto T\propto 1/a$ \cite{2000PhRvD..62.103512G}) and larger than
the Floquet exponent, the parametric resonance could be completely
suppressed. For wavelengths shorter than the plasma collision length, the
plasma frequency changes as $\omega _{k}^{2}\left( \eta \right) \rightarrow
\omega _{k}^{2}\left( \eta \right) +4\pi e^{2}n\left( \eta \right) /m$, with
$n\left( \eta \right) $ being the number density and $m$ the mass of the plasma
particles. This term plays the role of an effective mass that decays as $%
a^{-3}$, thus allowing for resonance, especially for large coupling
constants.

Whether or not primordial magnetic fields are amplified by parametric
resonance during preheating, depends on the existence of an oscillating
scalar or axion field, which the e.m. field is coupled to. If, for example,
that charged scalar field is the inflaton, the maximum amplifying factor
obtained for the gauge field is $\sim 10^{12}$, which is not enough to
give the minimum seed fields that can sustain a dynamo action
\cite{2001PhLB..502..216F}.

Exponential growth of large scale magnetic fields could also be achieved by
considering the Lagrangian \cite{1988PhRvD..37.2743T}%
\begin{equation}
\mathcal{L}=\frac{R}{16\pi G}- \frac{1}{4}F_{ab}F^{ab}- \frac{\beta}{2}\,RA_aA^a+ \mathcal{L}_{inflation}\,, \label{TurWid-1}
\end{equation}%
with $R$ the scalar curvature and $\beta $ a real constant. From the action $%
S=\int d^{4}x\mathcal{L}$, it is obtained that the evolution equation for the
Fourier components of the magnetic field
is given by
\cite{2001PhRvD..63j3515B},%
\begin{equation}
\delta B_{k}^{\prime \prime }+\left[ k^{2}+\Theta \left( \eta \right) \right]
\delta B_{k}=0\,,  \label{BaPoTsuVi-1}
\end{equation}%
with%
\begin{equation}
\Theta \left( \eta \right) =6\beta \frac{a^{\prime \prime }}{a}\,.
\label{BaPoTsuVi-2}
\end{equation}%
When the inflaton enters the oscillatory regime, the scalar curvature,
given by%
\begin{equation}
R=\frac{8\pi }{M_{Pl}^{2}}\left[ 4V\left( \phi \right) -\frac{\phi ^{\prime
2}}{a^{2}}\right]\,,   \label{BaPoTsuVi-3}
\end{equation}%
also oscillates and this fact can lead to efficient amplification of the
magnetic fluctuations. Considering $V\left( \phi \right) =\left(
1/2\right) m^{2}\phi ^{2}$, we have that during reheating, the inflaton
condensate evolves as%
\begin{equation}
\phi =\frac{M_{Pl}}{\sqrt{3\pi }mt}\sin \left( mt\right)\,,
\label{BaPoTsuVi-4}
\end{equation}%
and therefore the scalar curvature as%
\begin{equation}
R\simeq \frac{4}{3t^{2}}\left[ 1-3\cos \left( 2mt\right) \right]\,.
\label{BaPoTsuVi-5}
\end{equation}%
Defining $\mathcal{B}_{k}=a^{1/2}B_{k}$, eq. (\ref{BaPoTsuVi-1}) recasts into \cite{2001PhRvD..63j3515B}%
\begin{equation}
\frac{d^{2}\delta \mathcal{B}_{k}}{dz^{2}}+\left[ C_{k}-2q\cos \left(
2mt\right) \right] \delta \mathcal{B}_{k}=0\,,  \label{BaPoTsuVi-6}
\end{equation}%
with%
\begin{equation}
C_{k}=\frac{2}{3}q+\frac{k^{2}}{m^{2}a^{2}},\quad q=12\frac{\beta }{%
m^{2}t^{2}}\,,  \label{BaPoTsuVi-7}
\end{equation}%
when $\beta >0$, and%
\begin{equation}
C_{k}=-\frac{2}{3}q+\frac{k^{2}}{m^{2}a^{2}},\quad q=12\frac{\left\vert
\beta \right\vert }{m^{2}t^{2}}\,,  \label{BaPoTsuVi-8}
\end{equation}%
for $\beta <0$. Parametric resonance occurs when $q\lesssim 1$ initially \cite%
{2001PhRvD..63j3515B}, but to get relevant growth of $\mathcal{B}_{k}$ one
needs $\left\vert \beta \right\vert \gtrsim 1$. However, for $\left\vert
\beta \right\vert \gg 1$ the growth of the magnetic fluctuations is
suppressed \cite{1999PhRvD..60f3515T}. In the former case, as
the super-horizon $\delta \mathcal{B}_{k}$ modes are exponentially suppressed
during inflation, we do not expected to obtain high intensities from
parametric amplification. On the other hand, for sub-Hubble scales, the
suppression is weaker and consequently magnetic fields can be amplified
during preheating. For the latter case, namely when $\beta <0$, the amplification is mainly due to inflation rather than parametric resonance.
When finite electric conductivity is taken into account, a term of the form $%
-\varsigma _{e}a\delta \mathcal{B}_{k}^{\prime }$ is added to the r.h.s of eq. (%
\ref{BaPoTsuVi-1}), which counteracts the parametric
amplification of the fields. In summary, despite the exponential growth of
the magnetic fluctuations due to parametric resonance, the main
amplification occurs during inflation \cite{2001PhRvD..63j3515B}.

The possibility of further amplification by parametric resonance during
reheating of a seed hypermagnetic field generated during inflation, was
investigated by Dimopoulos et al \cite{2002PhRvD..65f3505D}. However the authors
also concluded that the fields do not grow substantially during preheating.

\subsubsection{Magnetogenesis by stochastic electric currents}

Another possibility to induce magnetic fields during reheating is tied to
the fact that abrupt changes in the metric at that stage may result in the
abundant creation of charged particles. This could generate stochastic
currents, which would eventually decay into the Maxwell field \cite%
{1998PhRvD..57.7139C,2000PhRvD..62.103512G}. As the inflaton is a gauge
singlet, it will not decay directly into charged species, so this mechanism
assumes the existence of another field, a charged one, that is in its vacuum
state during inflation. It becomes a particle state by the gravitational
field, due to the changes in the equation of state of the inflaton \cite%
{1999PhRvD..59f3505P}. Spin $1/2$ particles, such as the electrons, would be
conformally invariant at the high energies prevailing during inflation, and
consequently are not created in large numbers. Therefore, we must seek for a
minimally coupled charged scalar field, of which none is included in the
standard model but only in its supersymmetric extensions \cite%
{2000PhLB..472..287K}.

The scalar field can be decomposed into its real and imaginary parts as $\Phi
=\left( \phi _{1}+\phi _{2}\right) /\sqrt{2}$, and the associated current as%
\begin{equation}
J^{a}=J_{1}^{a}+J_{2}^{a}\,,  \label{cal-kan-11}
\end{equation}%
where%
\begin{equation}
J_{1k}=ie\left( \phi _{1}\partial _{k}\phi _{2}-\phi _{2}\partial _{k}\phi
_{1}\right)  \label{cal-kan-12}
\end{equation}%
and
\begin{equation}
J_{2k}=-e^{2}A_{k}\left( \phi _{1}^{2}+\phi _{2}^{2}\right)\,.
\label{cal-kan-13}
\end{equation}%
For a crude estimate of the field we can neglect $J_{2k}$. Assuming that
Ohm's law holds, $J_{1}^{i}=\varsigma \left( \eta \right) E^{i}$, where $\varsigma
\left( \eta \right) $ is the electric conductivity. Then, the equation of
the magnetic field is%
\begin{equation}
\left[ \partial _{\eta }^{2}+a\left( \eta \right) \varsigma \left( \eta \right)
\partial _{\eta }-\nabla ^{2}\right] B^a= \left( \bar{\nabla}\times \bar{%
J}_{1}\right) ^a\,.  \label{cal-kan-14}
\end{equation}%
Because the current is stochastic, the induced field must be evaluated through
its  two-point correlation function,
\begin{equation}
\left\langle B_{i}\left( x^a\right) B_{i^{\prime }}\left( x^{\prime a}\right) \right\rangle =H^{-4}\int \frac{dy^a dy^{\prime a}}{\left(2\pi \right) ^{4}}D^{ret}\left( x^a,y^a\right) D^{ret}\left(x^{\prime a},y^{\prime a}\right) N_{ii^{\prime }}\left( y^a,y^{\prime a}\right)\,,  \label{cal-kan-15}
\end{equation}%
where the ``noise kernel'' $N_{ii^{\prime }}$ is the Hadamard two-point
function
\begin{equation}
N_{ii^{\prime }}\left( y^a,y^{\prime a}\right) =\left\langle \left\{
\left( \bar{\nabla}\times \bar{J}_{1}\right) _{i}\left( x^a\right)
,\left( \bar{\nabla}\times \bar{J}_{1}\right) _{i^{\prime }}\left( x^{\prime a}\right) \right\} \right\rangle\,,  \label{cal-kan-16}
\end{equation}%
and $D^{ret}\left( x^a,y^a\right) $ is the retarded propagator of eq.
(\ref{cal-kan-14}). We are interested in fields coherent over a scale $%
\lambda $, so the spatial integral in eq. (\ref{cal-kan-15}) must be weighed
by a window function, $W\left( \lambda \right) $, that filters out
frequencies higher than the one associated to the field's scale of
coherence, $\lambda ^{-1}$. After weighing, the magnetic energy density
stored today in a region of size $k^{-1}$ can be directly inferred, giving%
\begin{equation}
\left\langle B_{\lambda }^{2}\right\rangle _{tod}\sim e^{2}\frac{%
H^{4}\lambda ^{-4}}{\varsigma _{0}^{2}m_{0}^{2}}\mathtt{ln}^{2}\left[ 2\frac{%
m_{0}}{H}\sqrt{\frac{m_{0}^{2}}{H^{2}}\left( 1+\tau _{1/2}\right) ^{2}+e%
\frac{T_{RH}^{2}}{H^{2}}}+2\frac{m_{0}^{2}}{H^{2}}\left( 1+\tau
_{1/2}\right) ^{2}\right]\,.  \label{cal-kan-17}
\end{equation}%
where $m_{0}$ is the bare mass of the scalar field, $T_{RH}$ the reheating
temperature of the universe (i.e., the temperature at the beginning of
radiation dominance), $\varsigma _{0}=e^{2}T_{RH}$ the electric conductivity at
the beginning of radiation dominance, $\tau _{1/2}$ the mean lifetime of the
current and $H$ the Hubble parameter, treated as time
independent during inflation. Observe that the field intensity depends very weakly on $\tau
_{1/2}$. Assuming instantaneous reheating, $T_{RH}=\sqrt{HM_{Pl}}$ an
estimate of $\left\langle B_{\lambda }\right\rangle _{tod}$ on a comoving
galactic scale $\lambda _{gal}\simeq 1$ Mpc is%
\begin{equation}
\left\langle B_{\lambda }\right\rangle _{tod}\sim e^{3}\frac{H^{3/2}\lambda
_{gal}^{-2}}{M_{Pl}^{1/2}m_{0}}\simeq 10^{-51}\text{ Gauss}
\label{cal-kan-18}
\end{equation}%
which is about 15 orders of magnitude weaker than the minimum required to feed the galactic dynamo.

Calzetta et al have considered the effect of the ``London current'', eq. (\ref{cal-kan-13}) \cite%
{2002PhRvD..65f3004C}. In this case the evolution equation of the magnetic two-point function shows two kernels, a local and a non-local one. Of
these, the local (non-dissipative) one dominates over the non-local (dissipative) one by several orders of magnitude throughout reheating, which
means that dissipation in this system is not due to ordinary electric
conductivity.

The equations for the magnetic field can be recast in the form a Langevin equation, which due to the local kernel looks like the London
equation for a superconducting medium:
\begin{equation}
\left[ \frac{d^{2}}{d\tau ^{2}}+C^{2}F^{2}\left[ z\left( \tau \right) \right]
\right] B_{s}\left( \tau \right) =F^{2}\left[ z\left( \tau \right) \right]\,.
\label{kandus-1}
\end{equation}
Here $F\left( z\right) =z^{1/2\gamma }J_{3/2\gamma }\left( z\right) $, $%
C^{2}=\tilde{e}^{2}z_{0}^{-4/\gamma }$, with $\tilde{e}^{2}=e^{2}2^{3/\gamma
-1}\Gamma ^{2}\left[ \left( 2\gamma +3\right) /2\gamma \right] \ln\left(
\Delta /\Upsilon\right) $, $\gamma $ being a parameter that determines the
temperature evolution during reheating, $\Delta =g^{1/2}T_{RH}/H$, (with $0\leq
g\leq 1$ being a coupling constant of the mass to the thermal bath and $T_{RH}$ being the
reheating temperature), and $\Upsilon$ is the dimensionless wavenumber
corresponding to the original inflationary patch. Finally, $J_{3/2\gamma }\left(
z\right) $ is a spherical Bessel function and $\Gamma ^{2}\left( \cdots
\right) $ a Gamma function.

Due to this current, the heavily amplified long-wavelength modes of the
scalar field act as a Landau-Ginzburg order parameter in a superconductor,
and as in the Meissner effect, the photon acquires a time-dependent mass.
This allows for an exponential growth of the Maxwell field during reheating.
The obtained intensities, however, are too weak ($\sim 10^{-53}$ G) to seed
the galactic dynamo. Besides, in this model the amplification is very
sensitive to the details of the reheating scenario, so it is not possible to
obtain generally valid estimates for the resulting magnetic intensity.

\subsubsection{Primordial magnetic fields from metric perturbations}

Amplification of electromagnetic vacuum fluctuations can also be achieved by scalar perturbations in
the metric during the transition
inflation-reheating, i.e., by breaking the conformal flatness of the
background geometry, rather than the conformal invariance of the Maxwell
field equations \cite{2001PhRvD..63j3515B,2001PhRvD..64h3006M}. The main
effect is due to super-horizon scalar perturbations, specially by those modes
that reenter the horizon right after the end of inflation. These
fluctuations create an inhomogeneous background, in which the magnetic field
evolves in a non conformally invariant way: the mode-mode coupling between
electromagnetic and metric fluctuations mixes positive and negative
frequency modes of the former field, thus breaking its conformal invariance.

The line element of a flat FLRW model with scalar metric perturbations, in
the longitudinal gauge, reads \cite{1984PThPS..78....1K,2001PhRvD..64h3006M}
\begin{equation}
dS^{2}=a^{2}\left( \eta \right) \left[ -\left( 1+2\Phi \right) d\eta
^{2}+\left( 1-2\Phi \right) dx_{i}dx^{i}\right]\,,  \label{mp-1}
\end{equation}%
with $\Phi \left( \eta ,x^{i}\right) $ representing the gauge invariant gravitational
potential. To first order in the cosmological and e.m. perturbations, $%
\delta A_{i}$, the evolution equation of the Fourier transform of the
latter is \cite{2001PhRvD..64h3006M}%
\begin{equation}
\frac{d^{2}}{d\eta ^{2}}\delta A_{i}\left( \bar{k},\eta \right) +k^{2}\delta
A_{i}\left( \bar{k},\eta \right) -J_{i}\left( \bar{k},\eta \right) =0\,,
\label{mp-2}
\end{equation}%
with $J_{i}\left( \bar{k},\eta \right) $ being a source term that depends only on
the Fourier transform of the metric perturbations and on their time derivatives
\cite{2001PhRvD..64h3006M}. The resulting field strength depends on the
power spectrum for super-Hubble metric perturbations, which is given by
\cite{2001PhRvD..64h3006M}%
\begin{equation}
\mathcal{P}_{\Phi }\left( k\right) =A_{S}^{2}\left( \frac{k}{k_{C}}\right)
^{n-1}\,,  \label{maroto-1}
\end{equation}%
where $A_{S}\simeq 5.10^{-5}$ sets the normalization at the COBE scale ($\lambda
_{C}\simeq 3000$~Mpc). At decoupling, the field strength on a coherence scale
corresponding to a galaxy, $k_{G}$, is%
\begin{equation}
B_{k_{G}}^{dec}|\simeq \frac{2^{3/2}\left( 2\pi \right) ^{3/4}A_{S}}{\sqrt{3n%
}a_{dec}^{2}}\frac{k_{max}^{n/2}k_{G}^{3/2}}{k_{C}^{(n-1)/2}}\,,
\label{maroto-2}
\end{equation}%
where $k_{max}$ is a cut-off that must be introduced in the case of a blue
spectrum ($n>1$) to avoid excessive primordial black hole production, and
that for negative tilt, is related to the minimum size of the horizon ($%
k_{max}\leq a_{I}H_{I}$, $I$ denoting the end of inflation). Observe that
eq. (\ref{maroto-2}) is a function of $k_{max}$, i.e., of the mechanism that
generated the perturbations. The resulting magnetic field spectrum is
thermal ($B_{k}\sim k^{3/2}$) in the low-momentum tail. The relation between
the energy densities in magnetic field and photons, for a suitable
wavenumber $k_{G}$, at decoupling turns out to be%
\begin{equation}
\frac{\rho _{\gamma }}{\rho _{B}}\left( k_{G}\right) \simeq 1.4\times
10^{36}\left( \frac{k_{G}}{k_{max}}\right) ^{n}\,.  \label{maroto-3}
\end{equation}%
The obtained intensities are upper limits, as dissipative effects were not
taken into account when deriving expression (\ref{maroto-2}).

Scalar metric perturbations can grow exponentially during preheating \cite%
{1999NuPhB.561..188B,2000PhRvD..62d3507B}, thus inducing strong enhancement
of magnetic fields. Let us consider the Lagrangian \cite%
{2001PhRvD..63j3515B}%
\begin{equation}
L=\frac{R}{16\pi G}-\frac{1}{4}F_{ab}F^{ab}-\frac{1}{2}\partial _{a}\phi
\partial ^{a}\phi -V\left( \phi \right) -\frac{1}{2}\partial _{a}\chi
\partial ^{a}\chi -\frac{1}{2}g^{2}\phi ^{2}\chi ^{2}\,,  \label{bapotsuvi-1}
\end{equation}%
with $V\left( \phi \right) =\left( 1/4\right) \lambda \phi ^{4}$, $\phi $
being the inflaton and $\chi $ the scalar field it is coupled to. In
this case metric perturbations are expected to grow due to enhancement of
the scalar field perturbations, and in turn the former would stimulate the
growth of magnetic field perturbations through gravitational scattering.

Assuming that on super-Hubble scales $\Phi $ depends only on time, and
adopting the Coulomb gauge ($A_{0}=0$, $\partial ^{i}A_{i}=0$), the Fourier
component $A_{i}\left( k\right) $ of the gauge potential reads%
\begin{equation}
A_{i}^{\prime \prime }\left( k,\eta \right) +k^{2}A_{i}\left( k,\eta \right)
=2\Phi ^{\prime }A_{i}^{\prime }\left( k,\eta \right)\,.  \label{bapotsuvi-2}
\end{equation}%
Defining $\tilde{A}_{i}\left( k,\eta \right) =\left( 1-\Phi \right)
A_{i}\left( k,\eta \right) $, to eliminate the term in $A_{i}^{\prime }\left(
k,\eta \right) $, eq. (\ref{bapotsuvi-2}) recasts into
\begin{equation}
\tilde{A}_{i}^{\prime \prime }\left( k,\eta \right) +k^{2}\tilde{A}%
_{i}\left( k,\eta \right) =\Phi ^{\prime \prime }\tilde{A}_{i}\left( k,\eta
\right)\,,   \label{bapotsuvi-3}
\end{equation}%
with its solution given in integral form by%
\begin{equation}
\tilde{A}_{i}\left( k,\eta \right) =\tilde{A}_{i}^{\left( 0\right) }\left(
k,\eta \right) +\frac{1}{k}\int_{\eta _{0}}^{\eta }\Phi ^{\prime \prime }%
\tilde{A}_{i}\left( k,\eta ^{\prime }\right) \sin \left[ k\left( \eta -\eta
^{\prime }\right) \right] d\eta ^{\prime }\,.  \label{bapotsuvi-4}
\end{equation}%
Decomposing the scalar fields as $\varphi \rightarrow \varphi +\delta
\varphi $ (where $\varphi $ denotes either $\phi $ or $\chi $), the
evolution equation for the Fourier transformed metric perturbations
reads%
\begin{equation}
\dot{\Phi}\left( k,\eta \right) +H\Phi \left( k,\eta \right) =4\pi G\left[
\dot{\phi}\left( k,\eta \right) \delta \phi \left( k,\eta \right) +\dot{\chi}%
\left( k,\eta \right) \delta \chi \left( k,\eta \right) \right]\,.
\label{bapotsuvi-5}
\end{equation}%
When the fluctuations $\delta \chi \left( k,\eta \right) $, with low $k$, are
excited during preheating, the corresponding metric and inflaton
perturbations, $\Phi \left( k,\eta \right) $ and $\delta \phi \left( k,\eta
\right) $ respectively, grow on large scales and thus enhance the magnetic fluctuations
\cite{2001PhRvD..63j3515B}. However, with the increase of $g^{2}/\lambda $, the
long wavelength modes $\delta \chi \left( k,\eta \right) $ are suppressed
during inflation. Sub-Hubble fluctuations, on the contrary, do not suffer
from suppression and exhibit parametric amplification during reheating \cite%
{2001PhRvD..6313503B}. Therefore, the mode-mode coupling between small-scale
metric perturbations and large-scale magnetic fields in eq. (\ref%
{bapotsuvi-2}) can enhance the latter. This model, however, has many
uncertainties and complexities, which require further research because they make it difficult to obtain reliable estimates for the final magnetic intensities.

\subsection{Magnetogenesis in phase transitions}\label{ssMPT}

The actual state of the particles in our universe is the result of phase
transitions that occured in the early phases of the expansion. At least two phase transitions are believed to have taken place
 in that epoch: the
electroweak (EW -- at $T_{EW}\sim 100$ GeV) and the quantum chromodynamical (QCD -- at $T_{QCD}\sim 200$
MeV). In the former case, the transition is from a symmetric, high temperature
phase with massless gauge bosons to the Higgs phase, in which the $SU\left(
2\right) \times U\left( 1\right) $ gauge symmetry is spontaneously broken
and all the masses of the model are generated.

In the QCD case, the transition is from a quark-gluon plasma to a
confinement phase with no free quarks and gluons. At the same energy scale, it is
expected that the global chiral symmetry of QCD with massless fermions is
spontaneously broken by the formation of a quark pair condensate.

First order phase transitions occur via bubble nucleation. Domains of new
phase of broken symmetry form, whose sizes are at most of the order of the
horizon at that time. As the horizons grow, different domains come into causal
contact and bubble walls collide with each other. Magnetogenesis occurs
through violent processes that take place during these collisions: reconnection
of magnetic field lines carried by the walls, MHD dynamos induced by the
turbulent flows produced by the collisions, etc. In every case, the question is whether the generated fields can explain the
intensity and morphology of the observed magnetic fields, or to seed further
amplification mechanisms, such as turbulent dynamos that could operate within
galaxies. In general, the magnetic fields that are produced during phase transitions can be
very strong, but typically have very small coherence lengths.

Second order transitions occur in a smooth and regular way, with approximate
local thermal equilibrium being maintained throughout the process. In spite
of this, magnetogenesis can also be possible as shown below.

It was recently proved \cite{2006Natur.443..675A} that the QCD transition in the hot universe was
an analytic cross-over rather than a phase transition. In this sense, the results on magnetogenesis obtained by considering the QCD transition as first order are
invalid and new research needs to be done.
We shall therefore
review in this section only the EW phase transition, which also
seems to provide a very suitable scenario for magnetogensis, since it facilitates the separation between electric and magnetic fields
as classical fields. Besides, while the Standard Model predicts a
smooth cross-over for this transition, its
extensions can give a strong first-order phase transition, which is a fundamental
ingredient for electroweak baryogenesis and the generation of primordial
magnetic fields.\footnote{In the Standard Model of particle physics, the
electroweak phase transition is of first order if the Higgs boson mass has $m_H < 72$ GeV
\cite{1996PhRvL..77.2887K,1998NuPhB.532..283R,1999PhRvL..82...21C}. Also, to suppress sphaleron processes in the broken phase, would actually require that $m_H < 35$ GeV. However, the
current experimental limit of $m_H$ is well above these values, at $m_H > 114$ GeV \cite{2003hep.ex...12023T},
thus turning the Standard Model an inadequate theory for baryogenesis.} Supersymmetric extensions of the Standard Model have been the most intensively studied~\cite{1993PhLB..307..106E,%
1996PhLB..380...81C,1996NuPhB.481...43L,1997PhLB..406...60F,%
1999EPJC...10..473H,2001NuPhB.606..183H}, but it is also possible to get a strong transition from more generic two-Higgs doublet models~\cite{1997PhRvD..55.3873C,2001NuPhB.597...23L}, from technicolor theories~\cite{2008PhRvD..78g5027C}, etc.~[8-17]~\cite{1993NuPhB.403..749E,2005PhRvD..71c6001G,%
2007PhRvD..75h3522A,2007JHEP...08..010P}.

Phase transitions in the early universe lead to another class of mechanism for generating primordial magnetic
fields, based on the Kibble mechanism \cite{1976JPhA....9.1387K}, i.e., on the generation of cosmic strings.
If the vacuum manifold $\mathcal{M}$ of the broken gauge theory that exhibits a phase transition has a
nontrivial first homotopy group \cite{1995RPPh...58..477H,1994IJMPA...9.2117B}, then a cosmic string network
will form generically \cite{1976JPhA....9.1387K}. This network has a characteristic length scale $\xi \left(t\right)$,
which expands with the expansion of the universe. Infinitely long strings and loops are formed, the smallest of the
latter decaying away via gravitational radiation. The result is a scaling solution, in which the string properties
such as $\xi \left(t\right)$ are proportional to the time passed
\cite{1985PhRvL..54.1868A,1989PhRvD..40..973A,1988PhRvL..60..257B,%
1990PhRvD..41.2408B,1990PhRvL..64..119A}. This
means that, if cosmic strings can produce randomly oriented magnetic fields, these could be coherent
by the Vachaspati mechanism \cite{1991PhLB..265..258V} over galactic scales at the time of galaxy formation,
as required by the dynamo paradigm. In the last subsection we review some works done on this mechanism.

Recently, a new mechanism of early magnetogenesis was proposed by Dolgov et al \cite{2010JCAP...08..031D},
whereby ferromagnetic domains of condensed W bosons would form in the broken phase of the standard
electroweak theory. These domains could create large-scale magnetic fields that would survive after the decay
of the domains due to flux-freezing. Although the authors do not give estimates for the produced fields, their work points
towards a new direction that should be explored further.

\subsubsection{Magnetogenesis in the electroweak phase transition}

In his seminal work of 1983~\cite{1983PhRvL..51.1488H}, C. Hogan was the first to
propose a mechanism of magnetogenesis based on a small-scale dynamo induced by a first order
phase transition in the early universe. His aim, however, was not to explain
the fields observed in galaxies, but to study the effect of the induced
fields on structure formation. The dynamo that Hogan proposed would be induced in
the wall of the bubbles by the ordered release of free energy during the
transition. Each bubble would be an independent dynamo, producing fields
correlated only on the scale of the bubble radius. The result is a pattern
of randomly oriented field lines that, properly averaged, would produce a
large scale field spanning over regions that are not causally connected, i.e.
coherent on larger scales, whose spectrum is of the form $B_{l}\propto
l^{-3/2}$ i.e., a dipole field, but of weaker intensity.

Since then, several mechanisms for magnetic field generation by first order
phase transitions have been proposed in the literature. In 1995 Baym, B\"{o}deker
and McLerran \cite{1996PhRvD..53..662B} proposed an also dynamo-based
mechanism, whereby seed fields (produced by thermal fluctuations) are
amplified by a turbulent dynamo induced by the collision of supersonic shock
waves created by the expansion of the walls of the broken symmetry bubbles.
Their work was framed within the standard model, since at the time it was still
believed that a first order phase transition was possible in it. In this
sense, the resulting magnetic intensities would be incorrect.

Concretely, when the expansion of the universe supercooled the cosmos below a
critical temperature, of the order $T_{cr}\sim 100$ GeV, then (at random locations)
the Higgs fields tunnels from the unbroken $SU\left( 2\right) \times U\left(
1\right) _{Y}$ phase to the broken $U\left( 1\right) _{em}$ phase, forming bubbles that expand and convert the false vacuum energy into
kinetic energy. As the shock fronts collide, turbulence is developed in the
cones associated to the bubble intersection, with Reynolds number%
\begin{equation}
Re\simeq \frac{v_{w}R_{b}}{\lambda }\,,  \label{gbm-1}
\end{equation}%
where $v_{w}$ is the wall velocity, $R_{b}$ is the size of a bubble at
collision time and $\lambda $ is the scattering length of fluctuations in the
electroweak plasma. Baym et al found that for scales smaller than $R_{b}$, $%
Re\sim 10^{12}$. Assuming $v_{w}\sim v_{fluid}\sim 10^{-1}$, that the
typical bubble radius after the completion of the phase transition is $%
R_{b}\sim f_{b}H_{EW}^{-1}$ (with $f_{b}\sim
10^{-2}-10^{-3}$ being the fractional size and $H_{EW}^{-1}\sim M_{Pl}g_{\ast
}^{-1/2}T_{c}^{-2}\sim 10$~cm that the event horizon at the
electroweak scale -- $M_{Pl}$ is the Planck mass and $g_{\ast }\sim 10^{2}$ is
the number of massless degrees of freedom in the matter) and that $\lambda \sim
Tg_{EW}\alpha ^{2}\left\vert \log \alpha \right\vert $ (with $\alpha $
the fine structure constant and $g_{EW}\sim g_{\ast }$ the number of degrees
of freedom that scatter by EW processes), Baym et al obtained%
\begin{equation}
Re\sim 10^{-3}\frac{M_{Pl}}{T_{c}}\alpha ^{2}\left\vert \log \alpha
\right\vert \sim 10^{12}\,.  \label{gbm-2}
\end{equation}%
Such a large Reynolds number means that turbulence is fully developed on
scales smaller than $R_{b}$. Assuming that the electric conductivity is large at
that epoch [49], strong magnetic turbulence should exist and in that
situation kinetic and magnetic energies are in equipartition, allowing us to
estimate that%
\begin{equation}
B^{2}\left( R_{b}\right) \sim g_{\ast }T_{c}^{4}v_{fluid}^{2}\,.  \label{gbm-3}
\end{equation}%
To obtain the intensity of the large-scale field, Baym et al assumed that the small-scale field formed a pattern of continuously distributed
dipoles, with distribution being a Gaussian. Therefore the correlation
function of the dipole density is%
\begin{equation}
\left\langle \nu ^{i}\left( 0\right) \nu ^{j}\left( \mathbf{r}\right)
\right\rangle =\kappa \delta ^{ij}\delta ^{3}\left( \mathbf{r}\right)\,,
\label{gbm-4}
\end{equation}%
while the one of the magnetic for $r\gg f_{b}H_{EW}^{-1}$ reads%
\begin{equation}
\left\langle B^{i}\left( 0\right) B^{i}\left( \mathbf{r}\right)
\right\rangle \simeq \frac{e^{2}\kappa }{r^{3}}\log \left( \frac{H_{EW}r}{%
f_{b}}\right)\,.  \label{gbm-5}
\end{equation}%
From eqs. (\ref{gbm-3}) and (\ref{gbm-5}), Baym et al. obtained that%
\begin{equation}
\left\langle B^{2}\right\rangle _{R}\sim v_{fluid}^{2}g_{\ast
}T_{c}^{4}\left( \frac{f_{b}}{H_{EW}R}\right) ^{3}\log ^{2}\left( \frac{%
H_{EW}R}{f_{b}}\right)\,,  \label{gbm-6}
\end{equation}%
where $\left\langle \cdots \right\rangle _{R}$ means averaging on a scale $R$. The present-time estimate for this magnetic field on a galactic scale, $%
l_{gal}\sim 10^{9}$~AU, is $B\left( l_{gal}\right) \sim 10^{-17}-10^{-20}$.

In 1991 T. Vachaspati~\cite{1991PhLB..265..258V} proposed a mechanism of
magnetogenesis based on second order cosmological phase transition. These transitions would produce domains of different vacuum
expectation values for the Higgs field, with these differences amounting to
gradients in the field. The latter would ultimately lead to electromagnetic fields
after the completion of the transition. This mechanism can produce fields
associated to other (unbroken) symmetries (like $SU(3)$) as well. When
applied to the electroweak transition, and assuming that the initial
correlation scale, $\chi _{i}\sim 2\left( gT_{i}\right) ^{-1}$, is of the
order of the inverse mass of the $W$ boson, with $T_{i}\simeq 10^{2}$ GeV,
an initial intensity of $B\sim gT_{i}^{2}/2\simeq 10^{23}$ G is obtained for
that correlation length. The initial energy density of the field is
comparable to that of the universe, $\Omega _{B}\left( t_{i}\right) $.
For a region of size $\ell _{i}=N\chi _{i}$, with $N\gg 1$, the Higgs field
is randomly oriented. Consequently the initial magnetic intensity on that
scale would be $B_{N}\sim gT_{i}^{2}/4N$, which at the electroweak scale
gives $B_{N}\left( t_{EW}\right) \sim 10^{23}N^{-1}$ G. At the QCD scale, we finds $%
B_{N}\left( t_{QCD}\right) \sim 10^{18}N^{-1}$ G and today $B_{N}\left(
t_{today}\right) \sim 10^{-6}N^{-1}$ G (with $%
N>10^{13}$ in all cases). For a scale of 100 Kpc today, $N=10^{24}$ and thus $B\sim
10^{-30}$ G.

The work of Vachaspati was questioned by Davidson in Ref. \cite%
{1996PhLB..380..253D}. She computed the electric current due to the dynamics
of the Higgs field and showed that it vanishes during the EW phase
transition. Her conclusion was that no large-scale magnetic fields are
generated by the classical rolling of the Higgs vacuum expectation value
during the electroweak phase transition. Later, in 1998, Grasso and Riotto
\cite{1998PhLB..418..258G} reanalysed the generation of magnetic fields
during the EW phase transition and found that the Vachaspati mechanism was plausible. Grasso and Riotto analyzed the two possibilities for the phase
transition: first order and second order. They showed that the magnetic
induction is connected to some semiclassical configurations of the gauge
fields, such as electroweak $Z$-strings and $W$-condensates. The initial
Higgs field configuration is%
\begin{equation}
\Phi _{in}\left( X\right) =\frac{1}{\sqrt{2}}\exp \left( -i\frac{\theta
\left( X\right) }{2}n^{a}\tau ^{a}\right) \left(
\begin{array}{c}
0 \\
\rho \left( X\right) e^{i\varphi /2}%
\end{array}%
\right)\,,  \label{gr-1}
\end{equation}%
with $\tau ^{a}$ representing the Pauli matrices, $n^{a}$ a unit vector in the $SU\left(
2\right) $ isospace, $\theta \left( X\right) $ the $U(1) $ Higgs
field phase and $\rho \left( X\right) $ the modulus of the Higgs field. The
equation of motion for the $SU\left( 2\right) $ gauge field in the adjoint
representation is%
\begin{equation}
D^cF_{cb}^{a}=-g\left\vert \rho \right\vert ^{2}\partial_b\theta \left(X\right) \left(n^{a}-n^{c}\hat{\phi}_c\hat{\phi}^a\right)\,,
\label{gr-2}
\end{equation}%
where it is assumed that the initial gauge fields $W_{\mu }^{a}$ and their
derivatives are zero at $t=0$. Also, $\hat{\phi}\equiv \Phi ^{\dag }\tau
^{a}\Phi /\Phi ^{\dag }\Phi =\cos \theta \hat{\phi}_{0}+\sin \theta ~\hat{n}%
\times \hat{\phi}_{0}+2\sin ^{2}\left( \theta /2\right) \left( \hat{n}\cdot
\hat{\phi}_{0}\right) ~\hat{n}$, with $\hat{\phi}_{0}^{T}\equiv -\left(
0,0,1\right) $. As $\hat{n}$ does not depend on the space coordinates, it is always
possible to assume that it is perpendicular to $\hat{\phi}_{0}$. In other words, $\hat{\phi}$ can be always obtained by rotating $\hat{\phi}_{0}$
at an angle $\theta $ in the $(\hat{\phi}_{0},\hat{n})$-plane. Then, eq. (\ref{gr-2}) becomes%
\begin{equation}
D^cF_{cb}^{a}=-g\left\vert \rho \right\vert ^{2}\partial_b\theta \left( X\right) n^{a}\,.  \label{gr-3}
\end{equation}%
This clearly shows that only the gauge field component along $\hat{n}$%
, namely $A_a=n^bW_{ab}$ is created by a nonvanishing gradient
of the phase between the two domains. When the full $SU\left( 2\right)
\times U\left( 1\right) _{Y}$ group is considered, it is no longer
possible to choose $\hat{n}$ arbitrarily, because the different orientations of $%
\hat{n}$, with respect to $\hat{\phi}_{0}$, correspond to different physical
situations. Setting $\hat{n}$ parallel to $\hat{\phi}_{0}$ and
assuming that the charged gauge field does not evolve significantly, Grasso
and Riotto found the following complete set of evolution equations, which is valid for
a finite (though short) time after the bubbles first contact:%
\begin{equation}
\partial^aF_{ab}^{Z}=\frac{g}{2\cos \theta _{W}}\rho ^{2}\left(
X\right) \left( \partial_b\varphi +\frac{g}{2\cos \theta _{W}}Z_b\right) ,  \label{gr-4}
\end{equation}%
and
\begin{equation}
d^ad_a\left( \rho \left( X\right) e^{i\varphi /2}\right) +2\lambda
\left( \rho ^{2}\left( X\right) -\frac{1}{2}\eta ^{2}\right) \rho \left(
X\right) e^{i\varphi /2}=0\,.  \label{gr-5}
\end{equation}%
Here $d_a=\partial_a+i\frac{g}{2\cos\theta_{W}}Z_a$, with
$\eta $ being the vacuum expectation value of $\Phi $ and $\lambda $ the
quartic coupling. Expressions (\ref{gr-4}) and (\ref{gr-5}) are the Nielsen-Olesen
equations of motion \cite{1973NuPhB..61...45N}. Their solution describes a $%
Z $-vortex with $\rho =0$ at its core \cite%
{1992PhRvL..68.1977V,1992PhRvL..69.216V}. The geometry of the system implies
that the vortex is closed, forming a ring whose axis coincides with the
conjunction of the bubble centers.

To determine the magnetic field produced during the process described above,
it is necessary to give a gauge-invariant definition of the electromagnetic
field in the presence of a non-trivial Higgs background. Grasso and Riotto
chose%
\begin{equation}
\partial ^aF_{ab}^{em}=2\tan \theta _{W}\partial^a\left[
Z_a\partial_b\log \rho \left( X\right) -Z_b\partial_a\log \rho \left( X\right) \right]\,,  \label{gr-6}
\end{equation}%
while Tornkvist used \cite{1998PhRvD..58d3501T}%
\begin{equation}
\mathcal{F}_{ab}^{em}\equiv -\sin\theta_{W}\hat{\phi}_cF_{ab}^c+ \cos \theta_{W}F_{ab}^{Y}+ \frac{\sin\theta}{g}\,\varepsilon_{cde}\hat{\phi}^{c} \left(D_a\hat{\phi}\right)^d\left(D_b\hat{\phi}\right)^e\,,  \label{gr-7}
\end{equation}%
obtaining no electric current and so no magnetic field. It must be stressed
that both definitions (\ref{gr-6}) and (\ref{gr-7}) fulfill the requirement
that they reproduce the standard definition in the broker phase with an
uniform Higgs background. Grasso and Riotto noted that the presence of an
inhomogeneous $W$-condensate, produced by string decay, gives rise to electric
currents that can sustain magnetic fields even after the $Z$-string has
disappeared.

An attempt to predict the strength of the magnetic field at the end of the
EW phase transition was done by Ahonen and Enqvist \cite{1998PhRvD..57..664A}
and by Enqvist \cite{1998IJMD...7..331E}, who analyzed the formation of
ring-like magnetic fields in collisions of bubles of broken phase in an
Abelian Higgs model. Under the assumption that magnetic fields are induced
by a process similar to the Kibble and Vilenkin mechanism \cite%
{1995PhRvD..52..679K}, it was concluded that a field of the order
of $B\simeq 2\times 10^{20}$ G, with a coherence length of about $10^{2}$ GeV$%
^{-1}$, could be induced. In addition, assuming that the plasma was endowed with
MHD turbulence, Ahonen and Enqvist found that the coherence scale could be
enhanced by the inverse cascade of the magnetic helicity, and so a field of $%
B_{rms}\simeq 10^{-21}$ G on a comoving scale of $10$ Mpc could be present
today. As stated earlier, however, the problem with first order phase transitions in the
standard model is that they are incompatible with the experimental lower limit
for the Higgs mass.

Grasso and Riotto also analyzed the creation of magnetic fields when the EW phase transition is of second order. In this case domains
where the Higgs field is physically correlated are formed near the critical
temperature. The formation of topological and non-topological vortices, is a
common phenomenon in second order phase transitions via the Kibble
mechanism. It is also known that the non-topological vortices share many common
features with the electroweak strings \cite{1996IJMPH..10..471V}. In this sense,
Grasso and Riotto argued that electroweak strings are formed during the
second order EW phase transition. To estimate the density of vortices (and
consequently the mean magnetic field), it is necessary to know the Ginzburg
temperature, $T_{G}$. This sets the threshold at which the thermal fluctuations of
the Higgs field, inside a given domain of broken symmetry, are no longer able
to restore the symmetry. The Ginzburg temperature was computed by the authors of
Ref. \cite{1998PhLB..418..258G}, after comparing the expansion rate of the
Universe with the nucleation rate per unit volume of sub-critical bubbles of
symmetric phase with size equal to the correlation length of the broken
phase. The latter is given by%
\begin{equation}
\Gamma _{ub}=\frac{1}{\ell _{b}^{4}}e^{-S_{3}^{ub}/T}\,,  \label{gr-8}
\end{equation}%
where $\ell _{b}$ is the correlation length in the broken phase and $%
S_{3}^{ub}$ is the high temperature limit of the Euclidean action \cite%
{1992PhRvD..45.3415E}. For the EW phase transition, $T_{G}\simeq T_{C}$, and
the corresponding size of a broken phase domain is determined by the
correlation length at $T=T_{G}$, i.e.,%
\begin{equation}
\ell _{b}\left( T_{G}\right) ^{-2}=V^{\prime \prime }\left( \left\langle
\phi \left( T_{G}\right) \right\rangle ,T_{G}\right)\,,   \label{gr-9}
\end{equation}%
where $V\left( \phi ,T\right) $ is the effective Higgs potential. Using the
fact that $\ell _{b}\left( T_{G}\right) ^{2}$ depends weakly on $M_{H}$,
Grasso and Riotto estimated the magnetic field strength, on a scale $\ell
_{b}\left( T_{G}\right) $ at the end of the EW phase transition, to be $%
B_{\ell }\sim 4e^{-1}\sin ^{2}\theta _{W}\ell _{b}^{2}\left( T_{G}\right)
\sim 10^{22}$ G. To obtain the intensity on cosmologically interesting
scales, the authors of Ref. \cite{1998PhLB..418..258G} followed the
procedure of line averaging sugested by Enqvist and Olesen, i.e., $%
\left\langle B\right\rangle _{rms,L}\equiv B_{\ell }/\sqrt{N}$, where $N$ is
the number of domains crossed by line, obtaining that a field coherent
on a scale of 1 Mpc today would have an intensity of $B_{0}\left( 1\text{ Mpc}\right)
\sim 10^{-21}$ G. It must be pointed out, however, that all these studies do not
take into account the dissipative effects of the primordial plasma. Consequently, the corresponding numerical results should be treated as upper limits.

The mechanism proposed by Vachaspati \cite{1991PhLB..265..258V} and later
analyzed by Grasso and Riotto \cite{1998PhLB..418..258G} (see also Cornwall
\cite{1997PhRvD..56.6146C}) was recently numerically confirmed and improved
by Diaz-Gil et al. \cite{2008PhRvL.100x1301D,2008JHEP...07..043D}. The
authors considered the full $SU\left( 2\right) \otimes U\left( 1\right) $
model in the framework of hybrid inflation. After a short period of hybrid
inflation that ends at the EW scale, where non-linearities in the Higgs
and gauge fields can be neglected, tachyonic preheating develops and
non-linearities in the fields cannot be neglected anymore. During this
period the Vachaspati mechanism operates, and magnetic string-like
configurations appear due to the gradients in the orientation of the Higgs
field. The important feature of the induced magnetic fields is that they are
helical, i.e., they posses a non null r.m.s. magnetic helicity. During the
subsequent phase of (first order) EW symmetry breaking, the magnetic fields are
squeezed in string-like structures localized in the regions between bubbles,
where the gradients of Higgs fields are still large. The evolution of the
coherence scale of these fields can be tracked for a short period of time
after the end of the phase transition. At that time it is important to track
the evolution of the low momentum part of the spectrum, which is the one
that can seed the fields for galaxies and clusters of galaxies. It is seen
that it carries a fraction of $\sim 10^{-2}$ of the total energy density,
which would be enough to explain the magnetic fields observed in clusters. The
correlation length grows as fast as the particle horizon (i.e., linearly in
time) and this behaviour is interpreted as an indication that an inverse
cascade of magnetic helicity is in operation. However, it is not possible to
extrapolate this behaviour to later times, due to our limited knowledge on the
primordial plasma features.

Stevens and Johnson \cite{2009PhRvD..80h3011S,2010arXiv1001.3694S} analyzed
the possibility of magnetogenesis by a first order EW phase transition,
possible for some choices of parameters in the minimal supersymmetric
Standard Model (see also \cite{2010PhRvD..81h5035H}). They considered the
Lagrangian%
\begin{equation}
\mathcal{L}=\mathcal{L}_{1}+\mathcal{L}_{2}+(\text{ leptonic, quark and
supersymmetric partner interactions})\,,  \label{sj-1}
\end{equation}%
with%
\begin{equation}
\mathcal{L}_{1}= -\frac{1}{4}W_{ab}^{i}W^{i\,ab}- \frac{1}{4}B_{ab}B^{ab}  \label{sj-2}
\end{equation}%
and
\begin{equation}
\mathcal{L}_{2}=\left\vert \left( i\partial_a-\frac{g}{2}\tau \cdot
W_a-\frac{g^{\prime }}{2}B_a\right) \Phi \right\vert ^{2}-V\left(
\Phi ,T\right)\,.  \label{sj-3}
\end{equation}%
Also, $T$ represents the temperature, while $W_{\mu \nu }^{i}$ and $B_{\mu \nu }$ are given by
\begin{equation}
W_{ab}^{i}= \partial_aW_b^{i}- \partial_bW_a^{i}- g\varepsilon^i_{jk}W_a^jW_b^k  \label{sj-4}
\end{equation}%
and
\begin{equation}
B_{ab}= \partial_aB_b- \partial_bB_a\,,  \label{sj-5}
\end{equation}%
respectively. In the previous equations, $W^{i}$ (with $i=+,-$) are the $W^{+}$, $W^{-}$
fields, $\Phi $ is the Higgs field and $\tau ^{i}$ is the $SU\left( 2\right)
$ generator (fermions are not considered in this model). In the framework of
the MSSM the bubbles that consist of a region of space filled by the
Higgs field with a cloud of the other constituents of the MSSM in the broken
phase. From this Lagrangian, one obtains the linearized equations of motion with $O\left(
3\right) $ symmetry, which are suitable to study collisions where the Higgs field is relatively unperturbed from its mean value within the
collision volume. Stevens and Johnson \cite{2010arXiv1001.3694S} found that
the coherent evolution of the charged $W$ fields within the bubbles is the
main source of the electric current that generates the magnetic field. In
their model, fermions are taken into account as a background that provides
dissipation through electric conductivity. They numerically integrated the
equations of motion of the model, paying special attention to the role of
the surface thickness of the bubbles, finding that the main sensitivity is
due to the steepness of the bubble surface: the steeper the transition, the
more enhanced the seed field becomes. Despite this, the authors of Ref.
\cite{2010arXiv1001.3694S} did not attempt to give the present-day value of
the generated magnetic field, because of uncertainties
in the properties of the host plasma.

\subsubsection{Magnetogenesis from cosmic strings}

The interaction between cosmic strings and magnetic fields was first discussed in
1986 by Ostriker et al \cite{1986PhLB..180..231O}, while their connection with primordial magnetogenesis was first suggested by Vachaspati in 1991
\cite{1991PhLB..265..258V}. Later the mechanism was further developed by Brandenberger
et al \cite{1992PhLB..293..287B}, also for superconducting cosmic strings, who showed
that these models are severely constrained by cosmological arguments: the only stable
confirgurations for those strings are springs and vortons, which produce matter
overdensities in the same manner primordial magnetic monopoles do. So, these models
had to be ruled out.

In 1999 Brandenberger and Zhang \cite{1999PhRvD..59h1301B} studied magnetogenesis by
anomalous global strings and discussed for the
first time the importance of the coherence length in these models.
The authors proposed a mechanism based on the realization that anomalous global strings
couple to electromagnetism \cite{1988NuPhB.302..280K}
through an induced $F_{\mu \nu }\tilde{F}^{\mu \nu }$ term in the low-energy
effective Lagrangian and therefore magnetic fields can be generated. The major
advantage of the mechanism is that the coherence scale of the induced magnetic fields is basically the curvature radius of the inducing string.
The mechanism is realized within QCD, namely there exists a class of
stringlike classical solutions of the linear sigma model, that describes
strong interactions below the confinement scale, called \textsl{pion strings}
\cite{1998PhRvD..58b7702Z}.
At low temperatures those strings are not topologically stable, decaying
at a temperature $T_{d}\sim 1$ MeV, but within the plasma they can stabilize
because the plasma interactions break the degeneracy among the three pions.

Since a pion string is made of $\sigma $ and $\pi ^{0}$ fields, it is
neutral under the $U_{em}\left( 1\right) $ symmetry. However, the $\pi ^{0}$
couples to photons via the Adler-Bell-Jackiw anomaly. In the linear sigma
model, the effective coupling of $\pi ^{0}$ to photons is obtained from the
contribution of the quark triangle diagram \cite{1996PhRvL..76.3084P}. At low energies only pions and photons are important, hence the
effective Lagrangian to leading order reads%
\begin{equation}
L_{low}= \frac{f_{\pi}^{2}}{4} Tr\left(\partial_a\Sigma^{\dag}\partial^a\Sigma\right)- \frac{1}{4}F_{ab}F^{ab}- \frac{N_{c}\alpha}{24\pi}\, \frac{\pi^{0}}{f_{\pi}}\,e^{abcd}F_{ab}F_{cd}\,,  \label{cs-1}
\end{equation}%
where $N_{c}=3$, $\Sigma =\exp \left( i\bar{\tau}\cdot \bar{\pi}/f_{\pi
}\right) $, $\bar{\tau}$ are the Pauli matrices and $\alpha $ is the
electromagnetic fine structure constant. From this Lagrangian one also obtains the classical equation of the electromagnetic field%
\begin{equation}
\partial_aF^{ab}= -\frac{\alpha}{\pi}\, \partial_a\left(\frac{\pi^0}{f_{\pi}}\right)\tilde{F}^{ab}\,.  \label{cs-2}
\end{equation}%
The key effect is due to the anomaly term in eq. (\ref{cs-2}). Charged
zero modes on the string will induce a magnetic field circling the string
that falls off less rapidly, as a function of the distance from the string,
than it is classically expected. Zero mode currents are automatically set up by
the analog of the Kibble mechanism \cite{1976JPhA....9.1387K} at the time of the
phase transition and therefore magnetic fields coherent in a scale of the
string size are automatically generated. The
coherent magnetic field, as a function of the distance $r$ from the string,
can be expressed as
\begin{equation}
B\left( r\right) =N_{c}\frac{en}{2\pi }\left( \frac{r}{r_{0}}\right)
^{\alpha /\pi }\frac{1}{r}  \label{cs-3}
\end{equation}%
with $n$ being the number density of charge carriers on the string, $r_{0}$ giving the
width of the string and $\alpha \ll 1$. By dimensional analysis,
Brandenberger and Zhang obtained that at the time $t_{c}=\allowbreak t_{c}$,
when the strings form, $r_{0}\sim T_{c}^{-1}$ and $n\sim T_{c}$. The initial
correlation length of the string network, $\xi \left( T_{c}\right) $
increases rapidly, approaching a scaling solution of the form $\xi \left(
t\right) \sim t$. During this evolution the charge density is diluted as the
strings stretch, while at the same time the merger of small strings into larger
ones leads to an increase of charge. Assuming that the initial separation of
the strings is microscopic, and that they decay during radiation dominance,
Brandenberger and Zhang obtained%
\begin{equation}
n\left( t_{d}\right) \sim \left( \frac{T_{d}}{T_{c}}\right) ^{p}n\left(
t_{c}\right)\,,   \label{cs-4}
\end{equation}%
with $p=5/4$ or $3/2$ \cite{1999PhRvD..59h1301B}. Also, the corresponding magnetic field at $t_{d}$
is%
\begin{equation}
B\left( t_{d}\right) \sim 10^{5}\frac{T_{c}}{1\text{GeV}}r_{m}^{-1}\left(
\frac{T_{d}}{T_{c}}\right) ^{p}\left( rT_{c}\right) ^{\alpha /\pi }~\text{%
Gauss}\,.  \label{cs-5}
\end{equation}%
Brandenberger and Zhang assumed that between $t_{d}$ and the present time,
$t_{0}$, the field propagates through a perfectly conducting plasma and found that today the magnetic intensity should be%
\begin{equation}
B\left( t_{0}\right) \sim 10^{-14}\frac{T_{c}}{1\text{GeV}}%
r_{kpc}^{-1}\left( \frac{T_{d}}{T_{c}}\right) ^{p}\left( \frac{T_{0}}{T_{d}}%
\right) \left( rT_{c}\right) ^{\alpha /\pi }~\text{Gauss}\,.  \label{cs-6}
\end{equation}%
Note that $T_{0}$ is the present time temperature, $r_{Kpc}$ is the present
distance from the original comoving location of the string expressed in Kpc
and $r$ the physical distance at $T_{d}$. Considering $t$ as an estimate of
the string separation, $T_{c}\sim 1$ GeV, $T_{d}\sim 1$ MeV and $p=1$,
Brandenberger and Zhang obtained%
\begin{equation}
B\left( t_{0}\right) \sim 10^{-26}\left( rT_{c}\right) ^{\alpha /\pi }~\text{%
Gauss}\,,  \label{cs-7}
\end{equation}
arguing that, if $r$ is of cosmological order and $T_{c}^{-1}\ll 1$, we can have $\left( rT_{c}\right) ^{\alpha /\pi }\gg 1$.

To analyze the coherence scale of the fields, the authors assumed that after $%
T_{d}$ the field lines are frozen in comoving coordinates, obtaining%
\begin{equation}
\xi \left( t_{d}\right) _{c}=\beta t_{d}z\left( t_{d}\right) =\beta 10^{-2}~~\text{Kpc}~T_{d}\left[\text{MeV}\right] ^{-1}\,,  \label{cs-8}
\end{equation}%
where $z\left( t_{d}\right) $ is the redshift at $t_{d}$ and $\beta \sim 1$
for scaling strings. On scales larger than $\xi \left( t_{d}\right) _{c}$
the fields have random orientation, yielding an average of%
\begin{equation}
\bar{B}=\frac{1}{\sqrt{N}}B\left( t_{0}\right)\,,   \label{cs-9}
\end{equation}%
with%
\begin{equation}
N=\left[ \frac{d}{\xi \left( t_{d}\right) _{c}}\right] ^{2}=d\left[ \text{kpc%
}\right] ^{2}\beta ^{-2}10^{4}T_{d}\left[ \text{MeV}\right] ^{2}\,,
\label{cs-10}
\end{equation}%
where $d$ is the scale the coherente field is calculated. The authors
found that $\bar{B}\simeq 10^{-2}B\left( t_{0}\right) $. Obviously, if the resistivity
of the host plasma is accounted for,
the suppression will be larger.

Recently, Gwyn et al \cite{2009PhRvD..79h3502G} extended the mechanism to heterotic cosmic
strings arising in \textsl{M} theory. Those strings, being stable, would produce
even stronger fields. This work is reviewed in the following section. Another possible way that cosmic strings could produce primordial magnetic fields
was proposed by Dimopoulos \cite{1998PhRvD..57.4629D}, by Dimopoulos and Davies \cite{1999PhLB..446..238D}
and by Battefeld et al \cite{2008JCAP...02..001B}. In these scenarios, the magnetic fields are
be induced by vortices produced by cosmic strings via the Harrison-Rees \cite{1970MNRAS.147..279H}
effect. This mechanism, however, was recently strongly criticized by Hollenstein et al
\cite{2008PhRvD..77f3517H}, who showed
that the Harrison-Rees effect is quite inefficient in producing cosmologically
intersting magnetic fields.

\newpage


\section{Magnetogenesis beyond the standard model}\label{sM-GBSM}

In this section we are going to review different types of magnetic field generation mechanisms involving theories beyond the standard model.
Primordial magnetic fields will be generated during inflation.
As it was shown in \S~\ref{sMCFRWMs}, on a non flat background perturbations in the electromagnetic field can be efficiently amplified during inflation within the standard model, that is within the standard linear theory of electrodynamics. On the contrary, on a flat background the amplification during inflation is not sufficient in order to be cosmologically relevant.
Following \cite{1988PhRvD..37.2743T} (see also \S~\ref{ssTIPMFs} earlier) the ratio $r$ of the energy density in the magnetic field $\rho_{\rm B}$ over the
energy density $\rho_{\gamma}$ in the background radiation is introduced, thus
\begin{eqnarray}
r\equiv\frac{\rho_{\rm B}}{\rho_{\gamma}},
\end{eqnarray}
where $\rho_{\rm B}=\frac{B^2}{8\pi}$.
In the case of linear electrodynamics on a flat background the magnetic energy density
decays as $a^{-4}$, where $a$ is the scale factor. Hence the ratio $r$ is a constant as the universe evolves.
This is also true  in the radiation dominated era when the universe is dominated by a highly
conducting plasma.
The interstellar magnetic field in our Galaxy is of the order of a few $\mu$G.
Assuming that a galactic dynamo, contributing an exponential factor in time, is operating since the time of the formation of the Galaxy requires an initial seed magnetic field at the time of galaxy formation of at least $B_s\simeq 10^{-20}$G \cite{1987QJRAS..28..197R}
which corresponds to a minimum magnetic to photon energy density ratio $r$ given by $r\simeq 10^{-37}$.
There is some controversy about the efficiency of such a galactic dynamo, thus working under the hypothesis that there is no efficient amplification of the initial magnetic seed field due to a galactic dynamo but only the amplification due to the collapsing protogalactic cloud requires $r$ to be at least of the order of $r=10^{-8}$ \cite{1988PhRvD..37.2743T}. These lower bounds on $r$ were derived assuming no cosmological constant.
In a flat universe with a large positive  cosmological constant assuming a galactic dynamo operating these bounds can be lowered significantly. In particular
for reasonable cosmological parameters
an initial seed magnetic field of at least $B_s\simeq 10^{-30}$ G is enough to explain the present day galactic magnetic field strength \cite{1999PhRvD..60b1301D}. This corresponds to $r=10^{-57}$.
In typical inflationary scenarios on a galactic scale of 1 Mpc $r\simeq 10^{-104}$ at the beginning of the radiation dominated era (cf. equation (\ref{adr1})) which is much below the required minimal value even in the presence of a cosmological constant.
Therefore, in the case of a flat background, it is necessary to go beyond the standard model.
There are different possibilities of modifying the standard four dimensional electromagnetic Lagrangian.

Currently models of modified gravity enjoy an intense activity due to the
fact that they can be used to describe the late time evolution of the universe at a global scale as well as, say,  the observed rotation curves of galaxies. Thus these models combine the effects of dark energy, which is used to model the accelerated expansion of the present universe, and
dark matter, which is postulated to exist in the form of halos around most galaxies.

The gravitational sector in theories of modified gravity is usually described by a Lagrangian of the form (e.g., \cite{sf})
\begin{eqnarray}
S=\frac{1}{8\pi G}\int d^4 x\sqrt{-g}f(R),
\end{eqnarray}
where $f(R)$ is a function of the Ricci scalar $R$, most often chosen to be of the form $f(R)\simeq R+\alpha R^n$, where $\alpha$ and $n$ are constants.
It should also be noted that one of the original realizations of inflation is given by $f(R)=R+R^2$ which can be shown to be equivalent to a conformally coupled scalar field
\cite{1980PhLB...91...99S}.
Usually it is argued that modified gravity theories are some kind of effective description resulting from taking into account quantum corrections to the classical Einstein-Hilbert action.

In order to study the generation of primordial magnetic fields in this type of theories the electromagnetic field has to be included in the Lagrangian \cite{lms}.
Considering flat space the conformal invariance of the Maxwell Lagrangian in four dimensions has to be broken in order to generate magnetic fields strong enough to seed the galactic magnetic field.
In the following a survey of models will be given which are used in this respect.

\subsection{Gravitational coupling of the gauge potential}

Models involving the gravitational coupling of the gauge potential are  described by Lagrangians of the form
$RA_{m}A^{m}$ and $R_{mn}A^{m}A^{n}$ where $R_{mn}$ is the Ricci tensor. Gauge invariance of the Lagrangian is broken explicitly and for that matter it does not seem very appealing.
The term $RA_{m}A^{m}$ describes a massive photon with its mass given by $m_{\gamma}\sim R^{1/2}$. Electromagnetism is then described by Proca theory and $A_{m}$ is the Proca field.

The strongest bound on the photon mass within our galaxy is obtained by assuming a Proca regime on all scales. The Proca field contributes to the magnetic pressure of the intergalactic medium which has to be counterbalanced by the thermal pressure of the plasma. Observations assuming standard electrodynamics indicate that within our galaxy the interstellar medium
is approximately in equipartition. From this it can be concluded that the magnetic pressure due to the Proca field has to be subleading with respect to the standard magnetic pressure \cite{1976UsFiN.119..551C}.
This implies the bound $m_{\gamma}<10^{-26}$ eV \cite{2007PhRvL..98a0402A}.
Furthermore, in \cite{2007PhRvL..98a0402A} it is pointed out that these limits depend on the mechanism on how the photon aquires mass. If it is via the Higgs mechanism then it is possible that large scale magnetic fields are effectively described by Maxwell's theory. In this case the strongest bound comes from the validity of Coulomb's law and is given by $m_{\gamma}<10^{-14}$ eV \cite{Williams,2007PhRvL..98a0402A}.
Using this type of Lagrangian on cosmologically scales, the typical scale is given by the
value of the Hubble parameter today, $H_0$, which results in the estimate,
$m_{\gamma}\sim R^{\frac{1}{2}}$ and $R^{\frac{1}{2}}\sim H$ then at present
the photon mass is given by $m_{\gamma}\sim H_0\sim 10^{-33}$eV which is well below the above mentioned  present limits on the photon mass \cite{1988PhRvD..37.2743T}.

The idea is that the initial magnetic seed field is created from the amplification of perturbations in the electromagnetic field during inflation. In \cite{1988PhRvD..37.2743T} the resulting magnetic field at the end of inflation in this type of theories was calculated using the assumption that at the time of horizon crossing during inflation the energy density in this mode is determined by the Gibbons-Hawking temperature. This was critically reconsidered in \cite{Mazzitelli} were it was found calculating the spectral energy density from first principle quantizing the corresponding canonical field that this assumption actually is an over-estimation of the actual energy density. However, here we follow the original calculation of \cite{1988PhRvD..37.2743T}.
The equations of motion are derived from the Lagrangian \cite{1988PhRvD..37.2743T}\footnote{Recall the notation convention: Latin indices take values between 0 and 3. Greek indices take values between 1 and 3.}
\begin{eqnarray}
\cll=-\frac{1}{4}F_{mn}F^{mn}-\frac{b}{2}RA^2-\frac{c}{2}R_{mn}A^{m}A^{n},
\end{eqnarray}
where $A^2\equiv A_{m}A^{m}$. Furthermore $b$ and $c$ are constants.
The equations of motion lead together with the parametrization of the Maxwell tensor in terms of the electric and magnetic field, $\hat{E}_{\alpha}$ and $\hat{B}_{\alpha}$ in the "lab" frame, respectively,
\begin{eqnarray}
F_{mn}=a^2\left(
\begin{array}{cccc}
0&-\hat{E}_x&-\hat{E}_y&-\hat{E}_z\\
\hat{E}_x& 0& \hat{B}_z&-\hat{B}_y\\
\hat{E}_y&-\hat{B}_z&0&\hat{B}_x\\
\hat{E}_z& \hat{B}_y & -\hat{B}_x &0
\end{array}
\right)
\label{feb}
\end{eqnarray}
to the equations
\begin{eqnarray}
\frac{1}{a^2}\frac{\partial}{\partial\eta}\left(a^2\hat{E}_{\alpha}\right)-{\rm curl}\hat{B}_{\alpha}-\frac{n}{\eta^2}\frac{A_{\alpha}}{a^2}&=&0
\label{e1}\\
\frac{1}{a^2}\frac{\partial}{\partial\eta}\Big(a^2\hat{B}_{\alpha}\Big)+{\rm curl}\hat{E}_{\alpha}&=&0,
\label{e2}
\end{eqnarray}
where $\eta$ is conformal time. Thus the line element is given by $ds^2=a^2(-d\eta^2+d\vec{x}^{\;2})$ and
\begin{eqnarray}
n\equiv\eta^2\left(
6b\frac{\ddot{a}}{a}+c\left[\frac{\ddot{a}}{a}+\Big(\frac{\dot{a}}{a}\Big)^2\right]
\right)\,,
\end{eqnarray}
where a dot indicates a derivative with respect to conformal time. Taking the curl of equation (\ref{e1}) and using it in equation (\ref{e2}) results in a wave-type equation for the magnetic field,
\begin{eqnarray}
\frac{1}{a^2}\frac{\partial^2}{\partial\eta^2}\Big(a^2\hat{B}_{\alpha}\Big)-D^2\hat{B}_{\alpha}+\frac{n}{\eta^2}\hat{B}_{\alpha}=0.
\end{eqnarray}
Expanding in Fourier modes using $F_{\alpha}(\vec{k}, \eta)\equiv a^2\int d^3x e^{i\vec{k}\cdot\vec{x}}\hat{B}_{\alpha}(\vec{x},\eta)$, yields
\begin{eqnarray}
\ddot{F}_{\alpha}(k)+k^2F_{\alpha}(k)+\frac{n}{\eta^2}F_{\alpha}(k)=0.
\label{Fk}
\end{eqnarray}
So that the additional terms in the Lagrangian can potentially act as pump terms, that is amplifying the spectral energy density of the magnetic field. The averaged magnetic field energy density is given by
$\rho_{\rm mag}(\eta)=\langle(\hat{B}_{\alpha}\hat{B}^{\alpha})(\vec{x},\eta)\rangle/(8\pi)$.
Using the correlation function
\begin{eqnarray}
\langle F_{\alpha}(\vec{k},\eta)F_{\beta}^*(\vec{q},\eta)\rangle=|F_{\mu}F^{\mu}|(k,\eta) \left(\delta_{\alpha\beta}-\frac{k_{\alpha}k_{\beta}}{k^2}\right)\,,
\end{eqnarray}
results in
\begin{eqnarray}
\rho_{\rm mag}(\eta)=\frac{1}{4\pi a^4}\int d^3k|F_{\mu}F^{\mu}|.
\end{eqnarray}
Thus, with $\rho_{\rm B}(k,\eta)=k\frac{d\rho_{\rm mag}}{dk}$, the spectral energy density is given by
\begin{eqnarray}
\rho_{\rm B}(k,\eta)=\frac{1}{4\pi a^4}k^3|F_{\mu}F^{\mu}|.
\end{eqnarray}
Therefore, solving equation (\ref{Fk}) for different types of scale factor, the  magnetic energy
density is estimated after the end of inflation using  $\rho_{\rm B}\propto |F_{\mu}F^{\mu}|/a^4$
\cite{1988PhRvD..37.2743T}.
Note that for standard electrodynamics, $n=0$, and thus the magnetic energy density simply scales at the usual  $a^{-4}$-rate of a frozen-in magnetic field.

\begin{figure}[ht]
\centerline{\epsfxsize=3.3in\epsfbox{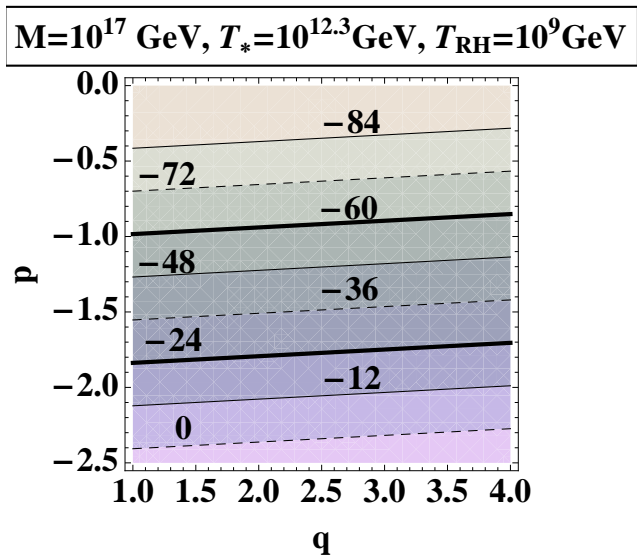}}
\caption{The logarithm of the ratio $r$, magnetic field energy density over photon energy density, is shown for different values of $p$ and $q$ in a model including $RA^2$ and $R_{\mu\nu}A^{\mu}A^{\nu}$ terms \cite{1988PhRvD..37.2743T}.
The numbers in the graph refer to the values of $\log_{10}r$ along the closest contour line. }
\label{fig5.1}
\end{figure}

As can be seen from equation (\ref{Fk}) modes well inside the horizon, that is those modes with comoving wave number $|k\eta|\gg 1$, simply oscillate, since the last term can be neglected and equation (\ref{Fk})  reduces to the equation of a harmonic oscillator.
In the opposite case, for modes well outside the horizon, satisfying $|k\eta|\ll 1$ the second term in equation (\ref{Fk}) becomes subleading and the resulting equation can easily be solved giving the solutions
\begin{eqnarray}
\sqrt{|F_{\mu}F^{\mu}|}\propto \eta^{m_{\pm}},
\end{eqnarray}
where $m_{\pm}\equiv \frac{1}{2}\Big(1\pm\sqrt{1-4n}\Big)$.
Calculating the behaviour of  $\rho_B$ during the different stages of the universe: de Sitter (inflation), radiation and matter dominated epoch, the fastest growing modes are determined by
$p\equiv m_{-}=\frac{1}{2}\Big(1-\sqrt{1-48b-12c}\Big)$ which is calculated during the de Sitter stage and
$q\equiv m_{+}=\frac{1}{2}\Big(1+\sqrt{1-48b-24c}\Big)$ calculated in the radiation dominated era
\cite{1988PhRvD..37.2743T}.
Thus, at the time of galaxy formation, the ratio of magnetic over photon energy density, $r$, is found to be \cite{1988PhRvD..37.2743T}
\begin{eqnarray}
r\simeq(7\times 10^{25})^{-2(p+2)}\Big(\frac{M}{M_{P}}\Big)^{4(q-p)/3}\Big(\frac{T_{\rm RH}}{M_{P}}\Big)^{2(2q-p)/3}\Big(\frac{T_*}{M_{Pl}}\Big)^{-8q/3}\Big(\frac{\lambda}{1 {\rm Mpc}}\Big)^{-2(p+2)},
\end{eqnarray}
where $T_*$ is the temperature at which plasma effects become important during reheating. In \cite{1988PhRvD..37.2743T} this is estimated to be of the order of $T_*\sim {\rm min}\Big[\Big(T_{\rm RH}M\Big)^{1/2},\Big(T^2_{\rm RH}M_{Pl}\Big)^{1/3}\Big]$.
Taking typical values for the physical parameters, there is a wide range for the exponents $p$ and $q$ such that $r$ is bigger than the minimal required value in order to seed the galactic magnetic field with a galactic dynamo operating, $r> 10^{-57}$, or without, $r> 10^{-8}$. In figure \ref{fig5.1} $\log r$ is shown for typical values of the parameters.

\subsection{Quantum corrections in QED in a curved background}

The QED one-loop vacuum polarization of the photon in a general curved background gives rise to terms coupling the Maxwell tensor to the curvature \cite{1980PhRvD..22..343D}.
Vacuum polarization describes the effect of virtual electron positron pair creation thus
giving the photon a size of the order of the electron Compton wave length which interacts with curvature. This leads to an interesting space-time structure including the phenomenon of
gravitational birefringence, where the photon propagation depends on its polarization and it can be faster than the speed of light  \cite{1980PhRvD..22..343D,Shore1,Shore2,Hollowood}.
In general the Lagrangian has the form \cite{1980PhRvD..22..343D}
\begin{eqnarray}
\cll&=&-\frac{1}{4}F_{mn}F^{mn}
\nonumber\\
&&-\frac{1}{4m_e^2}\Big(
b RF_{mn}F^{mn}+c R_{mn}F^{mk}F^{n}_{\;\; k}+ dR_{mnlk}F^{mn}F^{lk}+f(\nabla_{m}F^{mn})(\nabla_{a}F^{a}_{\;\;n})\Big),
\end{eqnarray}
where $b$, $c$ , $d$ and $f$ are constants.
The last term can be neglected with respect to the other terms since it leads to higher order derivative terms in the equations of motion \cite{kek1},
\begin{eqnarray}
\nabla^{m}F_{mn}&+&\frac{1}{m_e^2}\nabla^{m}\Big[bRF_{mn}+\frac{c}{2}\Big(
R^{l}_{\;\;m}F_{ln}-R^{l}_{\;\;n}F_{lm}\Big)+dR^{lk}_{\;\;\;mn}F_{lk}\Big]\nonumber\\
&+&\frac{f}{2m_e^2}
\left(
\nabla_{a}\nabla^{a}\nabla^{b}F_{bn}+
R^{a}_{\;\;n}\nabla^{b}F_{ba}
\right)=0.
\label{eef1}
\end{eqnarray}
In order to proceed here a different approach from the one used in the previous section will be followed. This will allow to actually determine the spectrum of the resulting magnetic field.
Instead of assuming that the energy density of the magnetic field at the time of horizon crossing corresponds to the  one calculated using the Gibbons-Hawking temperature, the spectrum of the resulting primordial field will be calculated by determining explicitly the Bogoliubov coefficients.
This gives the particle production and thus the spectral energy density of the primordial magnetic field. The background model will be described by two stages, an inflationary stage during which the  correction terms coupled to curvature are important and a radiation dominated stage determined by  standard Maxwell electrodynamics. Since the resulting field strength will be compared with observational values at the order of galactic scale corresponding to 1 Mpc today, it is not necessary to  include the evolution during the matter dominated period. Galactic length scales re-enter the horizon during the radiation dominated stage, which can  seen from the temperature $T$
\begin{eqnarray}
T\sim 78 \left(\frac{\lambda}{\rm Mpc}\right)^{-1} {\rm\; eV}\,,
\end{eqnarray}
at which a scale for $\lambda<\lambda_{eq}\sim 14\Omega_{\rm m}^{-1}h^{-2}$ Mpc crosses back into the horizon. Thus a galactic scale of order 1 Mpc enters the horizon at a time when the universe was at about 78 eV and thus inside the radiation dominated era, since radiation-matter equality occurs at $T_{eq}=5.6 \Omega_{\rm m} h^2 $eV.

The background cosmology is chosen to be such that
\begin{eqnarray}
a(\eta)=\left\{
\begin{array}{lr}
a_1\left(\frac{\eta}{\eta_1}\right)^{\beta}& \eta<\eta_1\\
&\\
a_1\left(\frac{\eta-2\eta_1}{-\eta_1}\right)&\eta\geq\eta_1.
\end{array}
\right.
\end{eqnarray}
In the following $a_1\equiv 1$. The matching between the inflationary phase and the radiation dominated era takes place at $\eta=\eta_1$. De Sitter inflation corresponds to $\beta=-1$ and for
$\beta<-1$ power-law inflation is taking place.

The  Maxwell tensor  is written in terms of the gauge potential $A_{m}$, that is $F_{mn}=\partial_{m}A_{n}-\partial_{n}A_{m}$. Furthermore, the radiation gauge,
$A_{0}=0$, $\partial_{\lambda}A_{\lambda}=0$ will be used.
Then the gauge potential is given in terms of the expansion in  Fourier modes by
\begin{eqnarray}
A_{\mu}(\eta,\vec{x})=\int \frac{d^3k}{(2\pi)^{\frac{3}{2}}\sqrt{2k}}\sum_{\lambda=1}^{2}\epsilon_{\vec{k}\;\mu}^{(\lambda)}\left[a^{(\lambda)}_{\vec{k}}A_k(\eta)e^{i\vec{k}\cdot\vec{x}}+a^{(\lambda)\,\dagger}_{\vec{k}}A^*_k(\eta)e^{-i\vec{k}\cdot\vec{x}}\right],
\end{eqnarray}
where the sum is over the two physical polarization
states and  $\epsilon_{\vec{k}\;\mu}^{(\lambda)}$ are the polarization vectors satisfying,
$\vec{\epsilon}_{\vec{k}}^{\;(\lambda)}\cdot\vec{k}=0$.
$a_{\vec{k}}^{(\lambda)}$ and $a^{(\lambda)\,\dagger}_{\vec{k}}$ are annihilation and creation operators, respectively, satisfying the standard commutation relations,
$[a_{\vec{k}}^{(\lambda)},a_{\vec{k}'}^{(\lambda')}]=0=[a^{(\lambda)\,\dagger}_{\vec{k}},a^{(\lambda')\,\dagger}_{\vec{k}'}]$ and $[a_{\vec{k}}^{(\lambda)},a^{(\lambda')\,\dagger}_{\vec{k}'}]=\delta_{\lambda\lambda'}\delta_{\vec{k}\vec{k}'}$. However, since the Lagrangian has additional terms coupling the electromagnetic field to the curvature, the commutation relations between the operator $A_j$ and its canonical momentum $\pi_{A_{\mu}}=\frac{\partial {\cll}}{\partial (\partial_{0}A_j)}$ are not the canonical ones. However, as will be done explicitly below, it is possible to define a canonical field which does satisfy the standard commutation relation. It is that field that is used to quantize the theory and calculate the production of particles.

\begin{figure}[ht]
\centerline{\epsfxsize=3in\epsfbox{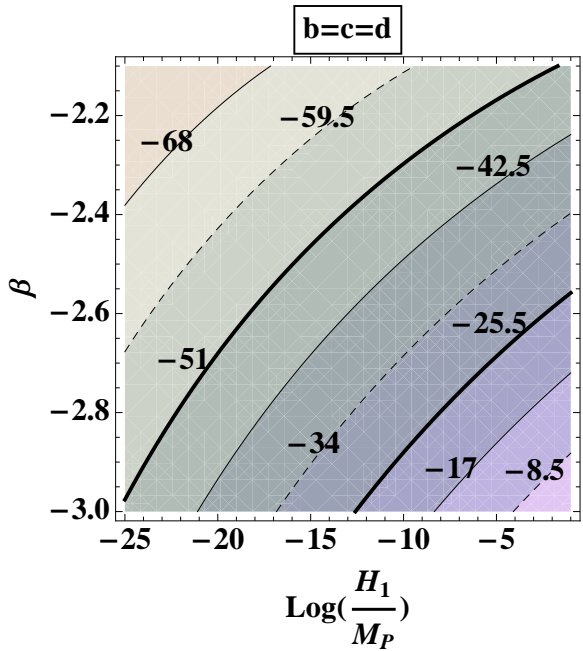}}
\caption{In the case $b=c=d$ the contour lines are shown for the maximum value of the logarithm of the ratio of magnetic to background radiation energy density. The values of $\log\left(\frac{H_1}{M_{Pl}}\right)$  correspond to
reheat temperatures between $10^7$ GeV and $10^{19}$ GeV.
The numbers within the graph refer to the value of $\log_{10}r_{\rm max}$ along the closest contour line.}
\label{fig5.3}
\end{figure}

The mode functions satisfy
 \begin{eqnarray}
F_1(\eta)\ddot{A}_k+F_2(\eta)\dot{A}_k+F_3(\eta)k^2A_k=0,
\end{eqnarray}
where a dot indicates $\frac{d}{d\eta}$ and
\begin{eqnarray}
F_1(\eta)&=&1+\frac{\mu_1}{m_e^2\eta_1^2}\left(\frac{\eta}{\eta_1}\right)^{-2(\beta+1)}
\hspace{2cm}
\mu_1=\beta\Big[6b(\beta-1)+c(\beta-2)-2d\Big]
\nonumber\\
F_2(\eta)&=&\frac{\mu_2}{\eta_1^3m_e^2}\left(\frac{\eta}{\eta_1}\right)^{-2\beta-3}
\hspace{2.85cm}
\mu_2=-2(\beta+1)\mu_1
\nonumber\\
F_3(\eta)&=&1+\frac{\mu_3}{\eta_1^2m_e^2}\left(\frac{\eta}{\eta_1}\right)^{-2(\beta+1)}
\hspace{2cm}
\mu_3=\beta\Big[6b(\beta-1)+c(2\beta-1)+2d\beta\Big].
\end{eqnarray}
In the case where the additional terms in the lagrangian are absent, that is $b=c=d=0$, the mode equation for  $A_k$ reduces to a simple harmonic oscillator equation. In this case $A_k$ itself can be used to implement the standard quantization scheme.
In general, however, it is necessary to use the canonical field $\Psi_{\mu}(\eta,\vec{x})$ and its Fourier amplitude $\Psi(\eta,\vec{k})$  defined by, respectively,
\begin{eqnarray}
\Psi_{\mu}(\eta,\vec{x})=F_1^{\frac{1}{2}}(\eta)A_{\mu}(\eta,\vec{x})\hspace{5cm}
\Psi=F_1^{\frac{1}{2}}(\eta)A_k.
\end{eqnarray}
With this the mode equation for $\Psi(\eta,k)$ is given by
\begin{eqnarray}
\Psi''+P\Psi=0,
\label{e3}
\end{eqnarray}
where a new dimensionless variable $z\equiv -k\eta$ has been defined and $' \equiv \frac{d}{dz}$.
Moreover,
\begin{eqnarray}
P=\frac{1}{4}\frac{\kappa_1z^{-4\beta-6}}{\Big[1+\kappa_2z^{-2(\beta+1)}\Big]^2}+\frac{1}{2}
\frac{\kappa_3z^{-2\beta-4}}{1+\kappa_2z^{-2(\beta+1)}}
+\frac{1+\kappa_4z^{-2(\beta+1)}}{1+\kappa_2z^{-2(\beta+1)}},
\label{P}
\end{eqnarray}
and
\begin{eqnarray}
\kappa_1&\equiv&\mu_2^2\kappa_0^2\hspace{1.5cm}
\kappa_2\equiv\mu_1\kappa_0\hspace{1.5cm}
\kappa_3\equiv(2\beta+3)\mu_2\kappa_0\hspace{1.5cm}
\kappa_4\equiv\mu_3\kappa_0\nonumber\\
{\rm where}\hspace{0.5cm}
\kappa_0&\equiv&\left(\frac{m_e}{H_1}\right)^{-2}\left(\frac{k}{k_1}\right)^{2(\beta+1)}.
\end{eqnarray}
Here the maximally amplified (comoving) wavenumber $k_1$ has been defined by $k_1\equiv\frac{1}{|\eta_1|}$. Furthermore $H_1$ is the value of the Hubble paramter at the beginning of the radiation dominated stage at $\eta_1$. It is related to $k_1$ by $k_1\sim H_1$.
Thus the canonical field satisfies the equation of a harmonic oscillator.  This is also the case of  a free scalar field in flat space-time. Therefore the canonical quantization procedure will be applied to the canonical field $\Psi$, which will be written as
\begin{eqnarray}
\Psi_{\mu}(\eta,\vec{x})=\int d^3k\;\epsilon_{\vec{k}\;\mu}^{(\lambda)}\left[a^{(\lambda)}_{\vec{k}}f_{\vec{k}}+a^{(\lambda)\,\dagger}_{\vec{k}}f_{\vec{k}}^*\right]\,.
\end{eqnarray}
Hence it will be required that the mode functions  $f_{\vec{k}}(x)\equiv\Psi e^{i\vec{k}\cdot\vec{x}}/(2\pi)^{\frac{3}{4}}$ and $f_{\vec{k}}^*$
form an orthonormal set, that is satisfying \cite{1982qfcs.book.....B},
\begin{eqnarray}
(f_{\vec{k}},f_{\vec{k'}})=\delta^{(3)}(\vec{k}-\vec{k}'),\hspace{1cm}
(f_{\vec{k}}^*,f_{\vec{k'}}^*)=-\delta^{(3)}(\vec{k}-\vec{k}'),\hspace{1cm}
(f_{\vec{k}},f_{\vec{k'}}^*)=0.
\end{eqnarray}
Furthermore, the scalar product is defined by
\begin{eqnarray}
(f_{\vec{k}},f_{\vec{k'}})=-i\int_{\Sigma}f_{\vec{k}}(x)\stackrel{\leftrightarrow}{\partial}_{m}f_{\vec{k'}}^*(x)d\Sigma^{m},
\end{eqnarray}
where $d\Sigma^{m}=n^{m}d\Sigma$ and $n^{m}$ is a future-directed unit vector orthogonal to the space-like hypersurface $\Sigma$ which is taken to be a Cauchy surface. Moreover, $d\Sigma$ is the volume element in $\Sigma$.
Also the notation $f_{\vec{k}}(x)\stackrel{\leftrightarrow}{\partial}_{m}f_{\vec{k'}}^*=f_{\vec{k}}\partial_{m}f_{\vec{k'}}^*-(\partial_{m}f_{\vec{k}})f_{\vec{k'}}^*$ was used.
Since $\Psi$ is quantized in flat space-time,
$d\Sigma^{m}=\delta^{m}_0d^3x$ and the normalization condition on the mode functions $f_{\vec{k}}$ reduces to
\begin{eqnarray}
\Psi(\eta,k)\partial_{\eta}\Psi(\eta,k')^*-\left[\partial_{\eta}\Psi(\eta,k)\right]\Psi(\eta,k')^*=i,
\end{eqnarray}
which is the Wronskian of the solutions of the differential equation (\ref{e3}).
The field equation in real space for  the Fourier transform $\Psi(\eta,\vec{x})$, assuming it to be real, can be derived from the Lagrangian
\begin{eqnarray}
\cll_{\Psi}=\frac{1}{2}\left(\eta^{ab}\Psi_{,a}\Psi_{,b}-m_{\rm eff}^2\Psi^2\right).
\end{eqnarray}
The effective mass can be determined by going back to $k$-space.
Using  equations (\ref{e3}) and (\ref{P})
\begin{eqnarray}
m_{\rm eff}^2=k^2(1-P).
\end{eqnarray}
It can be verified that $\Psi_{\mu}(\eta,\vec{x})$ and its canonical momentum satisfy the standard commutation relations.
The time-dependent effective mass $m_{\rm eff}(\eta)$ reflects the dynamics of the cosmological background. It also indicates that there is no unique vacuum state.
Having one set of orthonormal functions $f_{\vec{k}}$ another orthonormal set of mode functions $\tilde{f}_{\vec{k}}$ can be found.
Then the canonical field $\Psi_{\mu}$ has the expansion in terms of the annihilation and creation operators $\tilde{a} ^{(\lambda)}_{\vec{k}}$, $\tilde{a}^{(\lambda)\,\dagger}_{\vec{k}}$ and
$\tilde{f}_{\vec{k}}$
\begin{eqnarray}
\Psi_{\mu}(\eta,\vec{x})=\int d^3k\;\epsilon_{\vec{k}\;\mu}^{(\lambda)}\left[\tilde{a}^{(\lambda)}_{\vec{k}}\tilde{f}_{\vec{k}}+\tilde{a}^{(\lambda)\,\dagger}_{\vec{k}}\tilde{f}_{\vec{k}}^*\right]
\end{eqnarray}
and this defines a new vacuum state $|\tilde{0}\rangle$,
\begin{eqnarray}
\tilde{a}^{(\lambda)}_{\vec{k}}|\tilde{0}\rangle=0,
\end{eqnarray}
for all $\vec{k}$ and $\lambda$ and a new Fock space.
Since both sets of mode functions are complete, they
are related by the Bogolubov transformation \cite{1982qfcs.book.....B},
\begin{eqnarray}
\tilde{f}_{\vec{k}}=\sum_{\vec{q}}(\alpha_{\vec{k}\,\vec{q}}f_{\vec{q}}+\beta_{\vec{k}\,\vec{q}}f_{\vec{q}}^*),
\end{eqnarray}
where $\alpha_{\vec{k}\,\vec{q}}$ and $\beta_{\vec{k}\,\vec{q}}$ are the Bogolubov coefficients
satisfying
$
\sum_{\vec{k}}(\alpha_{\vec{q}\,\vec{k}}\alpha_{\vec{r}\,\vec{k}}^*-\beta_{\vec{q}\,\vec{k}}\beta_{\vec{r}\,\vec{k}}^*)=\delta_{\vec{q}\,\vec{r}}$.
Moreover, it is found that, suppressing the index $\lambda$, \cite{1982qfcs.book.....B},
\begin{eqnarray}
a_{\vec{k}}=\sum_{\vec{q}}(\alpha_{\vec{q}\,\vec{k}}\tilde{a}_{\vec{q}}+\beta_{\vec{q}\,\vec{k}}^*\tilde{a}_{\vec{q}}^{\dagger}).
\label{bd1}
\end{eqnarray}
Thus, equation (\ref{bd1}) implies that the vacuum state $|\tilde{0}\rangle$ is in general not annihilated by
$a_{\vec{k}}$, but rather gives
\begin{eqnarray}
a_{\vec{k}}|\tilde{0}\rangle=\sum_{\vec{q}}\beta_{\vec{q}\,\vec{k}}^*|\tilde{1}_{\vec{q}}\rangle\neq 0.
\end{eqnarray}
This means that the expectation value of the number operator $N_{\vec{k}}=a_{\vec{k}}^{\dagger}\,a_{\vec{k}}$ of $f_{\vec{k}}$-mode particles in the state $|\tilde{0}\rangle$ is given by
\begin{eqnarray}
\langle\tilde{0}|N_{\vec{k}}|\tilde{0}\rangle=\sum_{\vec{q}}|\beta_{\vec{q}\,\vec{k}}|^2.
\end{eqnarray}
In order to determine the particle production due to the time-dependent cosmological background the mode functions are matched at the transition time $\eta=\eta_1$.
Furthermore, on subhorizon scales corresponding to $z\gg 1$, the mode equation (\ref{e3}) reduces to the equation for a free harmonic oscillators and therefore does not give any important contribution. Only  modes on superhorizon scales are relevant, since in that case, for $z\ll1$, the mode equation can be approximated by
\begin{eqnarray}
\Psi''+(\xi_1z^{-2}+\xi_2)\Psi=0,
\label{e4}
\end{eqnarray}
where
\begin{eqnarray}
\xi_1=-(\beta+1)(\beta+2)
\hspace{2cm}
\xi_2=\frac{6b(\beta-1)+c(2\beta-1)+2d\beta}
{6b(\beta-1)+c(\beta-2)-2d}.
\end{eqnarray}
The particular choice $\beta=-1$ describes de Sitter inflation, and in this case $\xi_1=0$ and $\xi_2=1$  leading to  a plane wave solution which was also noted in
\cite{1980PhRvD..22..343D,1988PhRvD..37.2743T}. Furthermore,  $\beta=-2$  implies $\xi_1=0$, but $\xi_2=\frac{18b+5c+4d}{18b+4c+2d}$.
Equation (\ref{e4}) solved during power law inflation, $\beta<-1$ and $\beta\neq -2$, results in the following solution in terms of the Hankel function of the second kind, $H_{\nu}^{(2)}(x)$,
\begin{eqnarray}
\Psi^{\rm (I)}=\sqrt{\frac{\pi}{2k}}\sqrt{z}H_{\nu}^{(2)}(\sqrt{\xi_2}\,z),
\hspace{1cm}{\rm where}
\hspace{0.5cm}
\nu=\left|\beta+\frac{3}{2}\right|
 \end{eqnarray}
which gives the correctly normalized incoming wave function for $\eta\rightarrow-\infty$ for $\xi_2>0$ .
This means that the incoming vacuum solution at infinity is a plane wave solution and moreover approaches the positive frequency solution in Minkowski space-time.
It is assumed that electrodynamics becomes standard Maxwell electrodynamics at the beginning of the radiation dominated stage at $\eta=\eta_1$. Thus the terms due to the interaction between curvature and the electromagnetic field in the mode equation (\ref{e4}) can be neglected which leads to a free harmonic oscillator equation which is solved by the superposition of plane waves,
\begin{eqnarray}
\Psi^{\rm (RD)}=\frac{1}{\sqrt{k}}\left(c_{+}e^{-i(z-z_1)}+c_{-}e^{i(z-z_1)}\right),
\end{eqnarray}
where $z_1\equiv k|\eta_1|$ and $c_{\pm}$ are the Bogoliubov coefficients, corresponding to
$\alpha_{\vec{k}\,\vec{q}}=c_{+}\delta_{\vec{k}\,\vec{q}}$ and $\beta_{\vec{k}\,\vec{q}}^*=c_{-}\delta_{\vec{k}\,\vec{q}}$.
Therefore to determine the magnetic field energy spectrum, the Bogolubov coefficients are calculated by matching the solutions of the gauge potential and its first derivative at $\eta=\eta_1$ on superhorizon scales. Using the small argument limit of the Hankel functions
\cite{1965hmfw.book.....A}, this leads to
$\beta\neq-\frac{3}{2}$ \cite{kek1}
\begin{eqnarray}
|c_{-}|^2\simeq\frac{\left[\Gamma(\nu)\right]^2}{8\pi\mu_1}\left(\frac{1}{2}-\nu\right)^2
\left(\frac{m_e}{H_1}\right)^2\left(\frac{\xi_2}{4}\right)^{-\nu}\left(\frac{k}{k_1}\right)^{-1-2\nu}\,,
\end{eqnarray}
where it was used that in the approximation used here,
$F_1(\eta_1)\simeq\mu_1\left(\frac{m_e}{H_1}\right)^{-2}$. In the case $\beta=-\frac{3}{2}$
the limiting behavior of the mode function on superhorizon scales leads to a
a divergent  factor $\ln^2\left(\sqrt{\xi_2}\frac{k}{k_1}\right)$ in $|c_{-}|^2$. Thus, we will not pursue this case any further.
Including both polarization states the total spectral energy density of the photons is given by (cf., e.g., \cite{2002NewAR..46..659D})
\begin{eqnarray}
\rho(\omega)\equiv\frac{d\rho}{d\log k}\simeq
2\left(\frac{k}{a}\right)^4\frac{|c_{-}|^2}{\pi^2}\,.
\label{e5}
\end{eqnarray}
Since the electric field decays rapidly due to the high conductivity of the radiation dominated universe,  the spectral energy density (\ref{e5}) gives a measure of the magnetic field energy density, $\rho_{\rm B}$. Using  the density parameter of radiation,
$\Omega_{\gamma}= \left(\frac{H_1}{H}\right)^2\left(\frac{a_1}{a}\right)^4$,
the ratio of magnetic over background radiation energy density $r$ is given for $\beta\neq -2,-\frac{3}{2}, -1$, by \cite{kek1}
\begin{eqnarray}
r\simeq
\frac{2\left[\Gamma(\nu)\right]^2}{3\pi^2\mu_1}\left(\frac{1}{2}-\nu\right)^2\left(
\frac{m_e}{M_{Pl}}\right)^2\left(\frac{\xi_2}{4}\right)^{-\nu}\left(\frac{k}{k_1}\right)^{3-2\nu}\,,
\label{e6}
\end{eqnarray}
where $M_{Pl}$ is the Planck mass.
The magnetic field energy density can also be calculated using the two point function of the magnetic field, $\langle B_{\mu}(\vec{k})B_{\nu}^*(\vec{k}')\rangle$. This leads to an expression similar to (\ref{e6}).
Furthermore, the form of the magnetic field spectrum (\ref{e6}) imposes the constraint $\nu\leq\frac{3}{2}$ . This implies the range for $\beta$ given by  $-3<\beta<-1$ taking into account the constraint from power law inflation.
Using the constraint which was used to derive the mode equation (\ref{e4})
\begin{eqnarray}
\mu_1\left(\frac{m_e}{H_1}\right)^{-2}>1,
\label{mu}
\end{eqnarray}
the maximal value of $r$ which can be achieved within this model can be estimated.
It is found to be, for $\beta\neq -2,-\frac{3}{2}, -1$, at $\omega_G=10^{-14}$ Hz corresponding to a galactic scale of 1 Mpc, and using the maximal amplified frequency evaluated today,
$\omega_1(\eta_0)=6\times 10^{11}\left(\frac{H_1}{M_{Pl}}\right)^{\frac{1}{2}}$ Hz,
\cite{kek1}
\begin{eqnarray}
r_{\rm max}(\omega_G)=10^{-79+52\nu}\left[\Gamma(\nu)\right]^2\left(\frac{1}{2}-\nu\right)^2 \left(\frac{\xi_2}{4}\right)^{-\nu}\left(\frac{H_1}{M_{Pl}}\right)^{\nu+\frac{1}{2}}.
\end{eqnarray}
In figure (\ref{fig5.3}) $\log_{10}r_{\rm max}$ is shown for the case that the parameters determining the contributions due to the quantum corrections are all of the same order, $b\sim c\sim d$.
In this case the constraint on $\mu_1$ leads to a lower bound on the parameter $b$, given by
\begin{eqnarray}
b_{\rm min}\equiv \frac{10^{-45}}{\beta(7\beta-10)}\left(\frac{H_1}{M_{Pl}}\right)^{-2}.
\end{eqnarray}
As can be appreciated from figure (\ref{fig5.3}) there is a region in parameter space where the resulting  magnetic field strength is  larger than $B_s>10^{-20}$ G corresponding to $r>10^{-37}$, namely, $\beta<-2.4$, $\frac{H_1}{M_{Pl}}>10^{-18}$ bounded by the corresponding contour line.
In the case of  De Sitter inflation,  corresponding  to $\beta=-1$, there is no significant magnetic field generation since the mode functions during inflation as well as during the
radiation dominated stage are plane waves.
Furthermore, it can be checked that the resulting maximum magnetic field strength satisfies the bound due to gravitational wave production~\cite{2002PhRvD..65b3517C}. It was shown in
\cite{2002PhRvD..65b3517C} that for magnetic fields created before nucleosynthesis conversion of magnetic field energy into gravitational wave energy takes place. This leads to a maximal value of $r_{\rm GW}$ given by,
\cite{kek1}
\begin{eqnarray}
r_{\rm GW}\simeq 2\times 10^{-61+52\nu}
2^{\frac{5}{2}-\nu}\,\Gamma\left(\frac{5}{2}-\nu\right)
h_0^2\left(\frac{H_1}{M_{Pl}}\right)^{\nu-\frac{3}{2}},
\end{eqnarray}
at the galactic scale used here, $\lambda=1$ Mpc.
Thus, the requirement $r_{\rm max}\leq r_{\rm GW}$ leads to  an upper limit on $\frac{H_1}{M_{Pl}}$ , that is,
\begin{eqnarray}
\log_{10}\left(\frac{H_1}{M_{Pl}}\right)_{\rm max}\equiv 9+\frac{1}{2}\log_{10}\left[\frac{2^{\frac{7}{2}-\nu}h_0^2\Gamma\left(\frac{5}{2}-\nu\right)}{\Gamma^2(\nu)\left(\frac{1}{2}-\nu\right)^2}
\left(\frac{\xi_2}{4}\right)^{\nu}\right]\,.
\label{gw}
\end{eqnarray}
Hence  the allowed range is given by $\left(\frac{H_1}{M_{Pl}}\right)\leq\left(\frac{H_1}{M_{Pl}}\right)_{\rm max}$.  This is always satisfied since $H_1<M_{Pl}$.

It is also important to check that the fluctuation in the electromagnetic field during inflation are within the perturbative regime and thus there is no strong backreaction on the dynamics of inflation. This effect can be estimated by calculating the energy density in the electromagnetic field and comparing it with the total energy density during inflation given by,
\begin{eqnarray}
\frac{\rho}{M_{Pl}^4}=\frac{3}{8\pi}\left(\frac{H_1}{M_{Pl}}\right)^2\left(\frac{\eta}{\eta_1}\right)^{-2(\beta+1)}.
\end{eqnarray}
The average value of the electromagnetic field energy density is found to be \cite{kek1}
\begin{eqnarray}
\langle\rho_{\rm (em)}\rangle(\eta)&\simeq&\frac{2}{a^4\eta^2}\int_{0}^{k_*}dk k^2 |\Psi^{\rm (I)}|^2 \nonumber\\
&\simeq&\frac{1}{\pi}\frac{[\Gamma(\nu)]^2}{3-2\nu}\left(\frac{\xi_2}{4}\right)^{-\nu}H_1^4 \left(\frac{k_*}{k_1}\right)^{3-2\nu}\left(\frac{\eta}{\eta_1}\right)^{-4\beta-2\nu-1}\,,
\end{eqnarray}
where $k_*$ is the wave number corresponding to the scale which becomes superhorizon at the time $\eta$ during inflation. Thus   with $k_*\sim -\eta^{-1}$ the ratio $\langle\rho_{\rm (em)}\rangle(\eta)/\rho\sim\rho/M^4_{\rm P}$ which is always smaller than one in the classical domain.
Therefore, no backreaction effects have to be taken into account.
In other models of magnetic field generation during inflation backreaction does play a role~\cite{2009JCAP...08..025D,2009JCAP...12..009K}.

\subsection{Trace anomaly}

Linear electrodynamics is scale invariant at classical level. Taking into account quantum corrections it is known that this classical symmetry is broken. Defining the energy momentum tensor by (see for example, \cite{ps})
\begin{eqnarray}
\Theta^{ab}=2\frac{\delta}{\delta g_{ab}(x)}\int d^4x \cll_m,
\end{eqnarray}
then if the classical theory is  scale invariant there will be a conserved current $C^{a}=\Theta^{ab}x_{b}$ such that
\begin{eqnarray}
\partial_{a}C^{a}=\Theta^{a}_{a}.
\end{eqnarray}
A scale transformation is equivalent to a conformal transformation of the metric, such as
\begin{eqnarray}
g_{mn}(x)\rightarrow e^{2\sigma}g_{mn}(x).
\label{ct}
\end{eqnarray}
Since the trace vanishes of the electromagnetic field in linear electrodynamics, the current $C^{m}$ is conserved at the classical level.
However, when quantum corrections are included scale transformations are no longer a symmetry, since the renormalized coupling constant depend on the scale. Namely, the renormalized coupling constant changes under the conformal transformation (\ref{ct})
as \cite{ps}
\begin{eqnarray}
g\rightarrow g+\sigma\beta(g),
\end{eqnarray}
where $\beta(g)$ is the beta function.
The Lagrangian changes as $\sigma\beta(g)\frac{\partial}{\partial g}\cll$. Thus the current satisfies,
\begin{eqnarray}
\partial_{m}C^{m}=\Theta^{m}_{m}=\beta(g)\frac{\partial}{\partial g}\cll.
\end{eqnarray}
In massless QED the trace of the energy-momentum tensor can be found explicitly as \cite{ps}
\begin{eqnarray}
\Theta^{m}_{m}=\frac{\beta(e)}{2e^3}F_{ab}F^{ab}.
\end{eqnarray}
A similar expression for the trace of the energy-momentum is also found in QCD and other gauge theories.

The trace anomaly was used in \cite{Dolgov1,1993PhRvD..48.2499D} (see also \cite{2001astro.ph..6247P})  to study the generation of primordial magnetic fields during inflation. It induces a new term in Maxwell's equations, namely, in a flat Friedmann-Robertson-Walker background with scale factor $a$, \cite{Dolgov1,1993PhRvD..48.2499D},
\begin{eqnarray}
\partial_{m}F_{n}^{\;\;m}+\kappa\frac{\partial_{m}a}{a}F_{n}^{\;\;m}=0.
\end{eqnarray}
The constant $\kappa$ depends on the theory which is used. For example, for the SU$(N)$ gauge theory with $N_f$ generations of fermions in the fundamental representation \cite{Dolgov1,1993PhRvD..48.2499D},
\begin{eqnarray}
\kappa=\frac{\alpha}{\pi}\left(\frac{11N}{3}-\frac{2N_f}{3}\right),
\end{eqnarray}
where $\alpha$ is the fine structure constant taken at the time of horizon crossing of the scale $k^{-1}$ during inflation. Quantizing the gauge potential and finding the spectrum of the electromagnetic field in de Sitter inflation it is found that \cite{1993PhRvD..48.2499D}
\begin{eqnarray}
|A_k|\sim \left(\frac{H}{k}\right)^{\frac{\kappa}{2}}.
\end{eqnarray}
Thus for large values of $\kappa$ corresponding to a large number of light fermions during inflation the trace anomaly could provide an efficient mechanism to generate large magnetic fields during inflation
to serve as seed magnetic fields for a subsequent amplification by a galactic dynamo.

\subsection{Coupling to other fields and varying couplings}

In \cite{1988PhRvD..37.2743T} the coupling of a pseudoscalar axion  to electrodynamics was proposed which for
energy scales below the Peccei-Quinn symmetry breaking scale $f_a$ can be described by the effective Lagrangian,
\begin{eqnarray}
\cll=-\frac{1}{2}\partial_{m}\theta\partial^{m}\theta-\frac{1}{4}F_{mn}F^{mn}+g_a\theta F_{mn}\,^{*}F^{mn},
\end{eqnarray}
where $g_a$ is a coupling constant and the vacuum angle $\theta=\phi_a/f_a$, where $\phi_a$ is the axion field.
 In \cite{gfs} a similar model has been considered in detail, namely
the coupling of a pseudo Goldstone boson to electrodynamics. It is interesting to note that in these models  the created magnetic fields have non zero helicity  \cite{gfs,fc2,2009IJMPD..18.1395C,2010arXiv1005.5322D}. In \cite{gfs} the Lagrangian is assumed to be of the form
\begin{eqnarray}
\cll=-\frac{1}{4}\left(F_{mn}F^{mn}+g\phi F_{mn}\,^{*}F^{mn}\right),
\end{eqnarray}
where $g=\alpha/(2\pi f)$ and $f$ is the coupling constant and $\alpha$ the fine structure constant.
The field equations in Fourier space are found to be \cite{gfs}
\begin{eqnarray}
\frac{d^2 F_{\pm}}{d\eta^2}+\left(k^2\pm gk\frac{d\phi}{d\eta}\right)=0,
\end{eqnarray}
where $F_{\pm}=a^2(B_y\pm iB_z)$ are the two circular polarization modes. The electric field satisfies the equation, \cite{gfs}
\begin{eqnarray}
\frac{d^2 G_{\pm}}{d\eta^2}+\left(k^2\pm g\frac{d\phi}{d\eta}k\right)G_{\pm}=-g\frac{d^2\phi}{d\eta^2}F_{\pm},
\end{eqnarray}
where $G_{\pm}=a^2(E_y\pm iE_z)$.
Using a potential of the form $V(\phi)=\Lambda^4[1-\cos(\phi/f)]$
 for the scalar field, the resulting  amplification of the magnetic field during inflation is too weak in order to provide a seed field for the galactic dynamo \cite{gfs}. In \cite{2006JCAP...10..018A} a model  with  N pseudo Goldstein bosons has been investigated. It has been found that even in the case of one pseudo Goldstein boson due to the helical nature of the generated magnetic field the process of the inverse cascade will result in a strong enough seed field at the time of galaxy
formation.

A time-dependent  coupling of the electromagnetic field provides
another possibility of amplification of perturbations in the electromagnetic field during inflation.
This was first studied in \cite{1992ApJ...391L...1R} (see also, \cite{giov,2004PhRvD..69d3507B,2005PhRvD..71l3525B,2008JCAP...01..025M,2010PhRvD..81h3526E}) considering a lagrangian of the form \cite{1992ApJ...391L...1R}
\begin{eqnarray}
\cll\sim  e^{\alpha\phi}F_{mn}F^{mn},
\end{eqnarray}
where $\alpha$ is a constant. Perturbations in the electromagnetic field are amplified during de Sitter inflation. The resulting magnetic field for the choice of $\alpha=20$ is found to be as large as $6.5\times 10^{-10}$ G today.

In \cite{2008PhLB..659..661G} (see also \cite{2010JCAP...04..003G}) the generation of cosmologically relevant  magnetic fields and their subsequent signature in the CMB has been discussed in a model where
instead of the usual U(1)$_{\rm em}$  gauge field, the photon,
 a hypercharge field $Y_{m}$ \cite{1998PhRvD..57.2186G}, which is associated with the U(1)$_Y$ hypercharge group before the electroweak  phase transition when the SU(2)$\times$U(1)$_Y$ symmetry is still unbroken, is coupled to a spectator field during inflation.
After the electroweak phase transition the photon field is determined by the hypercharge field by $A_{m}=Y_{m}\cos\theta_{\rm W}$ which gives rise to a primordial magnetic field in the postinflationary universe.

\subsection{Magnetogenesis in string theory}

A natural candidate for a scalar field coupled to the electromagnetic field is provided within the low energy limit of string theory.
In the low energy limit string theory leads to Einstein gravity coupled to additional fields, such as the dilaton, which is a scalar field, and the antisymmetric tensor field which in four dimensions can be related to a pseudo-scalar field, the axion.
To lowest order in the inverse string tension $\alpha'$ and in the loop expansion controlled by the string coupling $g_{\rm s}$ the action in the so-called string frame
is given by
\begin{eqnarray}
S=-\frac{1}{16\pi G_D}\int d^Dx \sqrt{-g_D}e^{-\phi}\left[R_D+\partial_{m}\phi\partial^{m}\phi+\frac{1}{4}F_{mn}F^{mn}\right]\,,
\label{s1}
\end{eqnarray}
where $G_D$ is Newton's constant in $D$ dimensions, $R_D$ the Ricci scalar in D dimensions and $\phi$ is the dilaton. Indices take values between 0 and $D-1$. The string coupling is given by $g_{\rm s}=e^{\phi}$.
Superstring theory can be consistently quantized only in $D=10$ and M-theory predicts 11 space-time dimensions.
In order to reduce the resulting model to the four observed space-time dimensions, the extra space-time dimensions can either be treated as compactified to small extra dimensions, following the paradigm of Kaluza-Klein compactification, or one could model the observable universe as a four dimensional hypersurface embedded in a higher dimensional background space-time, which is the procedure followed in the models of brane cosmology \cite{2004LRR.....7....7M}.
Here the simplest model is used where the extra dimensions are compactified on static tori with small, constant radii. Thus the action (\ref{s1}) is used in $D=4$ dimensions \cite{ven,gv,lwc}.

It is difficult to implement the standard slow roll inflation paradigm in string cosmology, derived from the low energy limit of superstring theory. The reason for that is that the dilaton does not have an appropriate potential. The potential resulting from supersymmetry breaking is far too steep to allow for a slow roll phase in the evolution of the dilaton.  Inflation driven by the kinetic energy of the dilaton, however, can be realized. This is the pre-big-bang model \cite{ven,gv,lwc} where inflation takes place for negative times
({\it pre-big-bang phase}) and is matched to the standard radiation dominated stage for positive times ({\it post-big-bang phase}). Since in the low energy limit of superstring theory to lowest order  Einstein gravity
is recovered at cosmic time $t=0$ there is a space-time singularity, which follows from the theorems of Penrose and Hawking. Thus higher order corrections have to be included in order to regularize the transition between the pre- and post-big-bang era. In general, when calculating perturbations in  pre-big-bang inflation it is assumed that the background evolves from an asymptotically flat initial state at $t
\rightarrow-\infty$ to a high curvature phase at around $t=0$ which, however, never reaches a singular state. Only a few explicit, non-singular solutions are known and it seems difficult to determine the generic behaviour of pre-big-bang cosmologies  \cite{gamave}.
During pre-big-bang inflation in four dimensions the scale factor behaves in the string frame in which the universe is expanding and accelerating,
\begin{eqnarray}
a=a_1\left(\frac{\eta}{\eta_1}\right)^{-\frac{1}{1+\sqrt{3}}},
\end{eqnarray}
assuming the end of inflation at $\eta_1$.
The dilaton $\phi$ behaves in the low energy phase, $\eta<\eta_s$, as \cite{ggv1}
\begin{eqnarray}
\phi=-\sqrt{3}\ln|\eta| + {\rm const.}
\end{eqnarray}
After some time $\eta_s$ higher order corrections in the inverse string length $\alpha'$ become important and the universe enters into a string phase which lasts until the end of inflation at $\eta_1$.
During the string phase the dilaton evolves as \cite{ggv1}
\begin{eqnarray}
\phi=-2\beta\ln|\eta|+{\rm const.}
\hspace{2cm}\beta=-\frac{(\phi_s-\phi_1)}{2\ln z_s},
\end{eqnarray}
where $z_s\equiv a_1/a_s$ and $a_s$ and $a_1$ are the scale factors at the beginning of the string phase and at the end of inflation, respectively.

Maxwell's equations derived from the action (\ref{s1}) imply in the radiation gauge in Fourier space the mode equation for the gauge potential,
\cite{ggv1,ggv2}
\begin{eqnarray}
A_{k}''+[k^2-V(\eta)]A_k=0,\hspace{4cm} V(\eta)=g_{\rm s}(g_{\rm s}^{-1})''\,,
\end{eqnarray}
where a prime denotes the derivative with respect to conformal time $\eta$.
Matching the stage of pre-big-bang inflation to the radiation dominated era, quantizing the gauge potential and calculating the appropriate Bogoliubov coefficient results
in the following values of the ratio of magnetic energy density and background radiation energy density \cite{ggv1},
\begin{eqnarray}
r(\omega)&\simeq&\frac{g_1^2}{16\pi^2}\left(\frac{\omega}{\omega_1}\right)^{4-2\beta}
\hspace{4cm}\omega_s<\omega<\omega_1\\
r(\omega)&\simeq&\frac{g_1^2}{16\pi^2}\left(\frac{\omega}{\omega_1}\right)^{4-\sqrt{3}}z_s^{-\sqrt{3}}e^{-\Delta\phi_s}
\hspace{2.3cm}\omega<\omega_s,
\end{eqnarray}
where
$g_1$ is the value of the string coupling at the beginning of the radiation era,
$\Delta\phi_s=\phi_s-\phi_1$ and  $\omega_s\equiv\omega_1/z$.
\begin{figure}[ht]
\centerline{\epsfxsize=4in\epsfbox{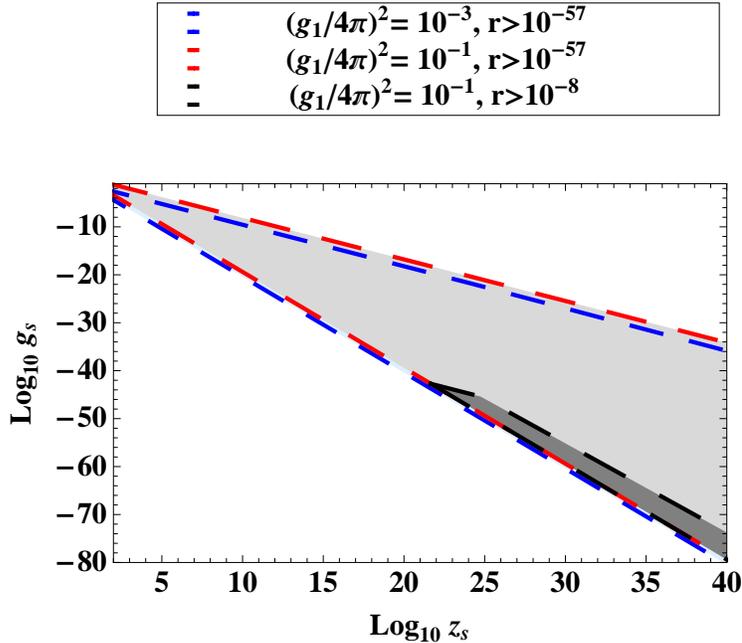}}
\caption{The logarithm of the ratio $r$, magnetic field energy density over photon energy density, is shown at galactic scale, corresponding to 1Mpc, for different values of the string coupling at the beginning of the radiation dominated era, $g_1$, and the minimal value $r_{\rm min}$ necessary to seed the galactic magnetic fields after amplification
by a galactic dynamo ($r(\omega_G)>r_{\rm min}=10^{-57}$) or without ($r(\omega_G)>r_{\rm min}=10^{-8}$).  $H_1\sim M_{Pl}$ is assumed. The shaded areas indicate the region in parameter space for which the two conditions $r(\omega_G)>r_{\rm min}$ and $r(\omega)<1$ are satisfied. The dark grey area corresponds to the domain  in  the  $z_s-g_s$ parameter space for which the galactic magnetic field could be seeded directly without the assumption of a galactic dynamo. }
\label{fig5.2}
\end{figure}
In figure \ref{fig5.2} the ratio $r(\omega)$ is plotted at galactic scale 1Mpc which corresponds to
 $\omega_G=10^{-14}$Mpc.
Imposing that $r(\omega)<1$ for all frequencies, leads to the condition \cite{ggv1,ggv2}
$z_s^{-2}<g_s/g_1$. This implies a lower bound on the value of the coupling at the beginning of the string phase, $g_s$. From figure \ref{fig5.2} it can be appreciated that
for a duration of the string phase determined by $z_s>10^{20}$ and a string coupling $g_s$ less than $10^{-42}$  the resulting magnetic field is strong enough to seed the galactic magnetic field directly.
Furthermore, figure \ref{fig5.2} shows that even for a very short string phase the resulting magnetic fields can be as strong as $10^{-30}$G which is the limiting value in case of action of a galactic dynamo in a universe with non vanishing cosmological constant.
Without taking neither the string phase nor the cosmological constant into account, corresponding to a minimal required value of $10^{-37}$  it was concluded in \cite{ll} that it is not possible to generate cosmologically relevant magnetic fields during pre-big-bang inflation.

Even though in this section we focus on mechanisms which rely on the amplification of electromagnetic perturbations during inflation, we will briefly comment on a different model of magnetogenesis in string theory.  In \cite{2009PhRvD..79h3502G} the generation of primordial magnetic fields from heterotic cosmic strings is studied. Heterotic fundamental cosmic strings were ruled out by Witten for stability reasons \cite{1985PhLB..153..243W}. However, as was shown in \cite{2004JHEP...06..013C} the presence of  branes offers a solution to the stability problem. In
\cite{2006PhRvD..74d5023B} heterotic cosmic strings are constructed by wrapping  M5 branes
around the 4-cycles of the Calabi-Yau manifold present in heterotic string theory.
In \cite{2009PhRvD..79h3502G} it was found that in a generalisation of the model of \cite{2006PhRvD..74d5023B}  the resulting heterotic strings are superconducting and as such can generate strong magnetic seed fields (see also section \ref{sM-GSM}).

\subsection{Magnetogenesis from extra dimensions}

Extra dimensions played a role in gravity ever since the proposal by Kaluza \cite{kal}  to explain gauge fields geometrically. Postulating a fifth dimension the  components of the metric involving the fifth coordinate can be interpreted as the components of the  gauge potential $A_{m}$ of electrodynamics and a scalar field $\phi$, whose effective coupling to electrodynamics in four dimensions is that of a dilaton (see, e.g.,\cite{duff}). Einstein's equations in vacuum in five dimensions imply Einstein's equations in four dimensions as well as Maxwell's equation for
the gauge potential  $A_{m}$ and the massless Klein-Gordon equation for the scalar field $\phi$,
if the dependence on the fifth coordinate is suppressed.  However, despite its successful unification of gravity and electrodynamics there is still something missing in this picture. The point is how to explain that we have not observed the fifth dimension and why there is no dependence on the extra dimension. These problems were solved by Klein \cite{klein}
assuming that the extra dimension is a circle of such a small radius that it is beyond observational limits. Using a Fourier expansion in the extra coordinate at each point in the four dimensional space-time there is an infinite number of four-dimensional fields. The zero mode  results in the original theory of Kaluza where the fields have no dependence on the extra coordinate. The remaining part of the spectrum corresponds to massive modes.

Extra dimensions appear naturally in models of string/M-theory which also admits solutions with large extra dimensions \cite{hw}. Contrary to the Kaluza-Klein picture in the case of large extra dimensions inspired by string theory, our observable four dimensional universe is described by a four-dimensional hyper surface ({\it brane}) embedded in a higher dimensional background space-time.
The cosmological solutions on the brane are influenced by the curvature of the higher dimensional space-time projected onto the brane. This leads for example to additional terms in the Friedmann equation \cite{2004LRR.....7....7M}.

In models derived from higher dimensional gravity the four dimensional Planck mass, which in this section will be denoted by  $M_4$, is no longer a fundamental parameter, but the  $D$-dimensional Planck mass $M_D$. Assuming for simplicity that all extra dimensions are of the same characteristic size $R$, the four-dimensional and the $D$-dimensional Planck masses are related by,
\begin{eqnarray}
M_4^2=R^nM_D^{n+2},
\end{eqnarray}
which follows directly using Gauss' law \cite{1998PhLB..429..263A,1998PhLB..436..257A,2001PhyU...44..871R}.
Newton gravity is observed to be valid downto length scales of the order of 1mm \cite{newt}. This can be used to put a bound on $M_D/M_4$.

Dynamical extra dimensions can generate a primordial magnetic field in four dimensions. This rests on the fact that the conformal invariance of Maxwell's equations is broken in four dimensions in the presence of higher dimensions of time-dependent size.
Assuming the metric to be of the form \cite{giov1,kek2}
\begin{eqnarray}
ds^2=-a^2(\eta)\left[d\eta^2-\delta_{\alpha\beta}dx^{\alpha}dx^{\beta}\right]
+b^2(\eta)\delta_{AB}dy^Ady^B,
\end{eqnarray}
where $\alpha,\beta=1,..,3$ and $A,B=4,..,3+n$, $n\geq 1$.
$a(\eta)$ and $b(\eta)$ are the scale factor
of the external, 3-dimensional space and the
internal, $n$-dimensional space, respectively.
Assuming that before a time $\eta=-\eta_1$ inflation takes place in the external dimensions while the extra dimensions are collapsing. After this time the universe enters the standard radiation dominated era with the extra dimensions frozen to a small size.
The first stage is described by a generalized vacuum Kasner solution. Thus the behaviour of the scale factors is determined by,
\begin{eqnarray}
a(\eta)&=&a_1\left(-\frac{\eta}{\eta_1}\right)^{\sigma},\hspace*{1.5cm}
b(\eta)=b_1\left(-\frac{\eta}{\eta_1}\right)^{\lambda},\hspace*{0.8cm}
{\rm for} \hspace*{0.5cm} \eta<-\eta_1
\label{bb1}\\
a(\eta)&=&a_1\left(\frac{\eta+2\eta_1}{\eta_1}\right),\hspace*{1.05cm}
b(\eta)=b_1, \hspace*{2.4cm}
{\rm for} \hspace*{0.5cm} \eta\geq-\eta_1\,.
\label{bb2}
\end{eqnarray}
In the following we set $a_1=1=b_1$.
The Kasner exponents $\sigma$ and $\lambda$ are given in terms of the number of extra dimensions by \cite{giov1},
\begin{eqnarray}
\sigma=-\frac{1}{2}\left(\sqrt{\frac{3n}{n+2}}-1\right),\hspace{2cm}
\lambda=\sqrt{\frac{3}{n(n+2)}}.
\end{eqnarray}
In $D$ dimensions Maxwell's equations are given by
$\nabla_{\tilde{A}}F^{\tilde{A}\tilde{B}}=0$
with $F_{\tilde{A}\tilde{B}}=\nabla_{[\tilde{A}}A_{\tilde{B}]}$,
$\tilde{A}, \tilde{B}=0,..,n+3$.
Assuming that $A_{\mu}=A_{\mu}(x^{\alpha},y^B,\eta)$ , $A_B=0$
and using the radiation gauge
$A_0=0$, $D_{\mu}A^{\mu}=0$,  Maxwell's equations imply
\begin{eqnarray}
-\frac{1}{b^n}\partial_0\left[b^n\partial_0 A_{\mu}\right]
+\sum_{\nu=1}^3\partial_{\nu}\partial_{\nu} A_{\mu}
+\left(\frac{a}{b}\right)^2\sum_{B=4}^{3+n}\partial_B
\partial_BA_{\mu}=0,
\end{eqnarray}
where $\partial_0\equiv\frac{\partial}{\partial\eta}$,
$\partial_{\mu}\equiv\frac{\partial}{\partial x^{\mu}}$ and
$\partial_B\equiv\frac{\partial}{\partial y^B}$.
Defining the canonical field $\Psi_{\mu}=b^{\frac{n}{2}}A_{\mu}$ the following expansion
is used
\begin{eqnarray}
\Psi_{\mu}(\eta,x^{\nu},y^A)=\int\frac{d^3 kd^nq}{(2\pi)^{\frac{3+n}{2}}}
\sum_{\alpha}e^{\alpha}_{\mu}({\bf l})\left[
a_{l,\alpha}\Psi_l(\eta)e^{i{\bf l}\cdot{\bf X}}
+a^{\dagger}_{-l,\alpha}\Psi^{*}_l(\eta)e^{-i{\bf l}\cdot{\bf X}}
\right],
\end{eqnarray}
where $l^{m}$ is a $(3+n)-$vector with components
$l^{\mu}\equiv k^{\mu}$, $l^A\equiv q^A$. Moreover, ${\bf l}\cdot{\bf X}=
{\bf k}\cdot{\bf x}+{\bf q}\cdot{\bf y}$.
$\alpha$ runs over the polarizations.
During inflation for $\eta<-\eta_1$ the mode equation is given by
\begin{eqnarray}
\Psi_l''+\left[k^2+\left(-\frac{\eta}{\eta_1}\right)^{2\beta}
q^2-\frac{N}{\eta^2}\right]\Psi_l=0,
\label{psi}
\end{eqnarray}
where $'\equiv\frac{\partial}{\partial\eta}$ and
$N\equiv\frac{1}{4}\left(n\lambda-1\right)^2-\frac{1}{4}$.
Moreover, $\beta\equiv\sigma-\lambda$.
$\beta<0$ since only solutions with contracting extra dimensions
are of interest here. $-1\leq\beta<-1/(1+\sqrt{3})$, where the
lower boundary corresponds to $n=1$ and the upper bound gives the
value for large $n$.
For $n=1$ and $n=6$ there are known solutions in closed form of equation (\ref{psi}).
Hence for $n=1$ the exact solution is used to find the spectrum of the primordial magnetic field and for $n>1$ approximate solutions are employed.

The solution for one extra dimension, $n=1$, is given by \cite{kek2}
\begin{eqnarray}
\Psi_l=\frac{\sqrt{\pi}}{2}e^{\frac{\pi}{2}q \eta_1}
\frac{\left(-k\eta\right)^{\frac{1}{2}}}{\sqrt{k}}
H_{i q\eta_1}^{(2)}(-k\eta),
\label{p-n1}
\end{eqnarray}
where $H_{\nu}^{(2)}(z)$ is the Hankel function of the second kind.

The approximate solution for $n>1$ is found by solving the mode equation (\ref{psi}) in two regimes determined by whether the term due to the modes $q$ in the extra dimensions, $(-\eta/\eta_1)^{2\beta}q^2$ is larger or smaller than $k^2$, where $k$ are the comoving wave numbers  in the observable three dimensional space.
When the contribution due to the wave numbers $q$ in the extra dimensions is subdominant
$\left(-\frac{\eta}{\eta_1}\right)^{2\beta}q^2<k^2$,
or $\omega_q<\omega_k$
in terms of the physical frequencies $\omega_k=k/a(\eta)$
and $\omega_q=q/b(\eta)$,
the canonical field is approximately given  by \cite{kek2},
\begin{eqnarray}
\Psi_l=\frac{\sqrt{\pi}}{2}\frac{\sqrt{-k\eta}}{\sqrt{k}}H_{\mu}^{(2)}(-k\eta),
\label{psi-mult1}
\end{eqnarray}
where $H_{\mu}^{(2)}$ is the Hankel function of the second kind and
$\mu^2\equiv\frac{1}{4}+N\Rightarrow \mu=\frac{1}{2}(n\lambda-1)$.
In the other case, that is for
$ \left(-\frac{\eta}{\eta_1}\right)^{2\beta}q^2>k^2$, or
$\omega_q>\omega_k$, it is found that \cite{kek2}
\begin{eqnarray}
\Psi_l=\frac{\sqrt{\pi}}{2}\left(-\kappa\eta\right)^{\frac{1}{2}}
H_{\mu\kappa}^{(2)}\left[\left(-q\eta\right)\kappa\left(-\frac{\eta}{\eta_1}
\right)^{\beta}\right],
\label{psi-mult2}
\end{eqnarray}
where $\kappa\equiv\frac{1}{\beta +1}$ and $\mu=\frac{1}{2}(n\lambda-1)$.

During the radiation dominated era $\eta>-\eta_1$ the mode equation is given by
\begin{eqnarray}
\Psi_l''+\left[k^2+\left(\frac{\eta+2\eta_1}{\eta_1}\right)^2q^2
\right]\Psi_l=0,
\end{eqnarray}
which is solved in terms of parabolic cylinder functions $E(\alpha,z)$ \cite{giov1},
\begin{eqnarray}
\Psi_l=\frac{1}{\sqrt{2}}\left(\frac{\eta_1}{2q}\right)^{\frac{1}{4}}
\left[c_{-}E(\alpha, z)+c_{+}E^*(\alpha, z)\right],
\label{psi-rad}
\end{eqnarray}
where $z\equiv\left(\frac{2q}{\eta_1}\right)^{\frac{1}{2}}
\left(\eta+2\eta_1\right)$ and $\alpha\equiv -\frac{\eta_1 k^2}{2q}$. Furthermore,
$|c_{\pm}|$ are the Bogoliubov coefficients satisfying the normalization
$|c_+|^2-|c_{-}|^2=1$.

The total magnetic energy density is given
by \cite{ks}
\begin{eqnarray}
\rho=2\frac{R^n}{(2\pi)^{n+3}}\int \left[\left(\frac{k}{a}
\right)^2+\left(\frac{q}{b}\right)^2\right]^{\frac{1}{2}}
|c_{-}|^2 dV,
\end{eqnarray}
where, assuming that the volume consists of two spheres,
$dV=\frac{1}{a^3b^n}\frac{2\pi^{\frac{3}{2}}}{\Gamma(\frac{3}{2})}
k^2dk\wedge\frac{2\pi^{\frac{n}{2}}}{\Gamma(\frac{n}{2})}q^{n-1}dq$.
At $\eta=-\eta_1$ the comoving wavenumbers $k$ and $q$ are equal
to the physical momenta, since $a_1=1=b_1$.
The spectral energy density $\rho(\omega_k)=d\rho/d {\rm log}\omega_k$
is then given by
\begin{eqnarray}
\rho(\omega_k)=16\frac{R^n}{(2\pi)^{n+3}}\frac{\pi^{1+\frac{n}{2}}}
{\Gamma(\frac{n}{2})}
\omega_k^{4+n}\int dY [1+Y^2]^{\frac{1}{2}}Y^{n-1}|c_{-}|^2,
\end{eqnarray}
where $Y\equiv\frac{\omega_q}{\omega_k}$, and $\omega_k=\frac{k}{a}$,
$\omega_q=\frac{q}{b}$.
To calculate the ratio of energy density in the magnetic field over the background radiation
energy density it will be assumed that $\omega_1=\frac{k_1}{a}$ and $k_1\sim H_1$ is the maximal wave number, leaving the horizon at the end of inflation and thus at end of  the dynamical higher dimensional phase  $\eta=-\eta_1$. Equally it is required that there is a maximal wave number $q_{max}$ in $q$-space corresponding to the modes in the extra dimensions.
These assumptions are justified by the sudden transition approximation used here. At the time of transition $\eta=-\eta_1$ the metric is continuous but not its first derivative.
For modes with periods much larger than the duration of the transition the transition can be treated
as instantaneous. However, in order to avoid an ultraviolet divergency an upper cut-off has to be imposed \cite{hu,gaver,gaba}. The following expressions for $r(\omega_k)$ are obtained \cite{kek2}:
\begin{enumerate}
\item For $q=0$, $n=1$
\begin{eqnarray}
r(\omega_k)\sim\frac{2}{3\pi^2}\left(\frac{H_1}{M_4}\right)^2
\left(\frac{\omega_k}{\omega_1}\right)^3
\left(\ln\frac{\omega_k}{\omega_1}\right)^2.
\label{rn1q0}
\end{eqnarray}

\item For $q>0$ and $n=1$
\begin{eqnarray}
r(\omega_k)\sim
\frac{1}{3\pi^3}\left(\frac{H_1}{M_4}\right)^3
\left(\frac{M_5}{M_4}\right)^{-3}
\left(\frac{\omega_k}{\omega_1}\right)^3
\left(\ln \frac{\omega_k}{\omega_1}\right)^2
\frac{\omega_{q_{max}}}{\omega_1},
\label{rn1q}
\end{eqnarray}
where $\omega_{q_{max}}(\eta)=\frac{q_{max}}{b}$
and it was assumed that $\omega_{q_{max}}>\omega_k$.

\item For $q=0$ and $n>1$
\begin{eqnarray}
r(\omega_k)\sim \frac{2^{n\lambda-2}}{3\pi^2}
\Gamma^2\left(\frac{n\lambda-1}{2}\right)
\left(2-n\lambda\right)^2\left(\frac{H_1}{M_4}\right)^2
\left(\frac{\omega_k}{\omega_1}\right)^{4-n\lambda}.
\label{roq0}
\end{eqnarray}
Furthermore, $n\lambda=\sqrt{\frac{3n}{n+2}}$.
Since $n\lambda<4$ the resulting spectrum
for $r(\omega_k)$ is increasing in frequency.

\item For $q>0$ and $n>1$
\begin{eqnarray}
r(\omega_k)&\sim& \cln
a^{1+2\mu\kappa-n}
\left(\frac{H_1}{M_D}\right)^{n+2}
\left(\frac{\omega_{q_{max}}}{\omega_1}\right)^{n-2\mu\kappa}
\left(\frac{\omega_k}{\omega_1}\right)^3
\label{ro}
\end{eqnarray}
where
\begin{eqnarray}
\cln&\equiv& \frac{16}{3}\frac{8\pi}{(2\pi)^{n+3}}
\frac{\pi^{1+\frac{n}{2}}}{\Gamma(\frac{n}{2})}
\frac{2^{2\mu\kappa-3}}{\pi(n-2\mu\kappa)}\Gamma^2(\mu\kappa)
\kappa^{1-2\mu\kappa}
\left(\mu-\frac{1}{2}\right)^2
\nonumber
\end{eqnarray}
where subleading terms have been omitted and
$\omega_{q_{max}}>\omega_k$ was assumed.
The resulting spectrum is growing in frequency.

\end{enumerate}

The resulting spectrum of the primordial magnetic field is characterized by
the Hubble parameter at the beginning of the radiation dominated era $H_1$, the $D$-dimensional Planck mass $M_D$ and the number of extra dimensions, $n$. In addition, in the case where the modes lying in the extra dimensions are taken into account, there is the maximal physical frequency  $\omega_{q_{max}}$ which is estimated assuming $q_{max}\sim  k_1$.
The spectrum is constrained by $r(\omega)<1$ for all frequencies.
The ratio of the $D$-dimensional over the four dimensional Planck mass is limited by the observation that Newtonian gravity is valid at least down to scales of the order of 1 mm \cite{newt}.
This leads to the lower bound
$\frac{M_D}{M_4}\geq (1.616\times 10^{-32})^{\frac{n}{n+2}}$.
Furthermore, with $T_1$ the temperature at the
beginning of the radiation epoch,
big bang nucleosynthesis requires that $T_1> 10$ MeV.
This imposes a bound on $H_1$ by using
$\frac{H_1}{M_4}=1.66 g_{*}^{\frac{1}{2}}(T_1)\left(
\frac{T_1}{M_4}\right)^2$, where for $T_1>300$ GeV the number of effective
degrees of freedom is given by $g_{*}(T_1)=106.75$ (see, e.g.,~\cite{1990eaun.book.....K}), namely, $\log\frac{H_1}{M_4}>-40.94$.

Imposing the various constraints leads to upper limits on $r(\omega_k)$ calculated at the galactic scale corresponding to 1 Mpc, that is,
$\omega_G=10^{-14}$ Hz and taking the maximally amplified frequency
evaluated today to be $\omega_1\sim 6\times 10^{11}{\rm Hz}\left(\frac{H_1}{M_{Pl}}\right)^{\frac{1}{2}}$ \cite{kek2}.
Not taking into account modes in the extra dimensions leads for one extra dimension to magnetic field strengths $B_s<10^{-39}$ G. However, taking into account these modes substantially increases the upper value of the magnetic field to upto $10^{-8}$G. Imposing the
constraint $T_1\sim M_5$ leads to magnetic seed fields $B_s<10^{-23}$ G.
In models with more than one extra dimension, $n>1$,
strong magnetic seed fields can be created if the
internal momenta are taken into account. In particular,
without the assumption that the temperature at the beginning of the
radiation epoch is of the order of the $D$-dimensional
Planck scale allows for the creation of seed magnetic fields
with strengths of upto $10^{-10}$ G. For more than three
extra dimensions, this also holds assuming $T_1\sim M_D$.
With this assumption for two and three
extra dimensions results in weaker magnetic seed fields,
with maximal field strengths, $B_s<10^{-18}$ G for two
extra dimensions and $B_s<10^{-13}$G for three extra
dimensions.

\subsection{Magnetogenesis in theories with broken Lorentz symmetry}

The spontaneous  breaking of Lorentz invariance is present in certain solutions of string field theory which leads to a non-vanishing photon mass described by the Lagrangian \cite{1999PhLB..455...96B}
\begin{eqnarray}
\cll=-\frac{1}{4}F_{mn}F^{mn}+M_L^2a^{-2\ell}A_{m}A^{m},
\end{eqnarray}
where $M_L^2\equiv\frac{m_L^2}{M_{Pl}^{2\ell}}$, $m_L$ is a light mass scale in comparison with the typical string energy scale and $2\ell$ is a positive integer. In \cite{1999PhLB..455...96B} cosmologically interesting magnetic field strengths are found for a diverse choice of parameters of the model.
In \cite{1998PhRvD..58k6002C} an extension of the standard model is presented in which due
to new physics at the Planck scale Lorentz symmetry is broken spontaneously. In the pure photon sector of the extended QED the Lagrangian is given by, \cite{1998PhRvD..58k6002C}
\begin{eqnarray}
\cll=-\frac{1}{4}F_{mn}F^{mn}-\frac{1}{4}(k_F)_{plmn}F^{pl}F^{mn}+\frac{1}{2}(k_{AF})^{p}\epsilon_{plmn}A^{l}F^{mn},
\label{lag692}
\end{eqnarray}
where the coupling $(k_F)_{plmn}$ is real and dimensionless and $(k_{AF})^{p}$ is real and has dimensions of mass.
In the context of the generation of primordial magnetic fields during inflation the Lagrangian (\ref{lag692}) has been investigated in \cite{2009PhLB..675..155C,2009PhLB..680..125C}. Analyzing the model resulting by taking into account only the first two terms in (\ref{lag692}) it has been shown in \cite{2009PhLB..675..155C} that magnetic fields of nanogauss field strength
on a megaparsec scale at present can be generated for a wide range of parameters.
In \cite{2009PhLB..680..125C} (see also, \cite{2009PhRvD..80f3006C}) primordial magnetogenesis during inflation has been discussed in a model resulting from considering only the first and third term in (\ref{lag692}). In this case the generated magnetic field is found to be maximally helical at the end of de Sitter inflation.
The subsequent inverse cascade of the magnetic field spectrum taking place in the turbulent plasma during the radiation dominated era results in a magnetic field with an interesting field strength and correlation length at the time of the protogalactic collapse.

Noncommutativity in space provides a different possibility of breaking Lorentz invariance which in this case is explicitly broken so that all the amplitudes are frame dependent. In the context of generation of primordial magnetic fields this was first discussed in \cite{2001PhRvL..87a1301M}.
Noncommutative spaces occur in string theory in the Seiberg-Witten limit \cite{1999JHEP...09..032S} and are described  by the commutation relation for the coordinate operators $\hat{x}^{m}$,
\begin{eqnarray}
[\hat{x}^{m},\hat{x}^{n}]=i\theta^{mn},
\end{eqnarray}
where $\theta^{mn}$ is a constant of dimension length which is conveniently parametrized in terms of the  noncommutativity scale $\Lambda_{NC}$, defined by
$\theta^{mn}\equiv\frac{c^{mn}}{\Lambda^2_{NC}}$ where $c^{mn}$ is an antisymmetric tensor with components of order unity  \cite{2001PhRvL..87a1301M}.
Moreover, in order to avoid problems with unitarity and causality, $\theta_{0\mu}=0$ is chosen such that only space is
noncommutative. In \cite{2000JHEP...08..045R} it was shown that the magnetic dipole moment of a charged massive particle, such as the electron, receives quantum corrections at one loop which are spin independent and proportional to $\theta_{\mu}\equiv\epsilon_{\mu\nu\kappa}\theta^{\nu\kappa}$.
This leads to a non vanishing magnetic field proportional to $\theta_{\mu}$  when summing
over all possible states. However, choosing the noncommutativity scale $\Lambda_{NC}\simeq 10^3$ GeV the authors find the resulting magnetic field to be too weak in order to successfully seed the galactic dynamo.
In \cite{2005PhRvD..71f7702G} noncommutative quantum field theory was used for the U(1) gauge field leading to a modified Lagrangian describing the photon which is of the form of the Lagrangian (\ref{lag692}) including the first and the third term. Moreover, in this case  $(k_{AF})_{m}$ is nonzero only for the spatial components and given by the noncommutativity parameter $\theta_{\mu}$.
Using the approach of \cite{2002PhRvD..66b3517B} to implement the stringy spacetime uncertainty relation which leads to an effective noncommutative space-time
\cite{2004PhRvD..70h3508B}  investigate primordial magnetogenesis in dilaton electromagnetism.
In \cite{2005PhRvD..71j3509A} the generation of primordial magnetic fields in inflation with a cut-off is investigated. The effect of the cut-off is to add extra terms to the action which in the model under consideration describes a photon with a mass term during inflation. The free parameter of the model can be chosen such that cosmologically relevant magnetic fields are obtained.

\subsection{Magnetogenesis and nonlinear electrodynamics}

So far the models in this section proposed to generate primordial magnetic fields are all situated within linear electrodynamics. In order to amplify  perturbations in the electromagnetic field during inflation the electromagnetic field is coupled to a scalar field or curvature terms,  quantum corrections resulting in the trace anomaly, symmetries are broken  or dynamical extra dimensions are taken into account.
All of these leading to the breaking of conformal invariance of Maxwell's equations in four dimensions.

Nonlinear electrodynamics provides yet another possibility of breaking conformal invariance of the  electromagnetic field. On large scales present day observations confirm the linearity  in the electric and magnetic fields of Maxwell's equations in vacuum. However, as smaller and smaller scales are approached one might expect deviations from linearity due to the fact that charges become more localized (see e.g. \cite{1975clel.book.....J}) and hence increases the energy density. This led to the hypothesis that there is some upper bound on the field strengths avoiding thus an infinite self-energy of  a charged particle.
A first example of a classical singularity-free  theory of the electron was proposed by Born and later by Born and Infeld \cite{b1,1934RSPSA.143..410B,1933Natur.132..970B,1934RSPSA.144..425B}.
The modified field equations can be derived from the Lagrangian of the form
\cite{1934RSPSA.143..410B}
\begin{eqnarray}
\cll=b^2\left(1-\sqrt{1-\left(\mathbf{E}^2-\mathbf{B}^2\right)/b^2}\right),
\label{eq691}
\end{eqnarray}
where $b$ is a maximal  field strength. In this section vector notation will be used to make expressions easier to read.
The electromagnetic field is modified at short distances and its energy density is finite. One of the problems with this type of theory is its quantization \cite{1975clel.book.....J}.
Nonlinear electromagnetism had been considered before by Mie \cite{mie}. However, it was discarded since it depended on the absolute values of the gauge potential
\cite{1934RSPSA.143..410B}.

Another place where nonlinear electrodynamics arises is in quantum electrodynamics.
Virtual electron pair creation induces a self-coupling of the electromagnetic field.
Heisenberg and Euler calculated the self-interaction energy for slowly varying, but arbitrarily strong electromagnetic fields \cite{he,schw,1970PhRvD...2.2341B}. It is described by the lagrangian \cite{he}
\begin{eqnarray}
\cll=-X+\frac{1}{8\pi^2}\int_0^{\infty}ds s^{-3}e^{-m_e^2s}\left[
(es)^2 Y \frac{Re\cosh esM}{Im \cosh es M}+1+\frac{2}{3}(es)^2X,
\right]\,,
\end{eqnarray}
where
\begin{eqnarray}
X\equiv\frac{1}{4}F_{mn}F^{mn}
\hspace{4cm}
Y\equiv\frac{1}{4}F_{mn}\,^*F^{mn}
\label{xy}
\end{eqnarray}
and $M^2\equiv2X-2iY$. Expanding the lagrangian leads to \cite{he,schw,1970PhRvD...2.2341B}
\begin{eqnarray}
L=X+\kappa_0X^2+\kappa_1Y^2.
\end{eqnarray}
This describes the Heisenberg-Euler theory for
the choice $\kappa_0=\frac{8\alpha^2}{45m_e^4}$ and
$\kappa_1=\frac{14\alpha^2}{45m_e^4}$, where $\alpha$ is the fine
structure constant and $m_e$ the electron mass.
Furthermore, the propagation of a photon in an external electromagnetic field
can be described effectively by the Heisenberg-Euler
langrangian. Moreover, the transition amplitude
for photon splitting in quantum electrodynamics is
nonvanishing in this case.
Photon splitting is a process in which  an electron-positron pair is created  and one of the particles emits a photon before annihilating with the other particle to generate the second photon. Thus
the initial one photon state transforms into a two photon final state.
In principle, this might lead to
observational effects, e.g., on the electromagnetic
radiation coming from neutron stars
which are known to have strong magnetic fields
\cite{1970PhRvD...2.2341B,1970PhRvL..25.1061A,1971AnPhy..67..599A}.
In particular, certain features in the spectra of pulsars
can be explained by photon splitting \cite{hbg,hl}.

Finally, Born-Infeld type actions
also appear as a low energy effective action of open strings
\cite{fra-tsey,tsey,gh1,gh2}.
As was shown in \cite{dbi} the low energy dynamics of D-branes
is described by the Dirac-Born-Infeld action.

To test whether nonlinear electrodynamics can lead to the generation of cosmologically relevant primordial magnetic fields the following model will be considered.
A stage of de Sitter inflation followed by reheating is matched to a standard radiation dominated era.
During inflation quantum fluctuations are excited within the horizon. Upon leaving the causal domain they become classical perturbations. It is assumed that electrodynamics is nonlinear during inflation and becomes linear once the universe enters reheating and subsequently the radiation dominated stage. This latter assumption ensures that the evolution during the radiation dominated era and subsequent stages of the universe are described by the standard model of cosmology. To study nonlinear electrodynamics in this setting was put forward in \cite{kek3} and independently in \cite{2008PhRvD..77d3001C}.

\subsubsection{Field equations}

The Born-Infeld or Heisenberg-Euler lagrangians are particular examples of theories of nonlinear electrodynamics.
In general the action of nonlinear electrodynamics coupled minimally
to gravity can be written as, see e.g. \cite{gh1,gh2,peb}
\begin{eqnarray}
S=\frac{1}{16\pi G_N}\int d^4x\sqrt{-g}R+\frac{1}{4\pi}
\int d^4x \sqrt{-g} L(X,Y),
\end{eqnarray}
where $X$ and $ Y$ are defined by equation (\ref{xy}).
Maxwell's electrodynamics corresponds to the choice $L=-X$.
The equations of motion are given by
\begin{eqnarray}
\nabla_{m}P^{mn}=0\,,
\label{p1}
\end{eqnarray}
where $P_{mn}=-\left(L_X F_{mn}+L_Y \;^{*}F_{mn}\right)$, where
the dual bi-vector
$^{*}F^{mn}$  is given by
\newline
$^{*}F^{mn}=\frac{1}{2\sqrt{-g}}\epsilon^{mnab}F_{ab}$,
and $\epsilon^{mnab}$ the Levi-Civita tensor with
$\epsilon_{0123}=+1$.
Furthermore $L_A$ denotes $L_A=\partial L/\partial A$,
and
\begin{eqnarray}
\nabla_{m} \, ^{*}F^{mn}=0,
\label{p2}
\end{eqnarray}
which implies that $F_{mn}=\partial_{m}A_{n}-\partial_{n}A_{m}$.
Assuming the background metric to be of the form,
\begin{eqnarray}
ds^2=a^2(\eta)\left[-d\eta^2+d{\mathbf x}^2\right].
\end{eqnarray}
Furthermore writing the Maxwell tensor in terms of  electric and magnetic fields
in the "lab" frame (cf. equation (\ref{feb}))
equations (\ref{p1}) and (\ref{p2}) imply \cite{kek3},
\begin{eqnarray}
\vec{D}\cdot\vec{\hat{E}}+\frac{(\vec{D} L_X)\cdot\vec{\hat{E}}}{L_X}
-\frac{(\vec{D} L_Y)\cdot\vec{\hat{B}}}{L_X}=0
\label{p3}\\
\frac{1}{a^2}\partial_{\eta}(a^2\vec{\hat{E}})-{\rm curl}\vec{\hat{B}}+\frac{\partial_{\eta}L_X}{L_X}
\vec{\hat{E}}-\frac{\partial_{\eta}L_Y}{L_X}\vec{\hat{B}}-\frac{(\vec{D} L_X)\times\vec{\hat{B}}}{L_X}
-\frac{(\vec{D} L_Y)\times\vec{\hat{E}}}{L_X}=0
\label{p4}\\
\vec{D}\cdot\vec{\hat{B}}=0
\label{p5}\\
\frac{1}{a^2}\partial_{\eta}(a^2\vec{\hat{B}})+{\rm curl} \vec{\hat{E}}=0.
\label{p6}
\end{eqnarray}
From these equations  two wave type equations can be derived which, however, contrary to the case of linear electrodynamics do not decouple the electric and magnetic field.
Taking the curl of equation (\ref{p4}) and using equations (\ref{p5}) and (\ref{p6})
a wave type equation for the magnetic field  $\vec{\hat{B}}$ can be found \cite{kek3}.
\begin{eqnarray}
& &\frac{1}{a^2}\frac{\partial^2}{\partial\eta^2}(a^2\vec{\hat{B}})
+\frac{1}{a^2}\frac{\partial_{\eta} L_X}{L_X}\partial_{\eta}(a^2\vec{\hat{B}})
+\frac{1}{a^2}\frac{\partial_{\eta}L_Y}{L_X}\partial_{\eta}(a^2\vec{\hat{E}})
+\frac{\partial_{\eta}L_Y}{L_X}\left(\frac{\partial_{\eta}L_X}{L_X}\vec{\hat{E}}-\frac{\partial_{\eta}L_Y}{L_X}
\vec{\hat{B}}\right)
\nonumber\\
&-&\D^2\vec{\hat{B}}
+\vec{\hat{E}}\times\vec{D}
\left(\frac{\partial_{\eta}L_X}{L_X}\right)
-\vec{\hat{B}}\times\vec{D}\left(\frac{\partial_{\eta}L_Y}{L_X}\right)
-\frac{\partial_{\eta}L_Y}{L_X}
\left[\frac{(\vec{D} L_X)\times\vec{\hat{B}}}{L_X}+\frac{(\vec{D} L_Y)\times\vec{\hat{E}}}{L_X}\right]
\nonumber\\
&+&{\rm curl}\left[\frac{(\vec{D} L_X)\times\vec{\hat{B}}}{L_X}\right]
+{\rm curl}\left[\frac{(\vec{D} L_Y)\times\vec{\hat{E}}}{L_X}\right]=0\,.
\label{b1}
\end{eqnarray}
Similarly, taking the time derivative of of equation (\ref{p4}) and using the
remaining equations results in
a wave type equation for the electric field $\vec{\hat{E}}$ \cite{kek3},
\begin{eqnarray}
& &\frac{\partial^2}{\partial\eta^2}\left(a^2\vec{\hat{E}}\right)
+\partial_{\eta}\left[\frac{\partial_{\eta} L_X}{L_X}a^2\vec{\hat{E}}\right]
-\partial_{\eta}\left[\frac{\partial_{\eta}L_Y}{L_X}a^2\vec{\vec{B}}\right]
\nonumber\\
&-&
D^2\left(a^2\vec{\hat{E}}\right)
-\partial_{\eta}\left[\frac{\left(\vec{D} L_X\right)\times \left(a^2\vec{\hat{B}}\right)}{L_X}\right]
-\partial_{\eta}\left[\frac{\left(\vec{D} L_Y\right)\times \left(a^2\vec{\hat{E}}\right)}{L_X}\right]
\nonumber\\
&-&\vec{D}\left[\frac{(\vec{D} L_X)\cdot(a^2\vec{\hat{E}})}{L_X}\right]
+\vec{D}\left[\frac{(\vec{D} L_Y)\cdot (a^2\vec{\hat{B}})}{L_X}\right]
=0\,.
\label{ef1}
\end{eqnarray}
In the long wavelength approximation spatial gradients can be neglected \cite{lowa}.
Thus neglecting spatial derivatives equation (\ref{b1}) implies,
\begin{eqnarray}
\Bc''+\frac{L_X'}{L_X}\Bc'+\frac{L_Y'}{L_X}\Ec'
+\frac{L_Y'}{L_X}\left[\frac{L_X'}{L_X}\Ec-\frac{L_Y'}{L_X}\Bc
\right]
=0,
\label{b2}
\end{eqnarray}
where $\Bc\equiv a^2\vec{\hat{B}}_k$, $\Ec\equiv a^2\vec{\hat{E}}_k$
and a prime denotes the derivative with respect to
conformal time $\eta$,  that is $' \equiv\frac{d}{d\eta}$.
Assuming that the lagrangian depends only on $X$, that is $L_Y=0$,
equation (\ref{b2}) implies
\begin{eqnarray}
\Bc'=\frac{\vec{K}_k}{L_X},
\label{b3}
\end{eqnarray}
where $\vec{K}_k$ is a constant vector and $L_X\neq 0$.
For $\vec{K}_k\equiv 0$ linear electrodynamics is recovered, for which
$\Bc=const.$
In the long wave length limit, the wave like equation for the electric field, equation (\ref{ef1}), yields to
\begin{eqnarray}
\frac{d}{d\eta}\left[
\Ec'+\frac{L_X'}{L_X}\Ec
-\frac{L_Y'}{L_X}\Bc\right]
\simeq 0.
\label{n1}
\end{eqnarray}
Thus integrating equation (\ref{n1}) results in
\begin{eqnarray}
\Ec'+\frac{L_X'}{L_X}\Ec
-\frac{L_Y'}{L_X}\Bc=\vec{P}_k,
\label{ec1}
\end{eqnarray}
where $\vec{P}_k$ is a constant vector.
The equations determining the magnetic and electric field,
 (\ref{b2}) and (\ref{n1}), are coupled nontrivially for $L_Y\neq 0$.
Thus in order to find solutions, the Lagrangian will be considered
to be  only a function of
$X$, $L=L(X)$.
Furthermore,  since $X=\frac{1}{2}(\vec{\hat{B}}^2-\vec{\hat{E}}^2)$ it is useful to find equations
for $\Ec^{\,2}$ and $\Bc^{\,2}$ which are given by, for $\vec{P}_k^2> 0$,
\begin{eqnarray}
\Ec^{\,2}\; ''+3\frac{L_X'}{L_X}\Ec^{\,2}\; '
+2\frac{L_X''}{L_X}\Ec^{\,2}&=&2\vec{P}^2_k
\label{E1}\\
\Bc^{\,2}\;  ''+\frac{L_X'}{L_X}\Bc^{\,2}\; '
-2\frac{\vec{K}^2_k}{L_X^2}&=&0.
\label{B1}
\end{eqnarray}
Assuming that the constant vector in equation (\ref{ec1}) vanishes, $\vec{P}_k=0$, leads to a significant  simplification. In this case, equation ({\ref{ec1}) for $L=L(X)$ can be solved immediately,
giving for the electric field
\begin{eqnarray}
\Ec=\frac{\vec{M}_k}{L_X},
\label{ecm1}
\end{eqnarray}
where $\vec{M}_k$ is a constant vector.
Thus for $\vec{P}_k=0$
equation (\ref{B1}) leads to an equation only
involving $X$ and $L_X$, namely,
\begin{eqnarray}
\frac{d^2}{d\eta^2}\left[2a^4X+\frac{\vec{M}^2_k}{L_X^2}\right]
+\frac{1}{L_X}\frac{dL_X}{d\eta}\frac{d}{d\eta}
\left[2a^4X+\frac{\vec{M}_k^2}{L_X^2}\right]-2\frac{\vec{K}_k^2}{L_X^2}=0.
\label{b4}
\end{eqnarray}

\subsubsection{A particular model}

In order to find explicit solutions of  equation (\ref{b4}) a particular lagrangian has to be chosen.
For simplicity the lagrangian is chosen to be of the
form
\begin{eqnarray}
L=-\left(\frac{X^2}{\Lambda^8}\right)^{\frac{\delta-1}{2}}X,
\label{L}
\end{eqnarray}
where $\delta$ is a dimensionless parameter and $\Lambda$ a dimensional
constant. This is the abelian Pagels-Tomboulis model \cite{2001AcPPB..32.2155A,sw}.
An effective model of low energy QCD is provided by its nonabelian version \cite{pt}. Clearly,
linear electrodynamics is
recovered for the choice $\delta=1$.
The lagrangian (\ref{L}) is chosen since it leads to a significant simplification of the equations,
but still allows to study the effects of a strongly nonlinear theory
of electrodynamics on the generation of primordial magnetic fields.
In general, the energy-momentum tensor derived from a lagrangian
$L(X)$ is given by
\begin{eqnarray}
T_{mn}=\frac{1}{4\pi}\left[L_Xg^{ab}F_{ma}
F_{bn}+g_{mn}L\right].
\label{T-PT}
\end{eqnarray}
Furthermore, for the lagrangian (\ref{L}) the trace of the energy-momentum tensor is given by
\begin{eqnarray}
T=\frac{1-\delta}{\pi}L,
\end{eqnarray}
which vanishes only in the case $\delta=1$ that is for
linear electrodynamics.
The energy-momentum tensor is calculated explicitly to check whether there
are any constraints on the parameter $\delta$.
Decomposing the Maxwell tensor with respect to a fundamental
observer  with 4-velocity $u_{m}$ into an electric field $\vec{E}$ and a magnetic field
$\vec{B}$, implies \cite{ellis,1997CQGra..14.2539T,1998CQGra..15.3523T,2007PhR...449..131B},
\begin{eqnarray}
F_{mn}=2E_{[m}u_{n]}-\eta_{mnks}u^{k}\;
B^{s},
\label{Fmunu}
\end{eqnarray}
where $\eta_{mnks}=\sqrt{-g}\epsilon_{mnks}$ and
$u_{m}u^{m}=-1$.
Thus the electric and magnetic field are given, respectively, by
$E_{m}=F_{nm}u^{n}$ and $B_{m}=\frac{1}{2}\eta_{mnkl}
u^{n}F^{kl}$.
The lab frame is defined by the proper lab coordinates $(t,\vec{r})$ determined by
$dt=ad\eta$, $d\vec{r}=ad\vec{x}$. Applying a coordinate transformation then
gives the relation between the fields measured by a fundamental
observer and the lab frame. Using the four velocity of the fluid
$u^{m}=(a^{-1},0,0,0)$ results in  the relation  \cite{1998PhRvD..58h3502S}
\begin{eqnarray}
\hat{E}_{\mu}=aE_{\mu},\hspace{2cm} \hat{B}_{\mu}=aB_{\mu}.
\end{eqnarray}
As shown in \cite{ellis,1997CQGra..14.2539T,1998CQGra..15.3523T,2007PhR...449..131B} the energy-momentum tensor of an electromagnetic
field can be cast into the form of an imperfect fluid.
The energy-momentum tensor of an imperfect fluid is of the form (see for example, \cite{ellis,1997CQGra..14.2539T,1998CQGra..15.3523T,2007PhR...449..131B}),
\begin{eqnarray}
T_{mn}=\rho u_{m}u_{n}+ph_{mn}+2q_{(m}u_{n)}+\pi_{mn},
\end{eqnarray}
where $\rho$ is the energy density, $p$ the pressure, $q_{m}$ the heat flux vector and
$\pi_{mn}$ an anisotropic pressure contribution of the fluid.
$h_{mn}=g_{mn}+u_{m}u_{n}$ is the metric on the space-like hypersurfaces
orthogonal to $u_{m}$.
With $q_{m}u^{m}=0$ and $\pi_{mn}u^{m}=0$,
\begin{eqnarray}
\rho&=&T_{mn}u^{m}u^{n}\hspace{2cm}
q_{a}=-T_{mn}u^{m}h^{n}_{a}\nonumber\\
Q_{ab}&\equiv &T_{mn}h^{m}_{a}h^{n}_{b}
\hspace{2cm}
Q_{ab}=ph_{ab}+\pi_{ab}.
\end{eqnarray}
Therefore using equations (\ref{T-PT}) and (\ref{Fmunu})  the energy density and the heat flux vector for the Pagels-Tomboulis model (\ref{L}) are found to be
\begin{eqnarray}
\rho&=&-\frac{1}{8\pi}\frac{L}{X}\left[\left(2\delta-1\right){E}_{a}
E^{a}+{B}_{a}
B^{a}\right]
\label{rho}\\
q_{a}&=&\frac{\delta}{4\pi}\frac{L}{X}\eta_{abmn}u^{b}
E^{m}B^{n}.
\end{eqnarray}
Imposing the condition that $\pi_{ab}$ is trace-free
then the pressure and $\pi_{ab}$ are given by
\begin{eqnarray}
p&=&\frac{1}{3}\rho-\frac{\delta-1}{3\pi}L\\
\pi_{ab}&=&-\frac{\delta}{4\pi}\frac{L}{X}\left[
\frac{1}{3}\left({E}_{m}{E}^{m}+{B}_{m}{B}^{m}\right)h_{ab}
-\left({E}_{a}{E}_{b}
+{B}_{a}{B}_{b}\right)\right].
\end{eqnarray}
Thus considering $\rho$ (cf. equation (\ref{rho})) in general there is a constraint
on $\delta$ which is $\delta\geq\frac{1}{2}$ required by
the positivity of the energy density $\rho$.

\subsubsection{Estimating the primordial magnetic field strength}

During de Sitter inflation electrodynamic is nonlinear and described by the Pagels-Tomboulis lagrangian (\ref{L}). Thus in the very early universe electrodynamics is highly nonlinear and very different from standard Maxwell electrodynamics.
At the end of inflation electrodynamics is assumed to become linear so that the description
of reheating and the subsequent radiation dominated stage are unaltered.

Recalling that the scale factor during de Sitter is given by
$
a(\eta)=a_1\left(\frac{\eta}{\eta_1}\right)^{-1},
$
where $\eta\leq\eta_1<0$.
The end of inflation is assumed to be at $\eta=\eta_1$.
The equations determining the electric and magnetic field in the long wave length limit, (\ref{E1}) and (\ref{B1}),  are coupled, since $X$ depends on $\vec{\hat{E}}^2$ and
$\vec{\hat{B}}^2$, in particular the invariant $X$ reads,
$2a^4X\simeq\Bc^{\,2}-\Ec^{\,2}$.
Therefore to find approximate solutions three different regimes will be considered.
\begin{enumerate}
\item $\Bc^{\,2}\simeq{\mathcal O}(\Ec^{\,2})$.
\item $\Bc^{\,2}\gg \Ec^{\,2}$. This implies the approximation
$2 a^4X\simeq \Bc^{\,2}$.
\item $\Ec^{\,2}\gg \Bc^{\,2}$. This implies the approximation
$2 a^4X\simeq -\Ec^{\,2}$.
\end{enumerate}
Following \cite{1988PhRvD..37.2743T} it is assumed that quantum fluctuations in the electromagnetic field
lead to initial electric and magnetic fields. The energy density at the time of first horizon crossing during inflation, say at a time $\eta_2$, is estimated to be of the order of the energy density of a thermal bath at the Gibbons-Hawking temperature of de Sitter space.
Furthermore it is useful to recall
the energy density in the magnetic field at the time of first horizon crossing corresponding to a value of the scale factor $a_2$ given by
$
\rho_{\rm B}(a_2)\simeq H^4\simeq\left(\frac{M^4}{M_{Pl}^2}\right)^2,
$
where $M^4$ is the constant energy density during inflation.
Assuming that initially the magnetic and electric energy densities are of the same order,
there is an equivalent expression for the electric energy density at the first horizon crossing.
While after the end of inflation, during the radiation dominated epoque, the electric field rapidly decays due to plasma effects, the magnetic field remains frozen-in.

\vspace{1cm}
\begin{enumerate}
\item $ \Bc^{\,2}\simeq{\mathcal O}(\Ec^{\,2})$:

In this case equation (\ref{b4}) can be approximately solved and leads to the magnetic field strength at the end of inflation corresponding to the value of the scale factor $a_1$ determined by
\cite{kek3},
\begin{eqnarray}
\frac{B^2_k(a_1)}{B^2_k(a_2)}\simeq e^{-4N(\lambda)}\frac{\cosh^2[m(x_1+(\delta-1)C_1)]}{\cosh^2[m(x_2+(\delta-1)C_1)]},
\end{eqnarray}
where $N(\lambda)$ is the number of e-folds before the end of inflation
at which $\lambda$ left the horizon, that is, $e^{N(\lambda)}=a_1/a_2$.
Moreover, $m\equiv\frac{|\vec{K}_k|}{M_{Pl}|\vec{M}_k|}$ and $x\equiv\frac{\eta}{M_{Pl}^{-1}}$.

Furthermore, the constant $C_1$ is chosen such that
$(\delta-1)C_1=-x_2$. Using that during de Sitter inflation, $a=a_1(\eta_1/\eta)$
and the number of e-folds, results in the magnetic energy density $\rho_B$ at the end of inflation,
\begin{eqnarray}
\rho_B(a_1)\simeq\rho_B(a_2)e^{-4N(\lambda)}\cosh^2[-mx_1(e^{N(\lambda)}-1)],
\end{eqnarray}
where $\rho_B=\frac{B^2}{8\pi}$.
Together with the expression for the number of $e$-folds remaining until the end of inflation after the comoving scale $\lambda$ has crossed the horizon during inflation (cf. equation (\ref{N})),
this results in the ratio of magnetic to background radiation energy density $r$ at the end of inflation \cite{kek3},
\begin{eqnarray}
r(a_1)&\simeq& 10^{-104}\left(\frac{\lambda}{\rm Mpc}\right)^{-4}
\left(\frac{M}{T_{RH}}\right)^{\frac{10}{3}}\nonumber\\
&&\times
\cosh^2\left[-9.2\times10^{25}\left(\frac{\lambda}{\rm Mpc}\right)
\left(\frac{M}{M_{Pl}}\right)^{\frac{2}{3}}\left(\frac{T_{RH}}{M_{Pl}}
\right)^{\frac{1}{3}}mx_1\right].
\label{r1}
\end{eqnarray}
There are bounds on the parameter $m$ coming from the requirements that $r$ should be larger than a  lower  value in order to be strong enough to seed the galactic magnetic field and smaller than an upper bound coming from the fact that $r$ has be less than one.
Therefore assuming $\lambda=1$ Mpc
and the inflationary energy scale $M=10^{17}$ GeV,
the following values for $-mx_1$ are found \cite{kek3}.
For a model with reheat temperature $T_{RH}=10^9$ GeV \cite{1988PhRvD..37.2743T}
the parameter $-mx_1$ has to be in the range $2.7\times 10^{-20}<-mx_1<5\times 10^{-20}$
in order to achieve a magnetic seed field with a field strength to be
at least $B_s\simeq 10^{-20}$G,
corresponding to $r_0=10^{-37}$. For a higher reheat temperature $T_{RH}=10^{17}$GeV \cite{1988PhRvD..37.2743T},
for the same magnetic seed field strength $-mx_1$ has to be in the range,
$9.5\times 10^{-23}<-mx_1<1.5\times 10^{-22}$.
And similarly, for the less conservative bound $r_0=10^{-57}$, for
$T_{RH}=10^9$ GeV, $-mx_1$ has to be in the range
$1.4\times 10^{-20}<-mx_1<5\times 10^{-20}$ and for
$T_{RH}=10^{17}$ GeV it is found that
$6.7\times 10^{-23}<-mx_1<1.5\times 10^{-22}$.

\vspace{0.5cm}

\item $  \Bc^{\,2}\gg \Ec^{\,2}$

For $\delta\neq \frac{1}{2}$ and $\delta\neq\frac{5}{4}$ the $r$ is found to be \cite{kek3}
\begin{eqnarray}
r(a_1)\simeq \left(9.2\times 10^{25}\right)^{-\frac{6}{2\delta-1}}
\left(\frac{\lambda}{\rm Mpc}\right)^{-\frac{6}{2\delta-1}}\left(
\frac{M}{M_{Pl}}\right)^{2\frac{6\delta-5}{2\delta-1}}
\left(\frac{T_{RH}}{M_{Pl}}\right)^{-\frac{4\delta}{2\delta-1}}.
\label{r_1}
\end{eqnarray}
For $\vec{P}_k^2>0$, solutions consistent with the assumption used in the approximation yield
cosmologically interesting magnetic fields  for
$\delta>1.9$ for $T_{RH}=10^{17}$ GeV and
for $\delta>3.0$ for $T_{RH}=10^9$ GeV. In these cases  the ratio of the
energy density of the magnetic field over
the energy density of the background radiation r is found to be
$r>10^{-37}$ corresponding to a primordial magnetic
field of at least $B_s=10^{-20}$G.

In the cases $\delta=\frac{1}{2}$ and $\delta=\frac{5}{4}$
the solutions show  a behaviour different from the power-law solutions for $X$ used in the former case (cf. equation (\ref{r_1})). On the one hand, for $\delta=\frac{5}{4}$ an implicit solution depending on Euler's $\beta$ function can be found which however makes it difficult to estimate explicitly the magnetic field strength at the end of inflation.
On the other hand, for $\delta=\frac{1}{2}$ a solution depending on simple functions can be found, namely the ratio $r$ at the end of inflation is given by, \cite{kek3}
\begin{eqnarray}
r(a_1)&\simeq& 10^{-104}\left(\frac{\lambda}{{\rm Mpc}}\right)^{-4}
\left(\frac{M}{T_{RH}}\right)^{\frac{10}{3}}
\nonumber\\
&&\times
\cosh^2\left[
-8\times 10^{77}x_1\left(\frac{\alpha_2}{18}\right)^{\frac{1}{2}}\left(\frac{\lambda}{{\rm Mpc}}\right)^3
\left(\frac{M}{M_{Pl}}\right)^2\frac{T_{RH}}{M_{Pl}}\right].
\label{r_1_2}
\end{eqnarray}
Here the constant $c_1$ has been chosen as $c_1\equiv-\left(\frac{\alpha_2}{18}\right)^{\frac{1}{2}}
x_1\left(\frac{\eta_2}{\eta_1}\right)^3$. Furthermore, $\alpha_2=\frac{4\vec{K}_k^2}{M_{Pl}^2\Lambda^4a_1^4}$.
 In this case the solutions found for the electric and magnetic field
are consistent with the approximation for $\vec{P}_k^2>0$ and
$\vec{P}_k^2=0$. Moreover, the resulting magnetic field is strong
enough to seed the galactic dynamo \cite{kek3}.

\vspace{0.5cm}

\item  $\Ec^{\,2}\gg \Bc^{\,2}$

In this case the  approximation implies that at the end of inflation the electric energy density is much larger than the magnetic energy density.

The ratio $r$ at the end of inflation is given by \cite{kek3}
\begin{eqnarray}
r(a_1)\simeq\left(9.2\times 10^{25}\right)^{-\beta}
\left(\frac{\lambda}{{\rm Mpc}}\right)^{-\beta}
\left(\frac{M}{M_{Pl}}\right)^{6-\frac{2\beta}{3}}
\left(\frac{T_{RH}}{M_{Pl}}\right)^{-2-\frac{\beta}{3}}.
\label{r_2}
\end{eqnarray}
For $\vec{P}_k^2>0$ the resulting magnetic fields very weak magnetic fields and $r(a_1)\ll 10^{-37}$ for typical values of the cosmlogical parameters.
However, in the case $\vec{P}_k^2=0$, for $\delta>19.5$ and a
reheat temperature $T_{RH}=10^9$ GeV  primordial magnetic fields
result which could successfully act as seed fields for
the galactic dynamo.

\end{enumerate}

In summary, using the Pagels-Tomboulis model an example of a theory of nonlinear electrodynamics has been provided which can lead during inflation to sufficient amplification of perturbations in the electromagnetic field in order to seed the galactic magnetic field.

In \cite{2008PhRvD..77d3001C} the resulting magnetic field  generated during inflation is estimated for Lagrangians of the form $\cll=\cll(X)$. The magnetic field strength is obtained by neglecting the magnetic field contribution to $X$ on superhorizon scales since equation (\ref{p6}) implies $B_{\mu}\sim k\eta E_{\mu}$ on these scales. Moreover, neglecting the spatial gradient terms in  equation (\ref{p4}), \cite{2008PhRvD..77d3001C} find the scaling
\begin{eqnarray}
\left({\mathcal L}_X\right)^2X\propto a^{-4}.
\end{eqnarray}
Using this relation for Lagrangians of the form
$
\cll=-X+\sum_{i=2}^n c_iX^i
$
and $\cll=-X\exp\left(-cX\right)$, where $c_i$ and $c$ are constants, and the Born-Infeld Lagrangian (\ref{eq691}), the resulting magnetic field strength at present is estimated. There is a range of parameters for which magnetic fields strong enough to directly seed the galactic magnetic field can be generated.

In \cite{2009PhRvD..80b3013M} the generation and evolution of primordial magnetic fields has been discussed during de Sitter inflation, reheating and the radiation dominated era in theories of  nonlinear electrodynamics described by Lagrangian densities $\cll\sim X+\gamma X^{\delta}$ and $\cll\sim X+\mu^8/X$, where $\gamma$, $\delta$ and $\mu$ are constants. It was found that not only primordial magnetic fields of interesting field strengths can be generated but also the baryon asymmetry by gravitational coupling between the baryon current and the curvature of the background. In  \cite{2010arXiv1007.3299V} primordial magnetic field generation has been discussed in DBI inflation.

\newpage

\section{Summary and outlook}\label{sSO}
Although the origin of cosmic magnetism is still the subject of debate, the ubiquitous presence of large-scale $B$-fields with similar (of $\mu$G order) strengths in galaxies, galaxy clusters and high-redshift protogalactic structures seems to suggest a common, primordial origin for them. Very recent reports indicating the presence of extragalactic magnetic fields close to $10^{-15}$~G in low density regions of the universe also point towards the same direction.

The possibility of a cosmological (pre-recombination) origin for all the large-scale magnetic fields is a relatively old suggestion and there have been numerous studies looking at the generation, the evolution and the potential implications of such primeval fields. There are still serious difficulties to overcome, however, especially when trying to produce the initial $B$-fields that will seed the galactic dynamo. Primordial magnetogenesis is still not a problem-free exercise, which probably explains the plethora of mechanisms proposed in the literature. Roughly speaking, magnetic seeds produced between inflation and recombination are too small in size, while those generated during inflation are generally too weak in strength. In either case, the galactic dynamo will not be able to operate successfully. The former of the aforementioned two problems is essentially due to causality, which severely constrains the coherence length of almost every $B$-field produced during the radiation era. The latter problem is attributed to the dramatic depletion suffered by typical inflationary magnetic fields. Primordial turbulence and magnetic-helicity conservation can in principle increase the initial coherence scale of magnetic seeds, especially of those generated during phase-transitions in the early radiation era. Considerable effort has also been invested in the search for viable physical mechanisms that could amplify weak inflationary magnetic fields. Solutions to the magnetic strength problem are typically sought outside the realm of classical electromagnetism and/or that of standard cosmology, although conventional amplification mechanisms can also be found in the literature. The aim of this review is to provide an up-to-date and as inclusive as possible discussion on the current state of primordial magnetogenesis.

Deciding whether the large-scale magnetic fields that we observe in the universe today are of cosmological origin, or not, would be a step of major importance for cosmology. If confirmed, such primordial fields could have affected in a variety of ways a number of physical processes that took place during the early, as well as the subsequent, evolution of the cosmos. Although the argument in favour of cosmological $B$-fields may not settle unless an unequivocal magnetic signature is detected in the CMB, their case gets stronger as more reports of magnetic fields at high-redshifts and in empty intergalactic space appear in the literature. Upcoming observations may also help in this respect. A new generation of radio telescopes, like the Expanded Very Large Array (EVLA), the Low Frequency Array (LOFAR), the Long Wavelength Array (LWA) and the Square Kilometer Array (SKA) have large-scale magnetic fields in their lists of main targets. If nothing else, the expected influx of new data should put extra constrains that may allow us to distinguish between the various scenarios of magnetic generation and evolution. Information of different type, but of analogous importance, may also come from CMB observations, like those associated with the ESA PLANCK satellite. At the same time, structure-formation simulations are becoming more sophisticated by the day and a number of research groups have started systematically incorporating magnetic fields into them. This in turn should help us understand and interpret better the non-thermal regime of galaxy formation. So, hopefully, we will soon have cosmological and structure formation models with fewer free parameters and more physics.\\

{\bf Acknowledgements:} A.K. wishes to thank the Brazilian Agency CNPq for financial support through project CNPq/471254/2008-8. K.E.K. is partially supported by Spanish Science Ministry grants FIS2009-07238, FPA2009-10612 and CSD2007-00042 and is grateful to the TH Division at CERN for their hospitality. The authors would also like to thank Amjad Ashoorioon, Kazuharu Bamba, Robert Brandenberger, Leonardo Campanelli, Hassan Firouzjani, Massimo Giovannini, Angela Lepidi, Dominik Schleicher and Lorenzo Sorbo for their comments and suggestions.

\newpage




\bibliography{ref-sec1,ref-sec2&3&4,ref-sec5,ref-sec6}
\bibliographystyle{h-elsevier}

\end{document}